%% file: main.tex
\documentclass{aa}
\usepackage{graphicx,txfonts,float,footmisc,natbib,twoopt} %,amsmath}
\usepackage[english]{babel}
\usepackage[dvipsnames]{xcolor}
\usepackage[colorlinks=true,linkcolor=Blue,citecolor=Blue]{hyperref}

\begin{document} 

%### Title
%%%%%%%%%%%%%%%%%%%%%%%%%%%%%%%%%%%%%%%%
\title{SPH modelling of wind-companion interactions in eccentric AGB binary systems}
% \thanks{Send offprint requests to jolien.malfait@kuleuven.be}

\author{J. Malfait \inst{1} \and  W. Homan \inst{2,1} \and S. Maes \inst{1} \and J. Bolte \inst{1} \and L. Siess \inst{2} \and F. De Ceuster \inst{3,1} \and L. Decin \inst{1,4}}

\institute{ Institute of Astronomy, KU Leuven, Celestijnenlaan 200D, 3001 Leuven, Belgium \and Institut d’Astronomie et d’Astrophysique, Universit\'e Libre de Bruxelles (ULB), CP 226, 1050 Brussels, Belgium
\and Department of Physics and Astronomy, University College London, Gower Place, London, WC1E 6BT, United Kingdom
\and School of Chemistry, University of Leeds, Leeds LS2 9JT, United Kingdom
}

\date{Received /
Accepted }

%### Abstract
%%%%%%%%%%%%%%%%%%%%%%%%%%%%%%%%%%%%%%%%
% \abstract{}{}{}{}{}
% 5 {} token are mandatory
\input{abstract}


\maketitle

\authorrunning{Malfait et al.}

%### Individual chapters
%%%%%%%%%%%%%%%%%%%%%%%%%%%%%%%%%%%%%%%%
%\listoftodos
\input{introduction}

\input{method}

\input{results}

\input{discussion}
\input{conclusion}

%### Acknowledgements
%###################################################

\begin{acknowledgements}
J.M., W.H., S.M., J.B., F.D.C. and L.D. acknowledge support from the ERC consolidator grant 646758 AEROSOL. Also, W.H. acknowledges support from the Fonds de la Recherche Scientifique (FNRS) through grant 40000307, J.B. acknowledges support from the KU Leuven C1 grant MAESTRO
C16/17/007, F.D.C. is supported by the EPSRC iCASE studentship programme, Intel Corporation and Cray Inc, and LS is senior FNRS researcher.
\end{acknowledgements}

%### Bibliography
%###################################################

\bibliographystyle{aa}
\bibliography{references.bib}

\newpage
\newpage

\input{appendix}

\end{document}

%% file: abstract.tex
\abstract
	% context heading (optional)
	% {} leave it empty if necessary  
	{The late evolutionary stages of low- and intermediate-mass stars are characterised by mass loss through a dust-driven stellar wind. Recent observations reveal complex structures within these winds, that are believed to be formed primarily via interaction with a companion. How these complexities arise, and which structures are formed in which type of systems, is still poorly understood. Particularly, there is a lack of studies investigating the structure formation in eccentric systems.}
	% aims heading (mandatory)
	{We aim to improve our understanding of the wind morphology of eccentric AGB binary systems by investigating the mechanism responsible for the different small-scale structures and global morphologies that arise in a polytropic wind with different velocities.}
	% methods heading (mandatory)	
	{Using the smoothed particle hydrodynamics (SPH) code \textsc{Phantom}, we generate nine different high-resolution, 3D simulations of an AGB star with a solar-mass companion with various wind velocity and eccentricity combinations. The models assume a polytropic gas, with no additional cooling. }
	% results heading (mandatory)
	{Compared to the zero-eccentricity situation, we find that for low eccentricities, for the case of a high wind velocity, and hence limited interaction between the wind and the companion, the standard two-edged spiral structure that dominates the shape of the wind in the orbital plane is only minimally affected. When the wind speed is lower, strong compression of the wind material by the companion occurs, causing a high-pressure region around the companion which shapes the wind into an irregular spiral. In extreme cases, with low wind velocity and high eccentricity, these instabilities grow to such proportion that they cause high-speed ejections of matter along the orbital plane, shaping the wind into a highly irregular morphology. In more eccentric orbits the amplitude of the phase-dependent wind-companion interaction increases significantly, introducing additional complexities that make the outbursts even more energetic, leading in some cases to high-speed polar flows of matter. Further, the orbital motion of the stars tends to flatten the global density distribution of the models with no instabilities. We distinguish global flattening from an equatorial density enhancement, the latter being formed by a strong gravitational interaction of the companion with the wind particles.
	We classify the resulting morphologies according to these new definitions, and find that (i) all low-velocity models have an equatorial density enhancement, and (ii) in general the flattening increases for decreasing wind velocity, until the low wind velocity results in high-energy outflows that clear away the flattening.}
	% conclusions heading (optional), leave it empty if necessary 
    { We conclude that for models with a high wind velocity, the short interaction with the companion results in a regular spiral morphology, that is flattened. In the case of a lower wind velocity, the stronger interaction results in the formation of a high-energy region and bow-shock structure that can shape the wind into an irregular morphology if instabilities arise. High-eccentricity models show a complex, phase-dependent interaction leading to wind structures that are irregular in three dimensions. However, the significant interaction with the companion compresses matter into an equatorial density enhancement, irrespective of eccentricity.}

%% file: introduction.tex
\section{Introduction}
\label{ch:intro}
Low- and intermediate-mass stars with masses between ${\sim}0.8$ and $8\,{\rm M_{\odot} }$ enter the asymptotic giant branch (AGB) phase near the end of their nuclear burning cycles. During this evolutionary stage, the star loses material from its extended envelope through a stellar wind at rates ranging from $10^{-8}$ to $10^{-5}\,{\rm{M_\odot\,yr^{-1}}}$ with terminal wind velocities between $5 - 30\,{\rm{km\,s^{-1}}}$ \citep{habing2004,Ramstedt2008}. These winds are generally believed to be driven by radiation pressure on dust grains that are formed in the cool outer atmosphere, where the atmospheric density scale height is increased due to the levitation of material by stellar pulsations \citep{Lamers1999,Hofner2018}.
These dust-driven pulsation-enhanced winds are key chemical laboratories and contribute to the chemical enrichment of ${\sim}85\%$ of the gas and ${\sim}35\%$ of the dust of the insterstellar medium \citep{Tielens2005}.
In order to obtain a correct understanding of the chemical evolution of the galaxy, it is of prime importance to achieve a good insight in the mechanics and properties of these stellar outflows. 

When up to $80 \%$ of the star's mass is expelled, it rapidly evolves through the post-AGB phase towards a planetary nebula (PN), where the expelled wind material is ionised by its stellar radiation. Observations of these post-AGB stars and PNe reveal complex morphologies, with only about $3.4 \%$ of all PNe possessing a circular symmetry \citep{Sahai2011}. 
In contrast, the overall geometry of about $80\%$ of the AGB outflows is found to be spherical \citep{Neri1998}. High-resolution observations performed in the last decade reveal that mid- to small-scale structures such as bipolarity, spirals, disks, arcs, etc. arise in the outflows of AGB stars \citep[e.g.][]{Ramstedt2014, Kervella2016, Decin2020,Homan2020(1),Homan2020(2)}. 

Over the last decades, particle-based (smoothed particle hydrodynamics; SPH) and grid-based (adaptive mesh refinement; AMR) 3D simulations have indicated that several of these structures can be formed in the AGB outflow when considering the gravitational interaction with a binary companion. The earliest of these studies by \cite{TheunsI,TheunsII} and \cite{MastrodemosI,MastrodemosII} already revealed that spirals, arcs, and accretion disks are naturally formed in the winds of such binary systems. Depending on the properties of the binary system and the AGB wind, various wind morphologies can result such as roughly spherical, flattened or bipolar morphologies \citep{MastrodemosII,Kim2012B}.
The hypothesis that many of the complexities in the AGB outflows result from the interaction with a companion star or planet is supported by binary population statistics \citep{Moe2017,Fulton2018}. Moreover, recently, \cite{Decin2020} confirmed with observations that binary companions are the shaping agents of AGB winds.

By the end of the AGB phase, tidal forces are expected to have circularized binary AGB systems with orbital periods shorter than a few thousand days \citep{Pols2004,Izzard2010,Saladino2019B}.
However, over $70 \%$ of the post-AGB binaries reside in eccentric orbits \citep{Oomen2018} and recent observations indicate that the eccentricity may be high already during the AGB phase \citep{Kim2015,Ramstedt2017}. Several mechanisms pumping the eccentricity and counteracting this circularisation have been proposed, such as tidal coupling of the binary to the circumbinary disk \citep{Dermine2013} and phase-dependent mass loss \citep{Soker2000}, but without a clear consensus. 
In the literature, only a few theoretical studies account for wind structures formed in eccentric binary systems \citep{Kim2017,Saladino2019B}. In order to improve our understanding of the impact of eccentricity on AGB outflows, we present and analyse the first set of 3D high-resolution hydrodynamic models of an AGB star with a solar-mass companion in orbits with different eccentricities.

In Section \ref{ch:method} we introduce the numerical setup and define the binary parameters of each model. Section \ref{ch:results} presents the wind structures and morphologies resulting from the various parameter combinations and a description of their formation mechanisms. In Section \ref{ch:discussion}, a classification is given of the global density distribution of the models according to their deviation from spherical symmetry. The main results are n in Section \ref{ch:conclusion}.

%% file: method.tex
\section{Method}
\label{ch:method}
\subsection{Numerical setup}
For the simulations, the same numerical setup is applied as described by \cite{Maes2021} and is shortly summarised here. The binary AGB systems are modelled with the 3D smoothed particle hydrodynamics (SPH) code \textsc{Phantom} \citep{phantom}. SPH codes use a Lagrangian method to solve the hydrodynamic Euler equations: 
\begin{eqnarray}
    \frac{{\rm d} \rho}{{\rm d} t} &=& -\rho (\vec{\nabla} \cdot \textit{\textbf{v}})  \quad {\rm  (continuity \ equation) } \\
    \label{consMom}
    \frac{{\rm d} \textit{\textbf{v}}}{{\rm d} t}& =& - \frac{\vec{\nabla} P}{\rho} + \textit{\textbf{a}}   \quad {\rm  (conservation \ of \  momentum)} \\
    \label{consEn}
    \frac{{\rm d} u}{{\rm d} t} &=&  -\frac{P}{\rho} (\vec{\nabla} \cdot \textit{\textbf{v}}) + \Lambda  \quad {\rm  (conservation \ of \ energy)}
\end{eqnarray}
where $\rho$ is the gas density, $\textit{\textbf{v}}$ the velocity, $P$ the pressure and $u$ the specific internal energy. The acceleration $\textit{\textbf{a}}$ denotes all accelerations from external forces and the parameter $\Lambda$ represents all the system's energy losses and gains. To close the above set of equations, the equation of state (EOS) for an ideal gas is used, given by
\begin{equation}
    \label{IdealGas}
    P = (\gamma - 1) \rho u ,
\end{equation}
with polytropic index $\gamma$.
Similar to \cite{Maes2021}, our wind thermodynamics are governed by a polytropic equation of state with an index of $\gamma=1.2$. It can be shown that this choice of $\gamma$ generates a temperature power-law of $T(r) \propto r^{-\epsilon}$ with $\epsilon \approx 0.6 - 0.7$, a typical value derived for AGB stellar winds \citep{Millar2004}. By regulating the temperature of the wind particles through Eq.~(\ref{IdealGas}), we do not take into account the additional cooling processes occurring in AGB winds, such as radiative cooling by collisionally excited molecules \citep{Decin2006}, which results in some limitations in the thermo-dynamics of our models. This approach is commonly used in the literature by \cite{Liu2017,ElMellah2020,Maes2021}.

The gas in the AGB wind is simulated through a distribution of SPH particles with fixed mass $m_i$, which is set by the input mass-loss rate. Self-gravity between the gas particles is neglected.
The necessary physics to describe the launch of a stellar wind through pulsation-enhanced formation of dust, and subsequent outward acceleration of this dust via coupling of the dust opacity with the incident stellar radiation, has yet to be coupled to the \textsc{Phantom} code. Since in our simulations the wind only consists of gas particles and not of dust grains, it is synthetically propagated by means of a parameter $\Gamma$, which reduces the gravitational pull of the AGB star on the gas through the additional effective potential term in the conservation of momentum (Eq.~(\ref{consMom}))  as follows:
\begin{equation}
    \frac{{\rm d} \textit{\textbf{v}}}{{\rm d} t} = - \frac{\vec{\nabla} P}{\rho} - \frac{G M_{\rm p}}{r_1^3} (1 - \Gamma)  \vec{r_1} - \frac{G M_{\rm s}}{r_2^3} \vec{r_2},
\end{equation}
in which $M_{{\rm p}}$ and $M_{\rm s}$ are the mass of the AGB star and companion, and $\vec{r_1}$ and $\vec{r_2}$ connect the considered gas particle to the centre of mass of the AGB star and companion, respectively.
To mimic a `free wind', in which the gravity of the AGB star is balanced by radiation pressure on dust, $\Gamma$ is set to $1$ \cite[e.g.][]{MastrodemosII,Kim2012A,Liu2017}.
By not including a chemical network, radiative transfer approach or other additional cooling and heating processes, the density and thereby mass-loss rate do not contribute to the dynamics of the wind particles \citep{Maes2021}.
The AGB mass-loss is modelled by initially defining five concentric shells, initiating at a given injection radius $R_{{\rm inj}}$ around the AGB star, on which the SPH particles are uniformly and equidistantly placed. The particles in the inner four shells serve as boundary condition to generate the pressure gradient at the edge of the stellar surface. The particles on the outer shell are launched radially with an initial speed $v_{\rm ini}$.
The AGB and companion star are assumed to be `gravity-only' SPH sink particles, whose internal structure is not modelled \citep{phantom}. They orbit their common center-of-mass (CoM) located at the origin of the system. 
The wind accretion mechanism adapted is `prompt accretion', indicating that wind particles within the accretion radius of a sink particle are accreted and removed from the simulation if a number of accretion checks are satisfied (for more details, see \cite{phantom}, Sect. 2.8.2). The accretion is modelled in such a way that the total mass, linear and angular momentum are conserved. 
We define an outer edge to the numerical domain in which the wind particles propagate to reduce the computation cost. Once the particles reach the outer boundary $R_{\rm bound}$, they are artificially removed from the simulation.
To ensure that the morphology of the models is self-similar (i.e. the global morphology does not change anymore during further evolution), the simulations are evolved until (i) a quasi-steady state is reached concerning the number of particles present in the system and (ii) the key parameters concerning the morphological shape have small variations in time. The second requirement is fulfilled after about $6$ orbital periods. From then on the simulations have a roughly constant flattening ratio (defined in Sect. \ref{flatteningSection}), arc densities and periodically reappearing wind structures. To illustrate this, the density, velocity and temperature profile in one radial direction are plotted for 2 simulations after $4$, $6$, $8$ and $10$ orbits in Fig.~\ref{selfSimPlot}. These plots show that the arc structures in the inner wind reappear periodically with the same width, density, velocity and temperature each orbital period.
To meet the first requirement, the models are evolved for $10$ orbital periods after which they typically contain about $1.2 \times 10^{\rm 6}$ SPH particles. 

\subsection{Binary setup}

The fixed binary setup of our models is summarised in Table~\ref{setupTable}. The models consist of an AGB star with accretion radius $R_{\rm p,accr}= 1.2 \, \rm{au}= 258\,{\rm R_{\odot}}$, temperature $T_{\rm p} = 3000$ K and mass $M_{\rm p} = 1.5\,{\rm M_{\odot}}$, which is the typical mass for bulge Miras of intermediate age \citep{Groenewegen2005}. The inner five shells, launching the wind, originate at the injection radius $R_{\rm inj} = 1.3 \,{\rm au}$. The AGB star is accompanied by a solar-mass companion ($M_{\rm s} = 1\,{\rm M_{\odot}}$) with an accretion radius of $R_{\rm s,accr} = 0.05 \, \rm{au} = 10.7\,{\rm R_{\odot}}$, located at a semi-major axis of $a = 6\, \rm{au}$, thereby the stars orbit their CoM with an orbital period of $P_{\rm orb} = 9.3$ yrs. The interaction of the wind with a companion in this long-period type binary ($P_{\rm orb} \geq 1$ yr) is found to be a dominant wind-shaping agent for asymmetrical AGB wind morphologies \citep{Decin2020}. The simulations are evolved for $t_{\rm max} = 93$ yrs, corresponding to the required $10$ orbital periods, and the outer boundary is set to $R_{\rm bound} = 200\, \rm{au}$.

Previous studies reveal that the four main parameters determining the wind structure are the wind velocity, semi-major axis, mass ratio and eccentricity \citep{MastrodemosII,Chen2017,Saladino2019B,Kim2019}. In this work, we focus on the impact of the wind speed and eccentricity through the nine models listed in Table~\ref{inputTable}. The impact of the semi-major axis and mass ratio on the morphology of similar models is studied by \cite{Maes2021}.
The input velocity $v_{\rm ini}$ varies between $5$, $10$ and $20\,{\rm{km\,s^{-1}}}$, corresponding to realistic terminal velocities $v_\infty$ of approximately $8.8$,  $12.3$ and $21.2\,{\rm{km\,s^{-1}}}$ in absence of a companion (see velocity distribution of single star simulations in Fig.~\ref{velProfileSS}). Although the mass-loss rates $\dot{M}$ do not affect the wind structures, they are set to respectively $2 \times 10^{-7}\,{\rm {M_{\odot}\,yr^{-1}}}$, $2 \times 10^{-6}\,{\rm{M_\odot\,yr^{-1}}}$ and $1 \times 10^{-4}\,{\rm{M_\odot\,yr^{-1}}}$ according to the quasi-linear relation to the wind velocity \citep{Ramstedt2009}. The corresponding fixed mass of the gas particles are $m_i = 5.83 \times 10^{-12} \, {\rm M_\odot}$, $5.76 \times 10^{-11} \, {\rm M_\odot}$ and $2.35 \times 10^{-9} \, {\rm M_\odot}$, respectively.
The three different velocity ranges, in combination with a fixed semi-major axis, are expected to represent different morphology classes \citep{MastrodemosII}. Next to the wind velocity, the eccentricity $e$ is varied between $0.00$, $0.25$ and $0.50$ to study the poorly understood wind structures that result from the binary interaction in eccentric systems.

\begin{table}
    \caption{Initial setup.}
    \begin{center}
    \begin{tabular}{lc}
    \hline 
    \hline
    Parameter           &  Initial value \\
    \hline 
    $M_{\rm p}$        [${\rm M_{\odot}}$] &  $1.5$\\
    $M_{\rm s}$         [${\rm M_{\odot}}$] & $1$ \\
    $R_{\rm p,accr}$    [${\rm R_{\odot}}$] &  $258$  \\
    $R_{{\rm s,accr}}$ [${\rm R_{\odot}}$] &  $10.7$ \\
    $T_{\rm p}$ [K] &  $3000$ \\
     $a$                [au] &  $6$ \\
    $R_{\rm bound}$     [au] &  $200$  \\
    $t_{\rm max}$       [yrs] &  $93$  \\
    \hline
    \end{tabular}
    \end{center}
    {\textbf{Notes.} \footnotesize{The setup for the following parameters is the same for all models: $M_{\rm p}$, $M_{\rm s}$, $R_{\rm p,accr}$ and $R_{\rm s,accr}$ are the initial masses and accretion radii of the primary (AGB) and secondary (companion) star, $a$ is the semi-major axis ogf the system and $R_{\rm bound}$ is the boundary radius beyond which gas particles are killed. The models run for $t_{\rm max} = 93$ yrs which corresponds to $10$ orbital periods.
    }}
    \label{setupTable}
\end{table}

\begin{table}
    \caption{Model characteristics.}
    \begin{center}
    \begin{tabular}{ccc}
    \hline
    \hline
    &&\\[-2ex]
    Model name &  $v_{\rm ini}$ [${\rm{km\,s^{-1}}}$] & $e$ \\
    &&\\[-2ex]
    \hline
    v05e00   & $5$  & $0.00$  \\
    v05e25   & $5$  & $0.25$  \\
    v05e50   & $5$  & $0.50$  \\
 
    v10e00   & $10$ & $0.00$ \\
    v10e25   & $10$ & $0.25$ \\
    v10e50   & $10$ & $0.50$ \\
 
    v20e00   & $20$ & $0.00$ \\
    v20e25   & $20$ & $0.25$ \\
    v20e50   & $20$ & $0.50$ \\
    \hline
    \end{tabular}
    \end{center}
    {\textbf{Notes.} \footnotesize{All models with their characteristic input values.
    The model names are set in such a way that the characteristics can be deduced from it, with `v...' denoting the input wind velocity in ${\rm{km\,s^{-1}}}$ and `e...' the value of the eccentricity of the system multiplied by a factor $100$.}}
    \label{inputTable}
\end{table}

%% file: results.tex
\section{Results}
\label{ch:results}

The impact of the gravitational influence of a binary companion on the wind morphology of an AGB star is twofold. On the one hand, it induces an orbital movement of the stars around their CoM. This generates a spiral shock and significantly beams the wind into the orbital plane resulting in a flattening \citep{Kim2012B,Kim2019,ElMellah2020}, as is discussed in Sect.~\ref{flatteningSection}. Secondly, the companion gravitationally attracts wind material such that it focuses the wind into a detached bow shock or accretion wake flowing behind the companion \citep{Saladino2018}. This can result in an equatorial density enhancement (EDE), on which we elaborate in Sect.~\ref{EDESection}.
Depending on the physical properties of the binary system, the relative importance of these effects determines the global morphology of the outflow. We here present the density distribution of the outflow of the models in 2D slices, aiming to investigate the influence of the wind velocity and eccentricity on the inner wind structure and on the resulting global morphology.
The radial density and velocity distribution of the models on three perpendicular axis through the CoM is provided in Sect.~\ref{radial1Dplots}, and provides a more detailed, quantitative view on the density and velocity distribution.' 

\begin{figure*} 
    \centering
    \includegraphics[width = \textwidth]{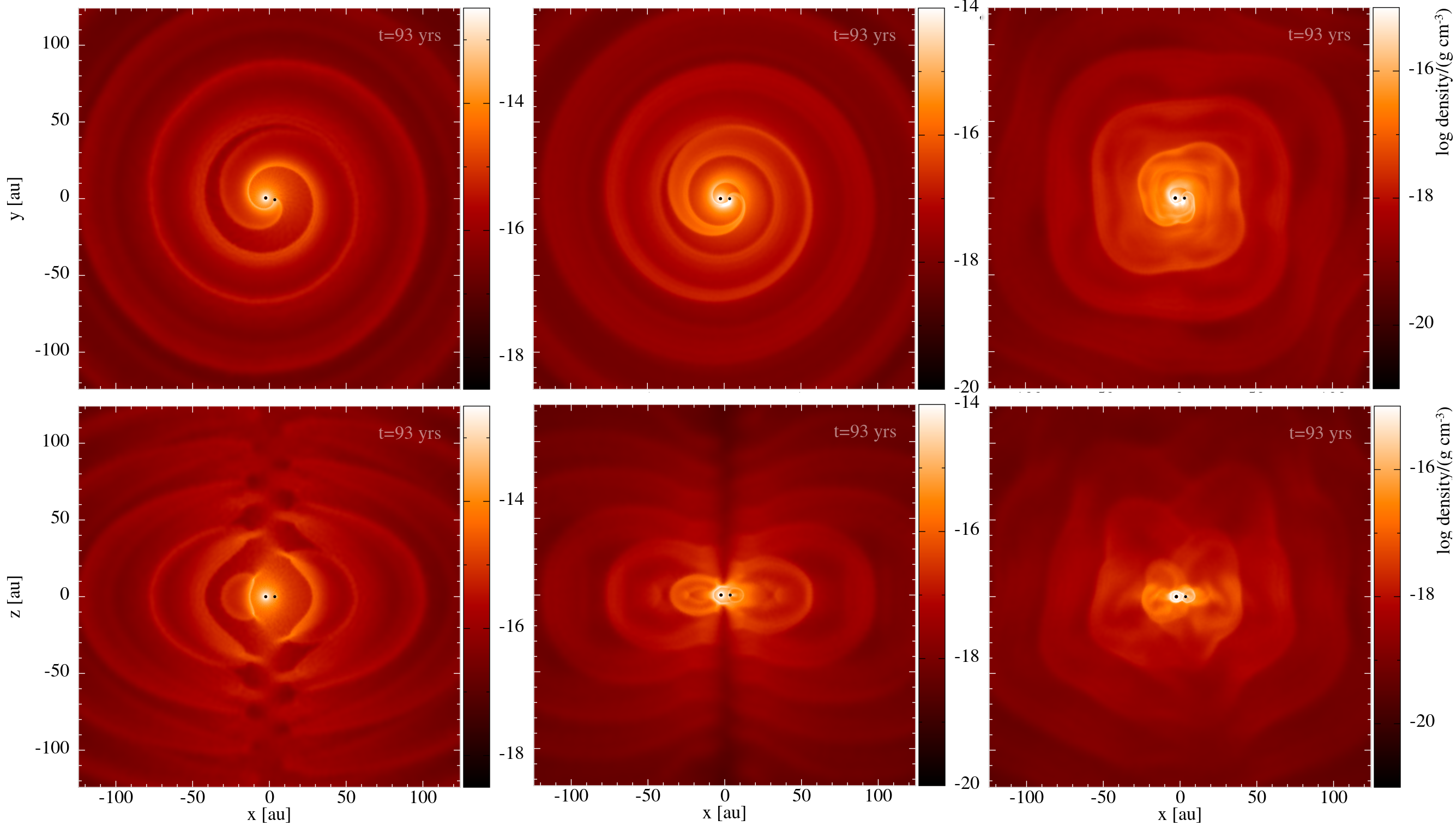}
    \caption{Density profile in slices through (top row) and perpendicular to (bottom row) the orbital plane of models v20e00 (left),  v10e00 (middle) and v05e00 (right). The AGB and companion star are annotated as the right and left dot, respectively, not to scale. The colour bars are scaled w.r.t. the input mass-loss rate for an optimal comparison.}
    \label{velMorph}
\end{figure*}

\subsection{Wind structure in circular systems}
To investigate the impact of the orbital eccentricity on the wind structures, we first need to understand the structure formation in the circular orbits, which are presented in this section arranged according to increasing complexity.
For each model we first analyse the wind structures in a slice through the orbital plane, and then the structures in a slice through the meridional plane, perpendicular to the orbital plane, through the two sink particles.

\subsubsection{ Model v20e00}
\begin{figure}
    \centering
    \includegraphics[width = 0.45\textwidth]{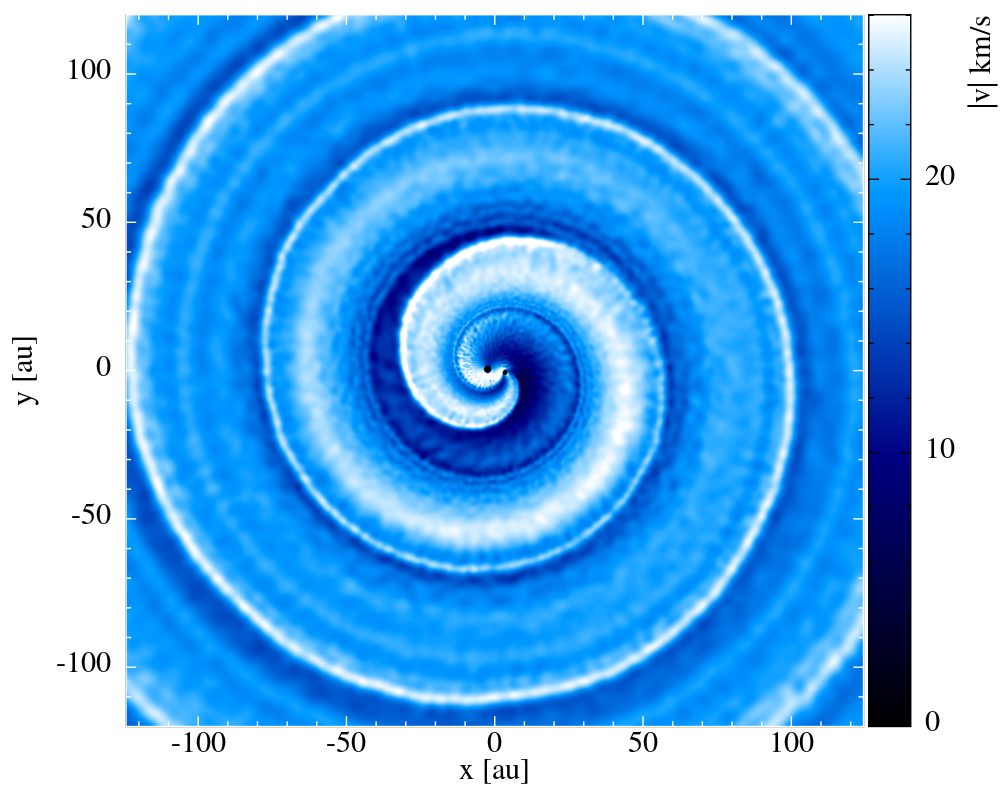}
    \caption{Velocity profile in slice through the orbital plane of model v20e00.}
    \label{velSliceM16c}
\end{figure}

The highest wind velocity model v20e00 with $v_{\rm ini} = 20\,{\rm{km\,s^{-1}}}$ and $e = 0.00$ consists of a spiral structure in the orbital plane and arcs in the meridional plane (see left panels of Fig.~\ref{velMorph}).
The orbital plane spiral structure is attached to the companion and delimited by a dense inner and a less dense outer edge. We define this inner edge as the `backward spiral edge' (BSE) and the outer one as the `frontward spiral edge' (FSE). These spiral structures are created as follows.
When wind particles approach the companion, its gravitational attraction deflects their path towards it. 
Material located closely behind the companion gets accelerated strongly in the direction of the orbital motion, whereas material that approaches the front of the companion experiences an acceleration in the opposite direction. 
Because the companion is in motion, conservation of energy and momentum induce a so-called gravitational slingshot. The impact and importance of this gravitational slingshot effect is studied by \cite{Maes2021}. 
The resulting acceleration in various directions introduces a velocity dispersion in the gas flowing behind the companion, such that the created wake widens. 
The material within the wake with the highest radial velocity component forms the outer FSE through collision with slower wind material. Similarly, the gas particles within the wake that have the lowest radial velocity component, are caught up by faster wind material close to the AGB star, and form the inner BSE. The FSE has a higher velocity ($v_{\rm FSE} \approx 25\,{\rm{km\,s^{-1}}}$) than the BSE ($v_{\rm BSE}\approx 10\,{\rm{km\,s^{-1}}}$) which is wrapped more closely around the AGB star, as can be seen from the velocity distribution in Fig.~\ref{velSliceM16c}. 
Because the spiral broadens, after about three quarters of an orbital period, the FSE catches up with the BSE originated one orbital period earlier (location of interaction at $x \approx 25\, \rm{au}$, $y \approx 50\, \rm{au}$ in the first panel of Fig.~\ref{velMorph}). This interaction results in one spiral shock that shapes the outer wind structure in the orbital plane (see Fig.~\ref{fullMorphCirc}).
This two-edged (FSE and BSE) spiral wind structure is most likely to form in case the wind velocity around the location of a stellar companion ($v_{\rm w}$) exceeds the orbital velocity ($ v_{\rm orb}$), following \cite{Saladino2018}. In this case, mass accretion onto the companion approaches the Bondi-Hoyle-Lyttleton (BHL) description \citep{BHL1,BHL2,Saladino2019A} and therefore we refer to this regime, where $v_{\rm w} \gg v_{\rm orb}$, as the BHL regime. 

The density distribution in the meridional plane consists of concentric arcs with one common centre (bottom left panel of Fig.~\ref{velMorph}). The side of the arcs closest to the stars is the cross section of the BSE and the outer side of these arcs the cross section of the FSE. This figure shows once more that the FSE catches up with the previous BSE (e.g. at $x\approx 50\, \rm{au}$, $|z| \lesssim 25\, \rm{au}$).  
The wind material that is gravitationally focused by the companion into a spiral structure reaches a smaller scale height than the wind particles that are only deflected by the orbital motion of the stars \citep{Kim2012B}. Some of these particles form the thin ends of the arcs near the polar axis. Unlike the broadened part of the arcs, resulting from material that was gravitationally attracted of the companion, the thin structures do not reveal a velocity dispersion.

\begin{figure}
    \centering
    \includegraphics[width=0.4\textwidth]{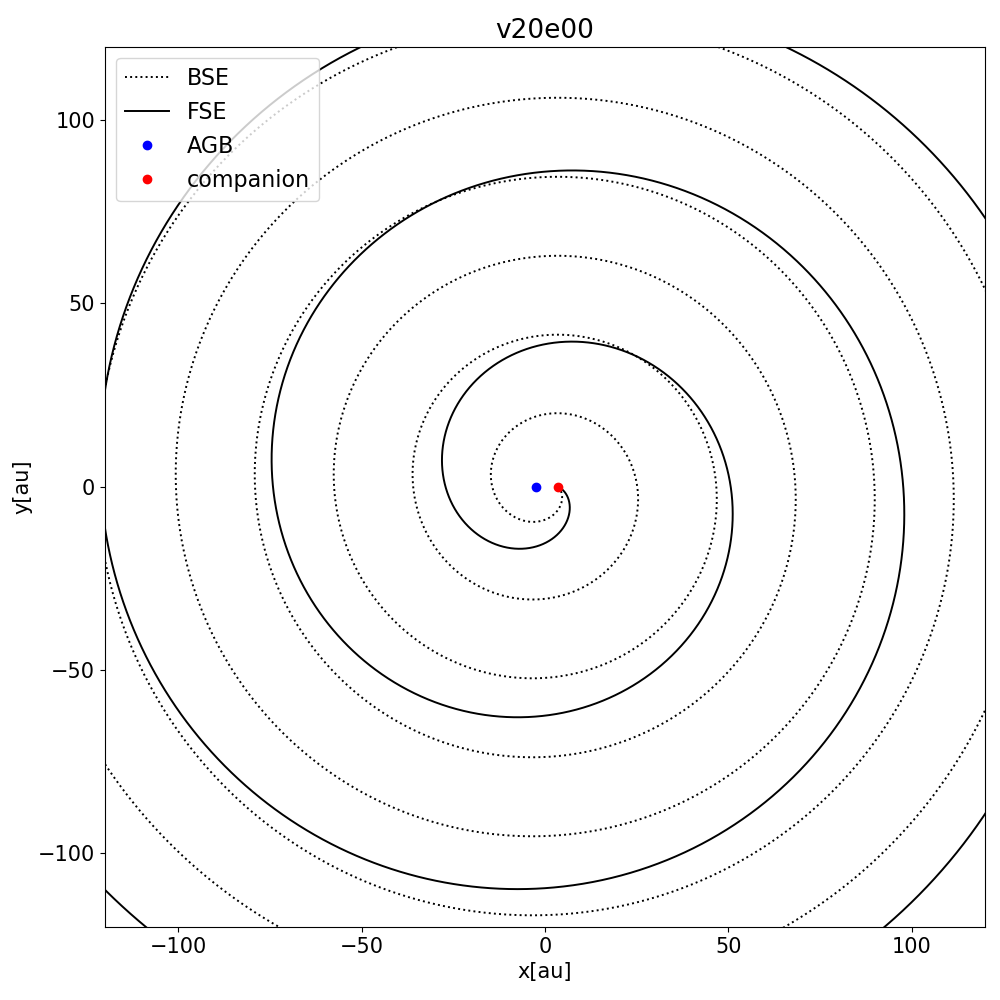}
    \caption{Analytical Archimedes spirals described by Eq.~(\ref{ArchimSpiral}), representing the orbital plane spiral structure of model v20e00 in Fig.~\ref{velMorph}.}
    \label{anal_e00}
\end{figure}

To improve our understanding of the thermo- and hydrodynamical interactions, we compare our numerical simulations to a mathematical model based on Archimedes spirals, which we refer to as the analytical model henceforth. The frontward and backward spiral edge of model v20e00 can be modelled as Archimedes spirals in the following way: 
\begin{equation}
    \frac{\partial r}{\partial \theta} 
     = \frac{\partial r}{\partial t}  \frac{\partial t}{\partial \theta} = \frac{v_r(t)}{\omega(\theta)} ,
\end{equation}
in which $\omega(\theta)$ can be derived from the orbital equation, 
\begin{equation}
    \omega(\theta) = 2 \pi \frac{ \left(1 + e \cos \theta\right)^2}{\left(P \left(1-e^2\right)\right)^{3/2}},
\end{equation}
where we adopt the polar coordinates $(r, \theta)$ with $r$ the radius w.r.t. the CoM and $\theta$ the angle to the axis connecting the sink particles. Here, $e$ is the eccentricity of the elliptical orbit of the stars, $v_r$ the radial velocity of the wind particles in a spiral edge assumed to be constant, $\omega$ the angular velocity and $P$ the orbital period of the system. Since both spiral edges originate from the location of the companion, their location of origin $r_i$ must equal the location of the companion at the initial angle $\theta_i$, i.e. $r_i = r(\theta_i) = r_{\rm comp}(\theta_i)$. For each value of $\theta \geq \theta_i$, $r(\theta)$ can be calculated as
\begin{equation}
    \label{ArchimSpiral}
    r(\theta)  = \int_{\theta_i}^{\theta} \frac{v_r}{\omega}  \, d\theta \ = 
    \int_{\theta_i}^{\theta} v_r \frac{ \left(P \left(1-e^2\right)\right)^{3/2}}{2 \pi \left(1 + e \cos \theta\right)^2} \, d\theta \ ,
\end{equation}
with boundary condition
\begin{equation}
    r(\theta_i) = r_{\rm comp}(\theta_i) = a \frac{M_p}{M_s+M_p} \frac{\left(1-e^2\right)}{\left(1+e \cos{\theta_i}\right)} .
\end{equation}
The analytical model for the two spiral edges of model v20e00, with approximated radial velocities $v_{r,{\rm BSE}} = 11\,{\rm{km\,s^{-1}}}$ and $v_{r,{\rm FSE}} = 24 \,{\rm{km\,s^{-1}}}$, is presented in Fig.~\ref{anal_e00} and resembles the orbital plane spiral structures in the first panel of Fig.~\ref{velMorph}.
The FSE catches up with the BSE after three quarters of an orbit. From then on, the analytical model strongly deviates from the simulation because of the importance of thermo- and hydrodynamical interactions. The single spiral structure that arises from the interaction between the FSE and BSE in the simulation resembles well the analytically modelled FSE, indicating that this frontward spiral edge dominates over the backward one in this model. 

\subsubsection{ Model v10e00}
\label{Sect:v10e00}

\begin{figure}
    \centering
    \includegraphics[width = 0.49\textwidth]{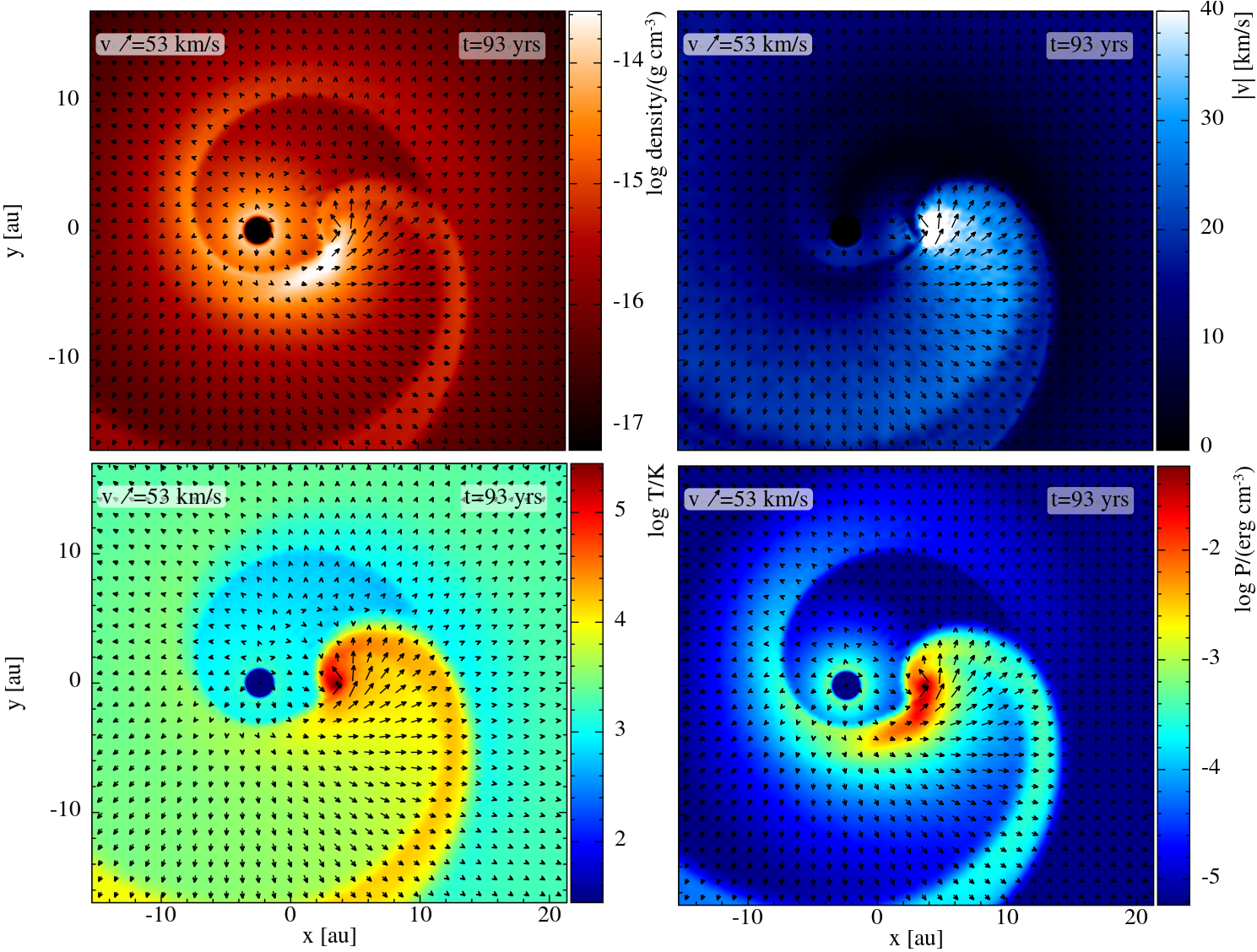}
    \caption{Circumstellar density, speed, temperature and pressure profile in the orbital plane of model v10e00, superimposed with the velocity vector profile which is shown as black arrows. In the top left panel, the USF can be identified as the high density arc-shape region around $x=2 \, {\rm au}$ and $-2 < y < 2 \, {\rm au}$.}
    \label{zoomedM19}
\end{figure}

For a lower wind velocity ($v_{\rm ini} = 10\,{\rm{km\,s^{-1}}}$), the interaction timescale of the wind particles with the companion increases, such that the gravitational impact of the companion is stronger. The spiral pattern revealed in the orbital plane slice of model v10e00 (middle top panel of Fig.~\ref{velMorph}) contains again a spiral structure. 
However, the inner wind structure differs from model v20e00, since here two separate spiral structures emerge from the location of the companion. A dense low-velocity spiral, which is comparable to the BSE of model v20e00, wraps closely around the AGB star. The second, outer, high-velocity spiral originates from a bow shock that flows in front of and around the companion, as can be seen on the detailed plots in Fig.~\ref{zoomedM19}. Hence we refer to this as the `bow shock spiral'.
The slow, inner spiral merges with this bow shock spiral after one revolution (at $x \approx 10\, \rm{au}$, $y\approx 4\, \rm{au}$ in Fig~\ref{zoomedM19}). 
The interaction between wind particles with different wind velocities induces a velocity dispersion and thereby broadening of the bow shock spiral structure such that, similar to the two-edged spiral structure in model v20e00, an inner backward spiral edge and outer frontward spiral edge result (see middle top panel Fig.~\ref{velMorph}).
Again, the FSE has a higher radial velocity and after less than $3/4$ of an orbit, it catches up with the BSE of the previous bow shock spiral.

To understand the formation of the bow shock spiral, Fig.~\ref{zoomedM19} presents the inner density, velocity, temperature and pressure distribution superimposed with the velocity vector profile.
The low wind velocity of this model implies that the wind particles can interact for a longer time with the companion, and thereby implies a stronger compression of wind material by the companion's gravity. 
The interaction of the wind particles with the moving companion generates a high-temperature and high-velocity region around the companion, which we will from now on refer to as the high-energy region (see $v$ and $T$ distribution in Fig.~\ref{zoomedM19}). 
The high energy induces a high local pressure (Eq.~(\ref{IdealGas})). In this way, the pressure ahead of the high-energy region is much lower, such that a strong pressure gradient is formed directed away from the companion. This pressure gradient bends the path of AGB wind material that is gravitationally deflected towards the companion. The resulting structure is called a bow shock, as it is similar to a bow shock in which the ram pressure equals the pressure of a stellar wind on the interstellar medium. 

The velocity vectors in Fig.~\ref{zoomedM19} show that the part of the bow shock located between the stars (around $-2\, \rm{au}$ $<y<2\, \rm{au}$, $x\approx2\, \rm{au}$) carries wind particles to a high-density and high-pressure region flowing behind the companion. 
We define this flow as the `umbrella stagnation flow' (USF), since the gas particles flow around the high-energy region similar to the stream of rain on an umbrella. The flow is named after the stagnation point flow in fluid dynamics, with the main difference that the high-energy region is not a solid surface, such that the no-slip condition is not valid here.
The wind particles collected through this USF in a high-density region move towards the companion and provide the wind-companion interaction, which on its turn keeps the temperature and wind velocity around the companion high. Since this process proceeds stable and continuous in this model, a regular self-similar bow shock results that rotates with the star around the CoM, shaping the outer regions into the regular spiral structure, shown in Fig.~\ref{velMorph}. This is unlike the next model, described in Sect.~\ref{Sect:v05e00}, where this process is subject to periodic instability.

The meridional plane density distribution of model v10e00 consists of bicentric rings, with two centers-of-origin, contrary to the concentric arcs in model v20e00 (see bottom row of Fig.~\ref{velMorph}). The rings represent the cross section in the meridional plane of the spiral generated by the bow shock. Unlike the arc structures with thin ends near the polar axis of model v20e00, the rings of this model (v10e00) reach up to the polar axis, forming an elongated peanut shape. This elongation or flattening is expected to occur due to the increased ratio of the orbital to the wind velocity around the location of the companion (see Sect.~\ref{globalMorph}).

\subsubsection{ Model v05e00}
\label{Sect:v05e00}
A more complex morphology is obtained when lowering the wind speed further, as in model v05e00. Because of the low kinetic energy and hence the long interaction timescale, the wind particles have a harder time escaping the companion's gravitational potential well. The orbital plane density distributions in the right panel of Figs.~\ref{velMorph} and~\ref{fullMorphCirc} reveal a squared spiral pattern. 

\begin{figure}
    \centering
    \includegraphics[width = 0.49 \textwidth]{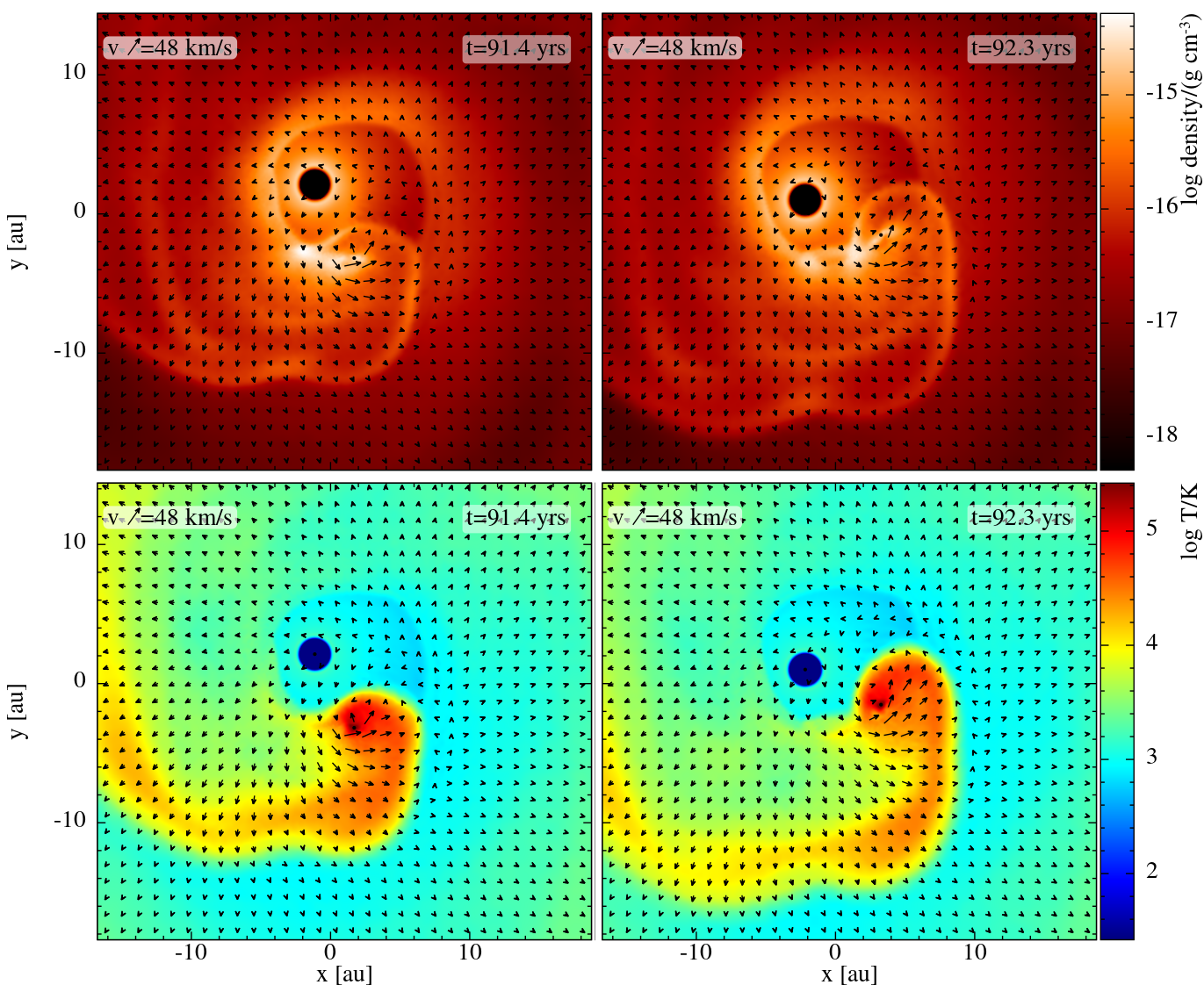}
    \caption{Circumstellar density and temperature profile in the orbital plane of model v05e00 at two different timesteps, superimposed with the velocity vector profile which is shown as black arrows.}
    \label{zoomedM7}
\end{figure}

The inner density and temperature distribution of this model at two different timesteps are presented in Fig.~\ref{zoomedM7}. Analogous to model v10e00 (Fig.~\ref{zoomedM19}), the strong wind-companion interaction results in a high-energy region delimited by a bow shock with an USF that creates a high-density region flowing behind the companion.
Unlike model v10e00, the supply of wind material through this USF is not steady, but is subject to periodic instability. Thereby, the amount of energetic wind particles collected in the high-energy region varies. This makes the rate at which such particles propagate radially outward change periodically. Regions of hot material propagate radially away from the star with the same period as with which it was supplied to the companion by the USF. 
This out-flowing material then deforms the bow shock and the resulting bow shock spiral.
Each time a blob of energetic material propagates away, the high-energy region shrinks. The periodic expansion and compression of this region goes along with a periodic variation in the distance at which the high-density region, to which particles are carried by the USF, is trailing behind the star.
This distance varies between ${\sim}2$ to $2.5\, \rm{au}$ and results in the periodically varying supply of wind material to the companion. 
The outward propagation of high-energy particles and the induced deformation of the bow shock spiral through this process is illustrated in Fig~\ref{zoomedM7}. In the first snapshot ($t = 91.4$ yrs), hot material (located around $-1 \, \rm{au} \leq x\leq5\, \rm{au}$, $-11 \, \rm{au} \leq y\leq-8\, \rm{au}$) is propagating radially outwards from the companion and in the second snapshot ($t = 92.3$ yrs), this hot-material region, that is located now further away from the companion, has expanded.
In this model, this described periodic process occurs about $4$ times per orbital period, indicating that $4$ injections of hot energetic material are launched during each orbit. The directions in which they are launched result in the `corners' of the squared global wind morphology in the right panel of Fig.~\ref{velMorph}. The clockwise rotation of the squared structure indicates that the period of this process is slightly larger than $1/4$ of the orbital period of the system. The exact cause of the instability and what determines its period are yet to be investigated, but are thought to be related to the total mass that is gravitationally compressed by the companion. 

The meridional plane density distribution of this model (v05e00) in the bottom right panel of Fig.~\ref{velMorph} harbours a rose-like pattern instead of the regular arc or ring structures of models v20e00 and v10e00. Similar to model v10e00, the meridional structures represent the 3D nature of the bow shock spiral. The described periodic launch of hot material that shapes the corners of the squared orbital plane structures, results in the extended lobes in the diagonal directions of the meridional plane.
The lower wind speed, and thereby increased ratio of orbital to wind velocity of this model, does not result in an elongated or flattened density distribution as in model v10e00, because the flattening is counteracted by the extended high-energy lobes. This is not predicted by classical theoretical expectations as we discuss in Sect.~\ref{flatteningSection}. 

\subsection{Wind structure in eccentric systems}

In eccentric systems the orbital separation varies between $r_{\rm a} = a (1+e)$ at apastron and $r_{\rm p} = a (1-e)$ at periastron. According to Kepler's second law, the induced variation in the orbital velocity $v_{\rm orb}(r) = \sqrt{G (M_{\rm s}+M_{\rm p})/r}$ throughout one orbital period makes the stars reside for a longer time around their apastron than around their periastron passage. Since the two stars lie on a straight line through the CoM, they are always separated by an angle $\pi$ in the orbital plane. We therefore define `apastron side' as the $x>0$ side at which the companion resides during apastron (and the AGB star during periastron). 

\subsubsection{ Model v20e25 \& v20e50}

\begin{figure}
    \centering
    \includegraphics[width=0.49\textwidth]{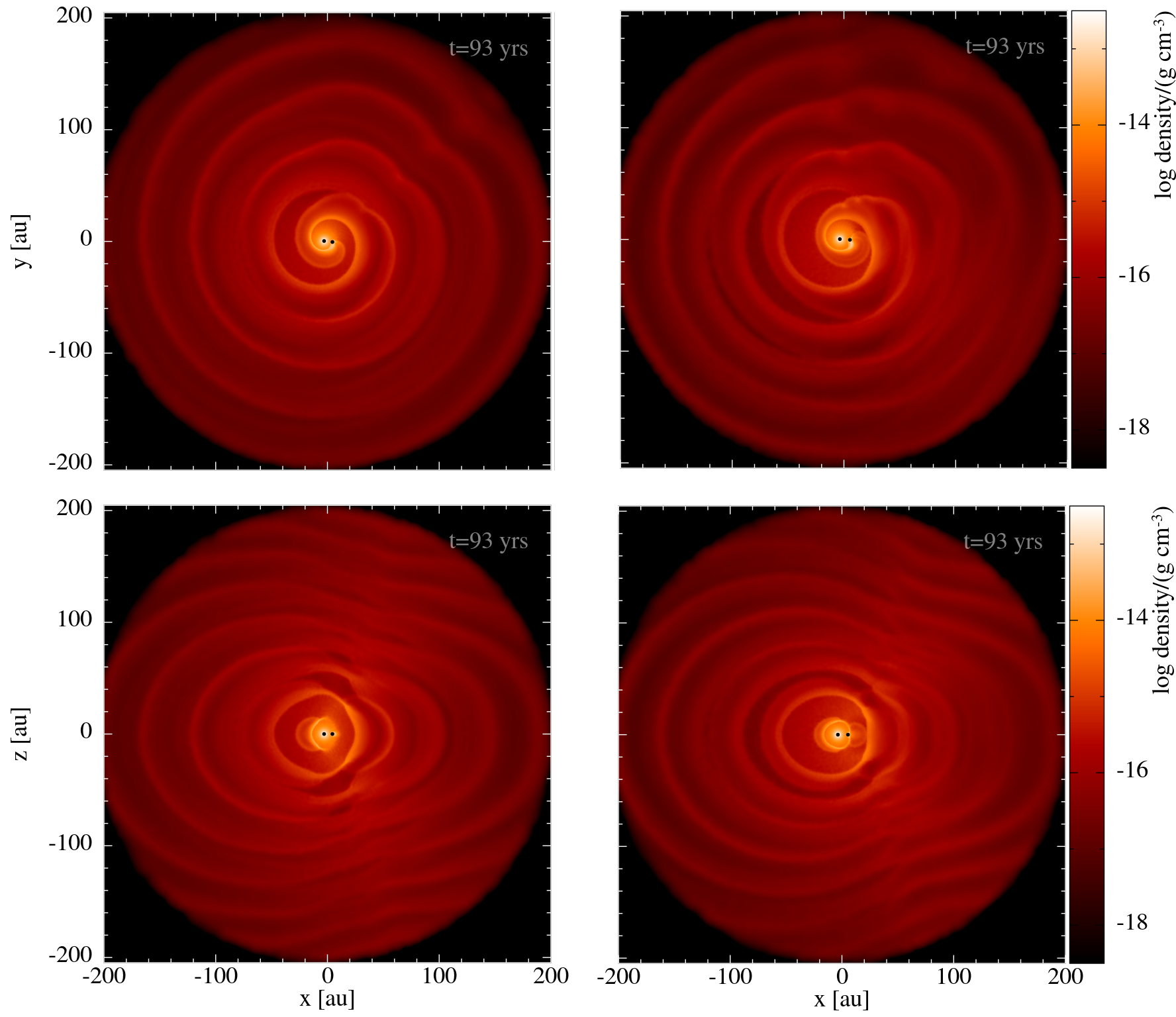}
    \caption{Similar plots as in Fig.~\ref{velMorph}, but for models v20e25 (left) and v20e50 (right).}
    \label{MorphFastEcc}
\end{figure}

The systems in which the eccentricity causes the least complexities are the ones in which the wind speed is highest, minimising the magnitude of the wind-companion interaction. 
The global morphology of models v20e25 and v20e50 is shown in Fig.~\ref{MorphFastEcc}. In these systems, the spiral structure in the orbital plane is no longer a regular Archimedes-like structure as in model v20e00, but is stretched towards the lower right ($x>y$) half of the orbital plane.
Besides this deformation of the global morphology, a kink appears in the $y>0$ apastron side region of the orbital plane of model v20e25, and a turbulent structure in the same region for model v20e50. At the periastron side of model v20e50, low-density inter-spiral gaps arise, whereas the apastron side consists of less distinct, overlapping structures. The meridional planes of both models display asymmetric ring-arc structures. 

\begin{figure}
    \centering
    \includegraphics[width=0.4\textwidth]{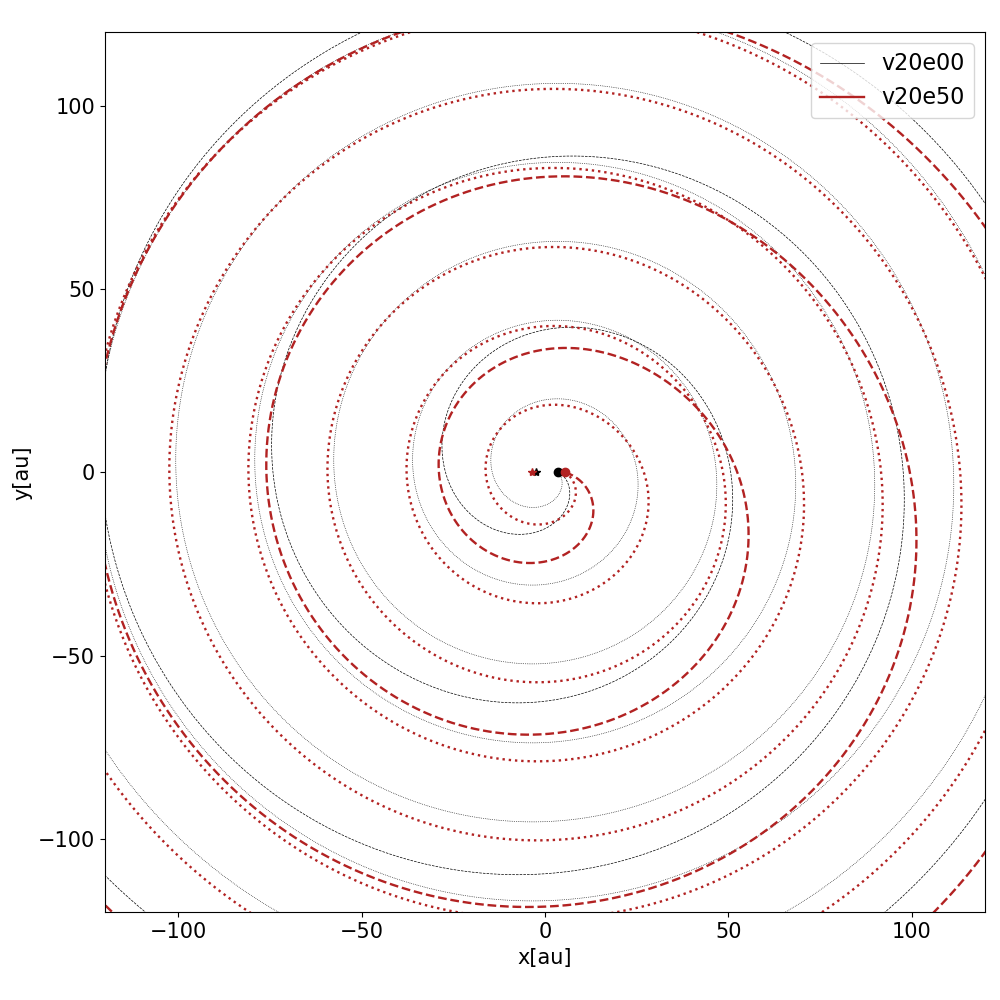}
    \caption{ Analytical Archimedes spirals described by Eq.~(\ref{ArchimSpiral}), representing the orbital plane spiral structure of model v20e50 in Fig.~\ref{MorphFastEcc}.}
    \label{anal_ecc}
\end{figure}

To distinguish which complexities result from the phase-dependent thermo- and hydrodynamical interactions, we solve Eq.~(\ref{ArchimSpiral}) for the eccentric case. The result is shown in Fig.~\ref{anal_ecc}, where the eccentric analytical spiral is drawn on top of the circular one.
Similar to the hydrodynamical simulations, the eccentric spirals of the analytical model are stretched towards the $x>0, y<0$ corner with respect to the circular ones, such that the shape of the BSE and FSE is compressed along the diagonal connecting the top right and bottom left of the orbital plane.
The kink, perturbation, and the difference in overlap between spiral structures at the apastron and periastron side of the hydrodynamical simulation do not arise in the analytical model, which indicates that these features are not a consequence of simple ballistic motion of gas, but rather of phase-dependent thermo- and hydrodynamical effects.

\begin{figure*} 
    \centering
    \includegraphics[width = \textwidth]{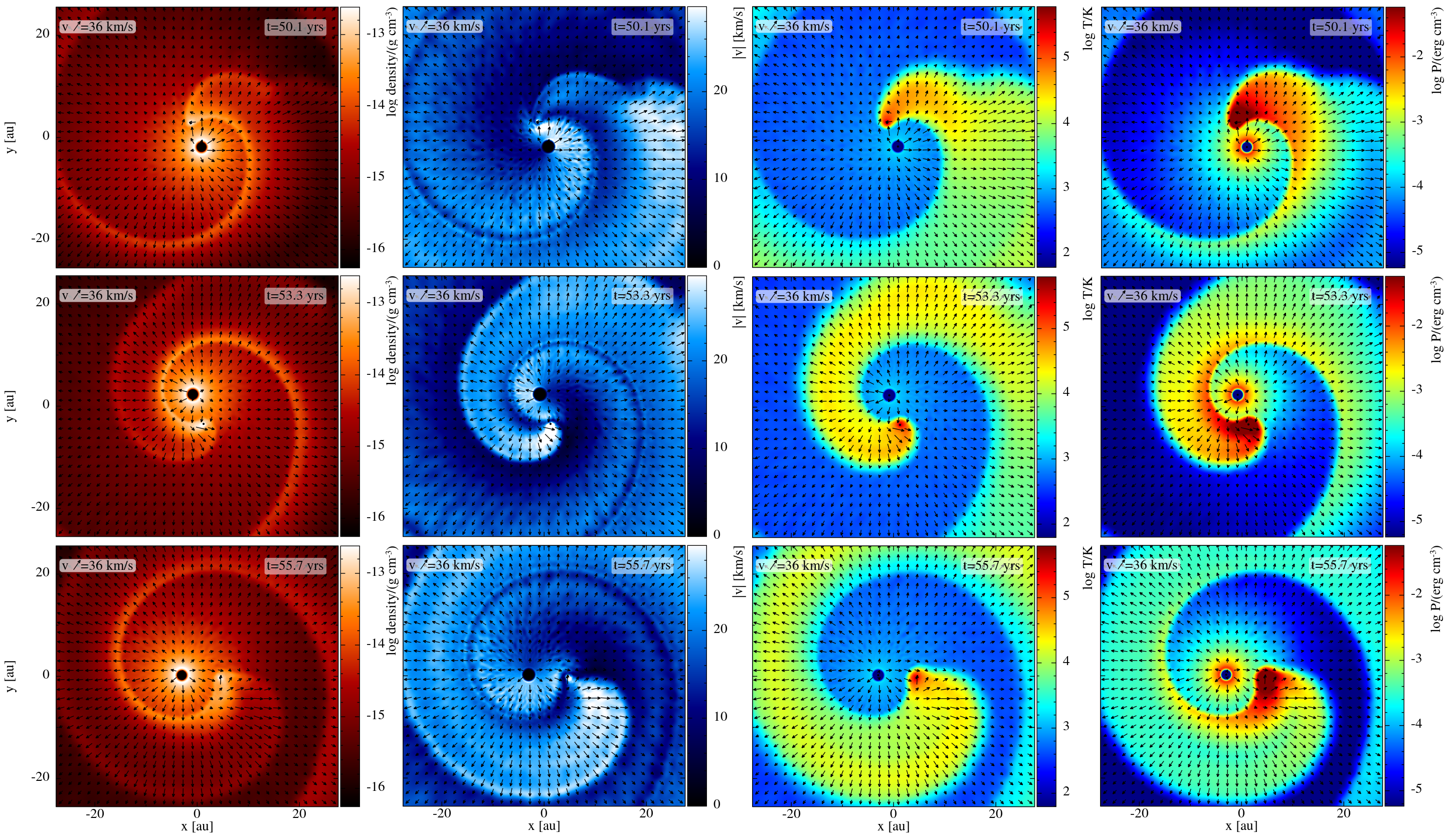}
    \caption{From left to right: density, speed, temperature and pressure distribution in the orbital plane of model v20e25 at 3 consecutive orbital phases. }
    \label{zoomM17}
\end{figure*}

To understand the mechanism shaping these perturbations, the inner wind density, speed, temperature and pressure distribution of model v20e25 are presented at 3 different timesteps throughout one orbital period in Fig.~\ref{zoomM17}. In the first snapshot ($t=50.1$ yrs), the stars are moving from apastron to periastron. In this regime $v_{\rm w}$ is still sufficiently larger than $v_{\rm orb}$, so that the morphology can be described by the BHL regime and a similar flow structure is observed to model v20e00, with a FSE and BSE (top left panel Fig.~\ref{velMorph}). At periastron, the orbital separation decreases to $4.5\, \rm{au}$ and the companion's orbital velocity increases to $14.9\,{\rm{km\,s^{-1}}}$. The BHL approach is no longer valid. Due to the higher gas density close to the AGB star, the companion influences a comparatively much larger mass of gas particles than around apastron. The stronger wind-companion interaction results in the compression of gas particles and the formation of a high-energy region, as explained in Sect.~\ref{Sect:v10e00}. 
As the stars move towards apastron (at $t= 53.3$ yrs in Fig.~\ref{zoomM17}) these energetic wind particles travel radially away from the position of the companion and push against the outer spiral edge. 
When the stars approach apastron, the path of the companion on its orbit takes a sharp turn as it curves back towards periastron, and the energetic wind particles, which are not bound to the companion, do not change their trajectory. 
The fast-moving, directional ejection of gas generates the kink in the outer spiral edge that travels radially outward and shapes the global morphology of the wind. This kink is clearly visible at $x\approx 15\, \rm{au}$, $y\approx 10\, \rm{au}$ in the first snapshot ($t= 50.1$ yrs) in Fig.~\ref{zoomM17} and in the global morphology (upper left panel of Fig.~\ref{MorphFastEcc}). The energy of the fast-moving gas dissipates as it traverses the obstructing slower, denser gas, until it reaches an outward velocity similar to the wind particles in the FSE of the BHL flow. As the orbital separation around apastron is relatively large, the companion again compresses less wind material and re-enters the BHL regime, with a regular FSE-BSE structure attached to it.

\begin{figure*} 
    \centering
    \includegraphics[width = \textwidth]{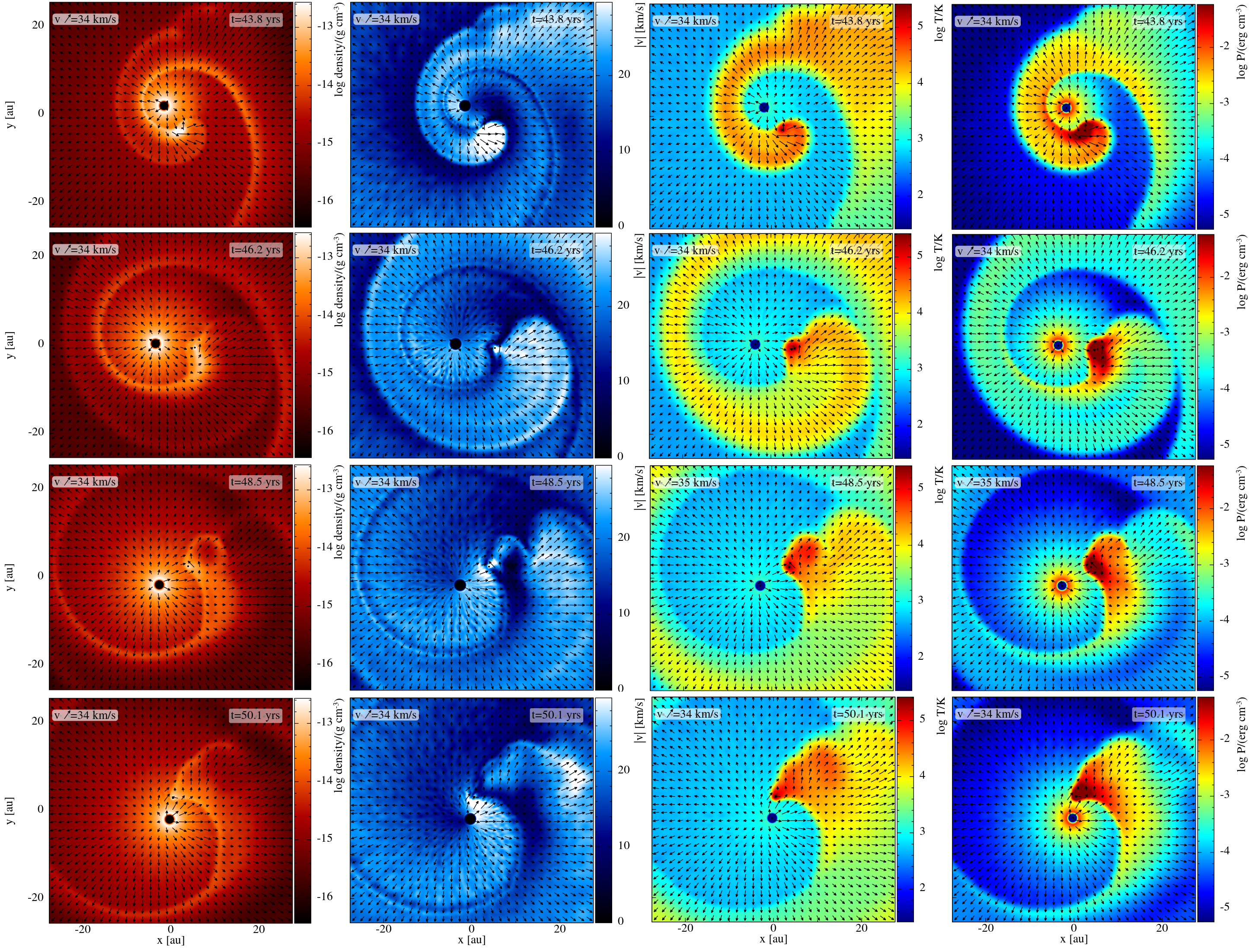}
    \caption{From left to right: density, speed, temperature and pressure distribution in the orbital plane of model v20e50 at 4 consecutive orbital phases. }
    \label{zoomM18}
\end{figure*}

The morphology of the same system with higher eccentricity, model v20e50, is shaped in a similar manner. The inner wind density, velocity, temperature and pressure profiles at four different timesteps of this model are shown in Fig.~\ref{zoomM18}. The higher eccentricity decreases the minimal orbital separation at periastron and thereby increases the amount of wind particles that interact with the companion. This results in a larger compression region as can be seen in the velocity and temperature profile in the first snapshot ($t=43.8$ yrs) in Fig.~\ref{zoomM18}. This region extends to the front of the companion, creating a bow shock and USF that delivers wind material to the high-density region behind the companion, as discussed in Sect.~\ref{Sect:v10e00}.  
Similar to model v20e25, high-energy material that is not attached to the companion travels radially outward around apastron (second snapshot, $t = 46.2$ yrs), creating a kink close to the companion. 
The rapidly propagating energetic gas particles traverse the previous inner spiral structure much earlier than in model v20e25. 
After apastron passage, the path of the companion changes, as it moves closer towards the AGB star, while a BSE-FSE structure forms behind it. The trajectory of the high-velocity gas within the bow shock that was surrounding the companion, diverges from the trajectory of the companion (at $t = 48.5$ yrs in Fig.~\ref{zoomM18}).
The transitions between the outflowing high-energy region, the bow shock and the newly forming FSE result in the turbulent structures appearing at the $y>0$ apastron side of the global morphology of Fig.~\ref{MorphFastEcc} (upper right panel).

Observations of the eccentric system AFGL3068 reveal spirals that seem to split up \citep{Kim2017}. These features are classified as `bifurcations' and have been shown to be characteristic patterns of eccentric systems \citep{Kim2019}. Such bifurcations also appear in the orbital plane morphology of model v20e50, in Fig.~\ref{MorphFastEcc}, and are locations that delimit low density gaps at the periastron side. 
The low-density regions and bifurcations originate from the variation in the inner wind structure at different orbital phases. Around apastron, the previously described energetic outflow makes the outer bow shock spiral catch up with the previous inner spiral structure, at $r \approx 20 \, {\rm au}$ (see $t=46.2 \, {\rm yrs}$ snapshot in Fig.~\ref{zoomM18}). 
However, around periastron, the inner wind structure is a 2-edged spiral with a limited difference in the radial velocity of the BSE and FSE. Thereby, the FSE only interacts with the previous BSE further out in the wind. This results in low-density regions between the FSE and previous BSE at periastron side, with a shrinking extent at increasing radii.
At the two points delimiting the low-density region, the spiral seems to converge or split up in two spirals, explaining why a bifurcation is observed. 
The low-density inter-arm regions only disappear when the spiral edges fully overlap after travelling more than $100\, \rm{au}$ in this model. 
This provides an explanation for the density asymmetry revealed in the models of \citet{Kim2019}. They proposed this asymmetry as a possible diagnostic for eccentric systems. However, our models show that for the selected set of orbital parameters this trend is only observed when the eccentricity is sufficiently large. For the low-density regions to survive up to the outer part of the envelope, the radial velocity of the FSE should be approximately as low as the radial velocity of the BSE.

\subsubsection{ Model v10e25 \& v10e50}
\label{Sectv10ecc}
\begin{figure*}
    \centering
    \includegraphics[width = \textwidth]{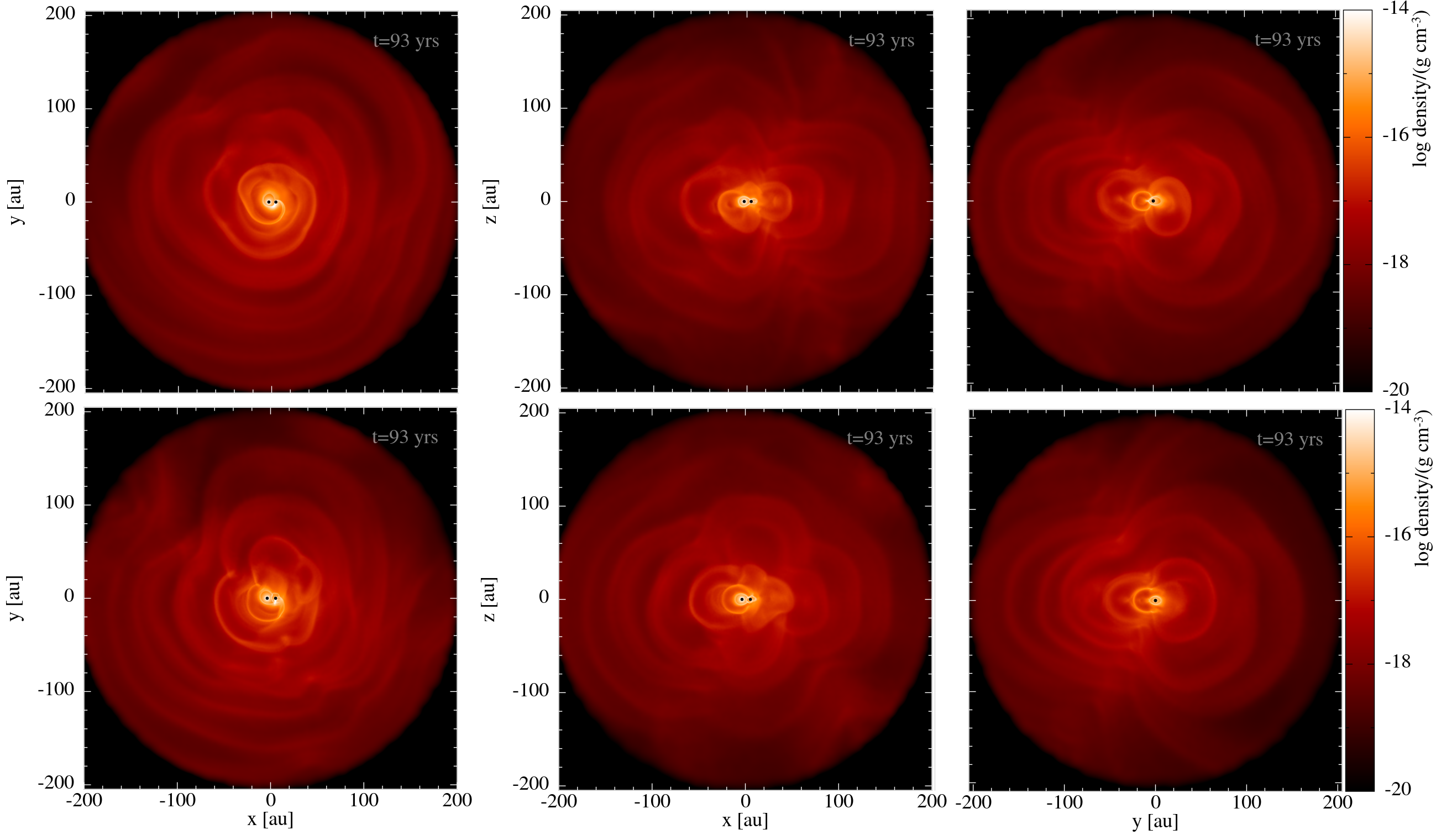}
    \caption{Density distribution in a slice through the orbital plane (left) and in two perpendicular meridional slices (middle and right) of models v10e25 (top) and v10e50 (bottom).}
    \label{MorphMedEcc}
\end{figure*}

\begin{figure} 
    \centering
    \includegraphics[width =0.49 \textwidth]{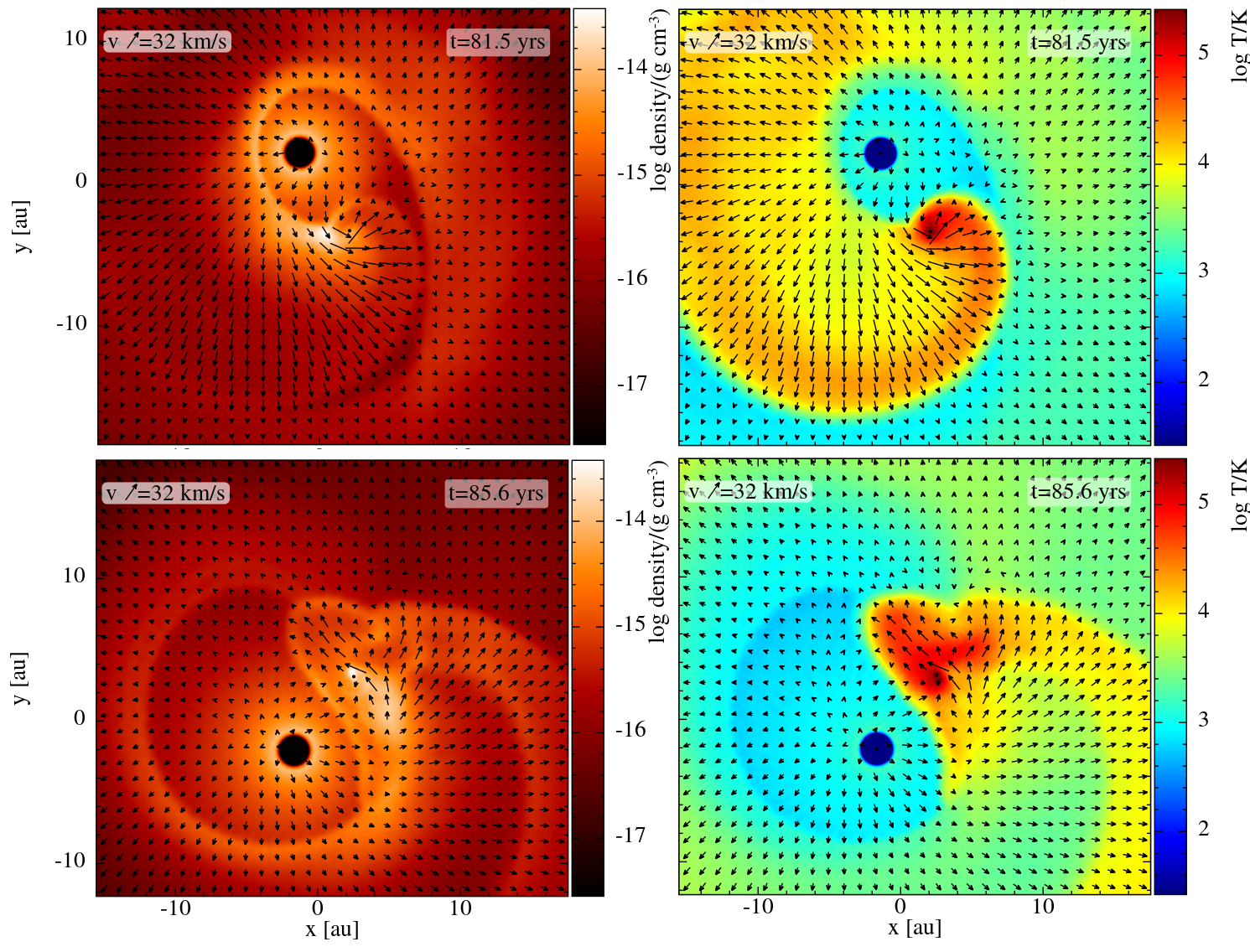}
     \caption{Density and temperature distribution in the orbital plane of model v10e25 at 2 consecutive orbital phases. }
    \label{zoomM20}
\end{figure}

\begin{figure} 
    \centering   
    \includegraphics[width =0.49 \textwidth]{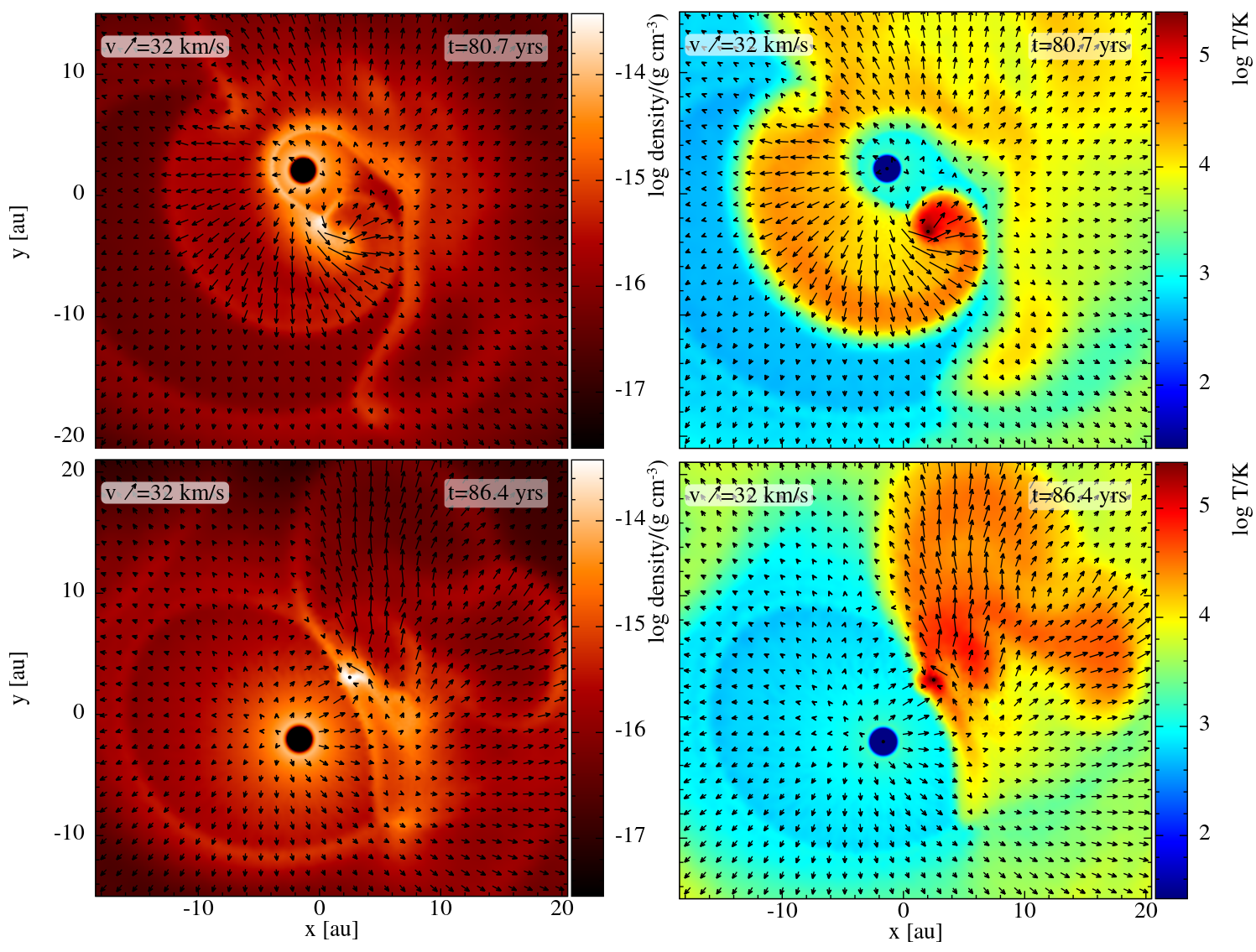}
    \caption{Density and temperature distribution in the orbital plane of model v10e50 at 2 consecutive orbital phases.}
    \label{zoomM21}
\end{figure}

For lower wind velocities, the stronger wind-companion interaction, which is enhanced around periastron passage in the eccentric systems, makes the inner wind structure more prone to periodic instabilities of the USF (see Sect.~\ref{Sect:v05e00}).
The 3D global morphology of models v10e25 and v10e50 is presented in Fig.~\ref{MorphMedEcc} through the density distribution in the orbital plane and in two meridional planes perpendicular to each other. The velocity distribution in these planes is shown in Fig.~\ref{VelMedEcc}.
In these models, we cannot identify an indication of regular spiral structures in the orbital plane and of arc or ring shapes in the meridional planes. The different density distribution in the two perpendicular meridional planes displays the 3D asymmetric nature and complexity of these models. To comprehend these complex morphologies, the inner wind density and temperature distribution in the orbital plane are displayed at two different timesteps for models v10e25 and v10e50 in Figs.~\ref{zoomM20} and \ref{zoomM21}, respectively. 

As the stars of model v10e25 (Fig.~\ref{zoomM20}) orbit from periastron to apastron, the companion is surrounded by a hot bow shock and USF that resemble the structure formed in the circular model v10e00, displayed in Fig.~\ref{zoomedM19}. When the stars approach apastron, this bow shock collides with the previous colder, slow inner edge structure. This structure is wrapped closer around the AGB star than in model v10e00, such that there is a more head-on collision with a larger impact on the irregularity of the morphology. This interaction results in an unstable USF and irregular bow shock structure that surrounds energetic wind material, as can be seen on the second snapshot plots ($t= 85.6$ yrs) in Fig.~\ref{zoomM20}. Only when the stars reach periastron, a more regular bow shock structure has formed again.

For model v10e50, at the first snapshot of Fig.~\ref{zoomM21}, the companion is once more surrounded by a bow shock and USF. Compared to model v10e25, the smaller orbital separation at periastron results in a stronger gas compression around the companion and thereby a larger pressure gradient. This makes the bow shock and high-energy region extend more to the front of the companion. 
Moreover, the higher eccentricity results in a more complex interaction of this bow shock with the low-energy structure of the previous orbit.
As the stars move towards periastron, part of the high-energy bubble, that was surrounded by the bow shock, escapes the stars and propagates radially outward in the positive $y$-direction (at $t=86.4$ yrs in Fig.~\ref{zoomM21}). 
While the energetic material propagates away from the companion, the star enters a low-energy, cold gas region and compresses new wind particles, resulting in the formation of a new high-energy region and bow shock. 
The escaping energetic material and the motion of the companion into the low-energy region result in the kink located at $x \approx -10\, \rm{au}$, $y \approx 10\, \rm{au}$ at $t=80.7$ yrs and in the energetic outflow in the $y>0$ region of the outer wind in the orbital plane, displayed in the upper middle plots in Figs.~\ref{MorphMedEcc} and \ref{VelMedEcc}.


\subsubsection{ Model v05e25 \& v05e50}
\label{sect:v05ecc}

\begin{figure*}
    \centering
    \includegraphics[width = \textwidth]{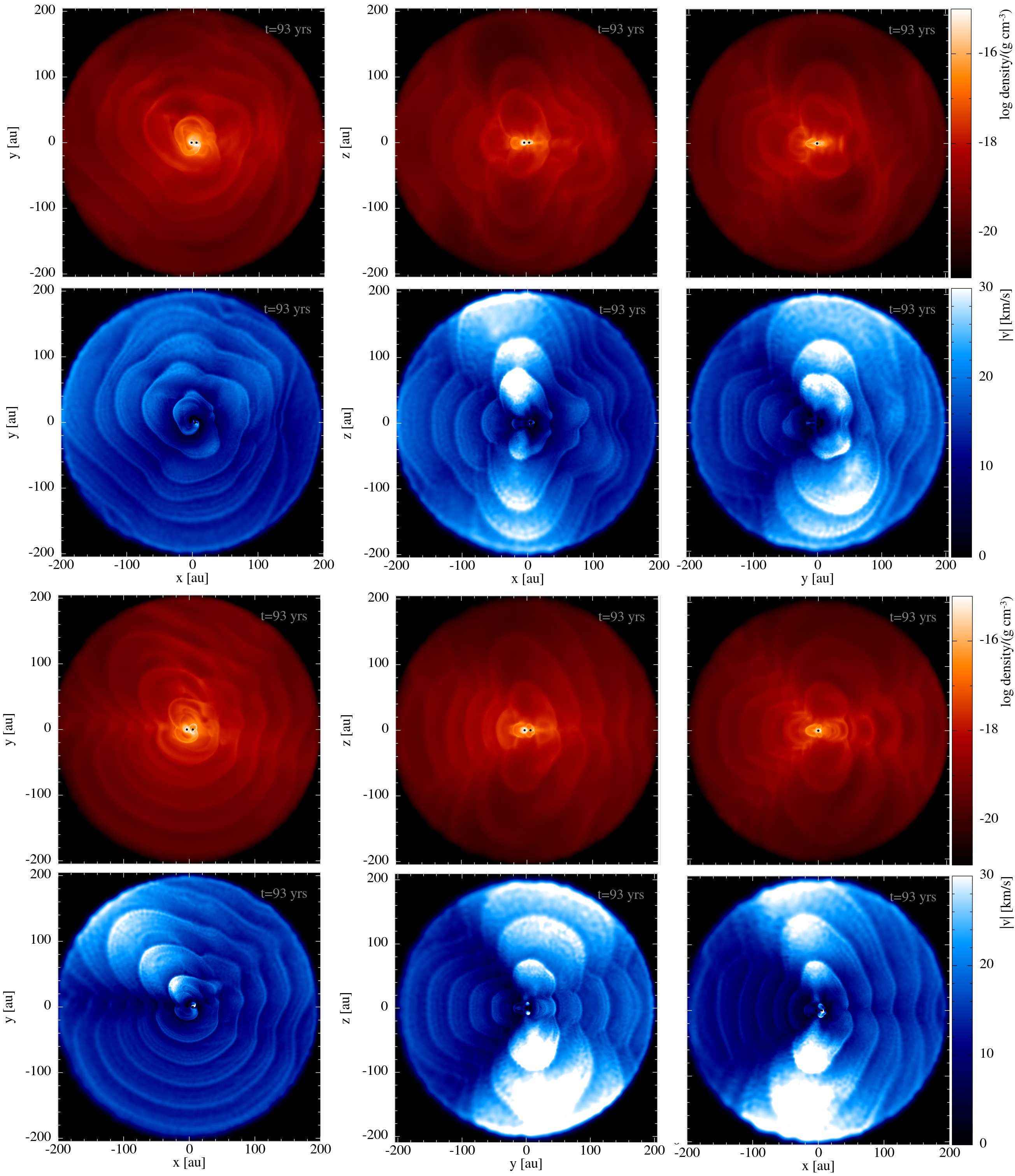}
    \caption{Density and velocity distribution in a slice through the orbital plane (left) and in two perpendicular meridional slices (middle and right) of models v05e25 (first two rows) and v05e50 (bottom two rows).}
    \label{MorphSlowEcc}
\end{figure*}

\begin{figure*} 
    \centering
    \includegraphics[width = \textwidth]{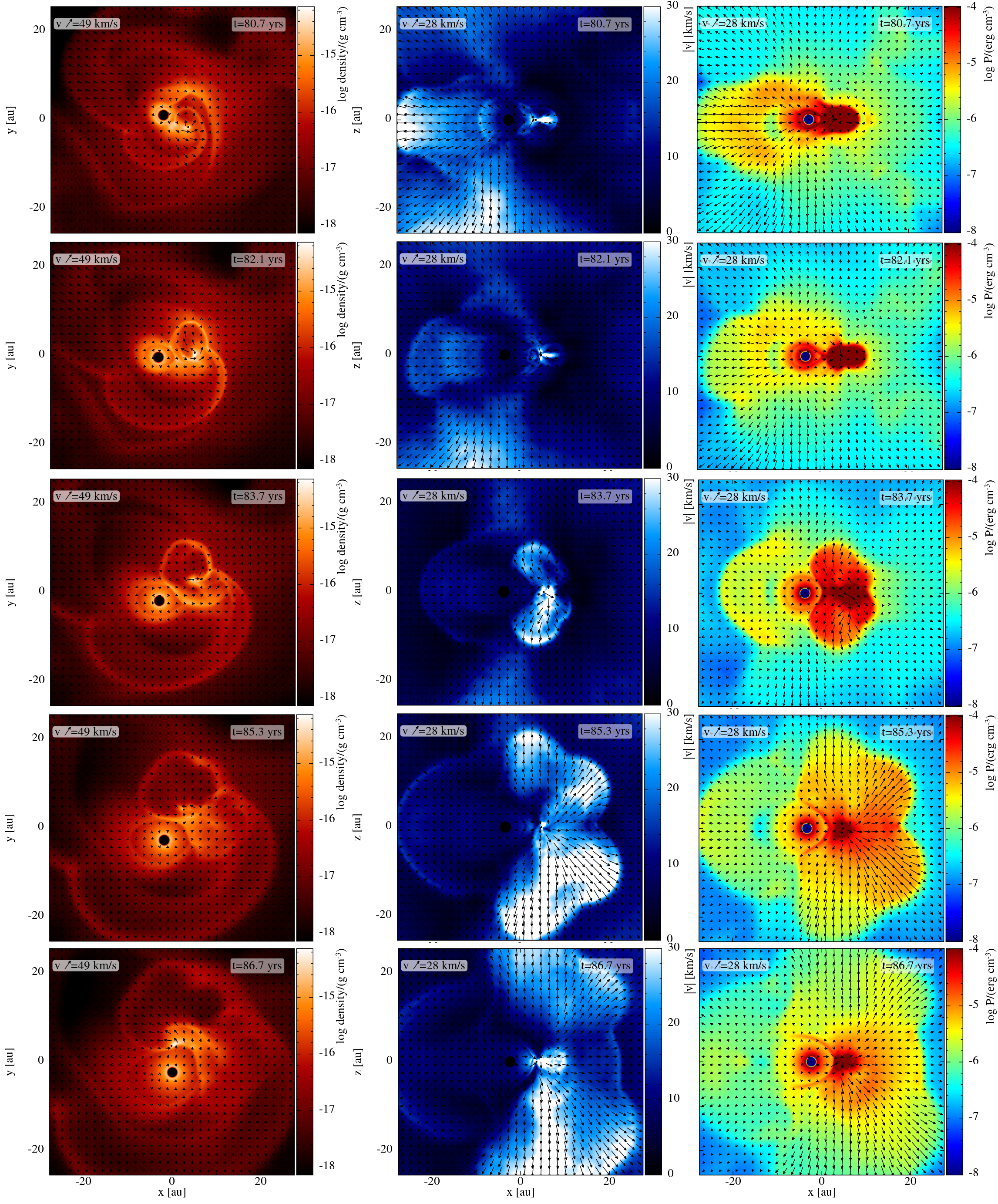}
    \caption{From left to right: the orbital plane slice density distribution, and the velocity and pressure distribution of model v05e50 in the meridional plane slice through the stars at 5 different timesteps.}
    \label{zoomM9}
\end{figure*}

The density and velocity distribution in the orbital plane and in two meridional planes perpendicular to each other of the eccentric lowest wind speed models v05e25 and v05e50 are displayed in Fig.~\ref{MorphSlowEcc}. The orbital plane structure of model v05e25 is similar to that of model v10e25 (upper left panel of Fig.~\ref{MorphMedEcc}). Model v05e50 has a similar, but more pronounced fast outflow and kink as model v10e50 in the $y>0$ region of the orbital plane. The meridional plane morphology of these models resembles those of models v10e25 and v10e50, except for the appearance of energetic polar outflows. 

The structure formation in these models is illustrated by the inner wind density in the orbital plane, and the velocity and pressure in the meridional plane, of model v05e50 at different timesteps in Fig.~\ref{zoomM9}. 
Although the wind-companion interaction is stronger, the mechanism shaping the wind structures of these models is similar as in models v10e25 and v10e50:  
(i) The companion is again surrounded by a high-energy region and relatively stable bow shock spiral in the orbital plane, when orbiting from periastron to apastron (e.g. at $t= 80.7$ yrs). This regular structure shapes the $x<0, y<0$ orbital plane region of model v05e50 in Fig.~\ref{MorphSlowEcc} into a regular spiral morphology and shapes the $x<0$ and $y<0$ sides of the two meridional planes in this figure into regular arc structures.
(ii) The regular bow shock again collides with another, irregular high-density structure and this results in a disturbed, unstable bow shock spiral (e.g. at $t = 82.1$ yrs and consequent snapshots in Fig.~\ref{zoomM9}). This bow shock spiral shapes the entire $x>0$ side of the orbital plane of model v05e50 in Fig.~\ref{MorphSlowEcc}.
(iii) In the same way as in model v10e50, the unstable bow shock surrounds a high-energy, low-density region that propagates away from the companion (e.g. region around $-5 \, \rm{au} \leq x \leq 10 \, \rm{au} $, $0 \, \rm{au} \leq y \leq 20 \, \rm{au}$ at $t=85.3 \, {\rm yrs}$ in Fig.~\ref{zoomM9}). Here, this results in a high-energy region in the $x<0, y>0$ part of the orbital plane (see lower left panels of Fig.~\ref{MorphSlowEcc}).
This high-energy region also emerges as a fast outflow in the $x<0$ direction in the meridional plane velocity distribution at the first snapshot ($t = 80.7$ yrs) of Fig.~\ref{zoomM9}.

To understand how the remarkable fast polar outflows in the meridional plane of models v05e25 and v05e50 are formed, the velocity and pressure in the meridional plane through the stars is presented next to the corresponding orbital plane density at different orbital phases in Fig.~\ref{zoomM9}.
The first and second snapshots reveal how the companion compresses material around it in a compact, high-pressure region. Around apastron ($t = 83.7$ yrs), the strong pressure gradient causes this high-energy region to expand towards the poles, where the surrounding density is substantially lower. This expansion proceeds with speeds up to $30\,{\rm{km\,s^{-1}}}$ (see lower right plots of Fig.~\ref{1Dplots}). The butterfly-shaped high-pressure region (see last two snapshots in Fig.~\ref{zoomM9}) shapes the energetic polar outflows of the meridional planes in Fig.~\ref{MorphSlowEcc}.
The cold meridional plane arc structures at the side of the AGB star (in the last three snapshots of Fig.~\ref{zoomM9}) create the regular arc structures on the $x<0$ and $y<0$ sides of the meridional planes in Fig.~\ref{MorphSlowEcc}.

%% file: discussion.tex
\section{Discussion}
\label{ch:discussion}
In Sect.~\ref{ch:results} we focused on the detailed structures in the wind morphology and the mechanisms shaping them. In this section we focus on the global density distribution of the models on a more quantitative basis.
Previous studies reveal that the global morphology of binary AGB outflows deviates from spherical symmetry, depending on the configuration of the system \citep[e.g. ][]{MastrodemosII,Kim2012B,ElMellah2020}. Observations of Post-AGB systems often show a circumbinary disk and bipolar outflows \citep{VanWinckel2003,Bujarrabal2013}. The study of flattened global morphologies, or density enhancements around the orbital plane, in the progenitor AGB phase may provide critical information about the origin of such disks.
\label{globalMorph}

\subsection{Flattened global morphology}
\label{flatteningSection}

\begin{figure}
    \centering
    \includegraphics[width =0.4\textwidth]{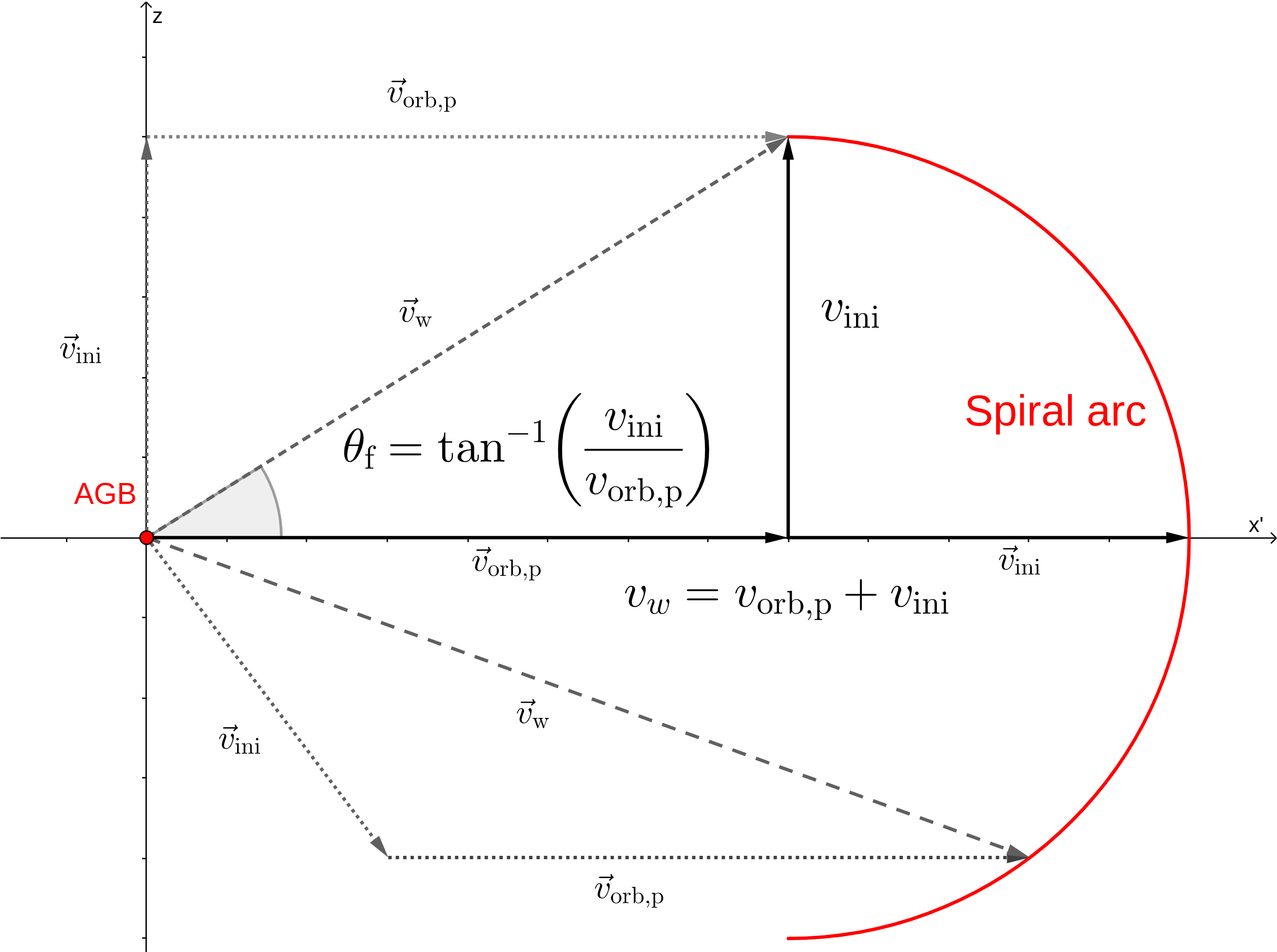}
    \caption{Spiral arc propagation in meridional view, only taking into account the orbital motion of the stars. 
    The ratio of the maximum height, reached by a wind particle sent out in the $z$-direction, to the maximal distance travelled by a particle sent out in the orbital plane direction is the flattening ratio $R_{\rm f}$ given by Eq.~(\ref{R_f}). The angle with the orbital plane of the estimated path of a particle that reaches a maximal height is $\theta_{{\rm f}}$ given by Eq.~(\ref{theta_f}).}
    \label{theorFlatteningFig}
\end{figure}

If the stellar wind is modelled in a frame co-rotating with the stars, the orbital motion of the stars results in (i) a Coriolis force $F_{\rm cor} = -2 m_i \vec{\Omega} \times \vec{v}_{i}$ and (ii) outward directed centrifugal force $F_{\rm cen} = - m_i \vec{\omega} \times (\vec{\Omega} \times \vec{r}_{i})$ on a particle with mass $m_i$, position $r_i$ and velocity $\vec{v}_{i}$, measured in the non-inertial reference frame that rotates with an orbital velocity $\vec{\Omega}$.
These forces depend on the gas velocity with respect to the location of the AGB star. Because these gas velocities possess a strong polar-angle dependence, these forces contribute to the asymmetric shaping of the wind (second law of Newton). In the non-rotating frame, the same effect is established by the orbital velocity of the AGB star that alters the entire wind velocity distribution. Based on this insight, we make an analytical prediction of the shape of the arcs in the meridional plane, illustrated in Fig.~\ref{theorFlatteningFig}. This provides a prediction of the flattening of the model, resulting solely from the effect of the orbital motion of the AGB star.
The wind particles leave the AGB star with a velocity 
\begin{equation}
    \vec{v}_{\rm w} = \vec{v}_{\rm ini}+\vec{v}_{{\rm orb,p}},
\end{equation}
which is the vector sum of the initial velocity $\vec{v}_{\rm ini}$ and the orbital velocity of the AGB star $\vec{v}_{{\rm orb,p}}$, which deflects the path of the particles to the orbital plane.
Wind particles that are sent out by the AGB star perpendicular to the orbital plane, follow a radial direction with an angle 
\begin{equation}
    \label{theta_f}
    \theta_{{\rm f}} = \tan^{-1}{\left(\frac{v_{\rm ini}}{v_{{\rm orb,p}}}\right)}
\end{equation}
with respect to the orbital plane. 
We quantify the degree of flattening as the ratio of the `vertical' travelled distance in the direction perpendicular away from the orbital plane to the `horizontal' travelled distance in the orbital plane direction. A lower ratio corresponds to a stronger predicted flattening.
From Fig.~\ref{theorFlatteningFig} we deduce the flattening ratio
\begin{equation}
    \label{R_f}
    R_{\rm f} = \frac{v_{\rm ini}}{v_{\rm ini} + v_{{\rm orb,p}}},
\end{equation}
with a lower $R_{\rm f}$ corresponding to a stronger predicted flattening.
This approach does not account for the gravitational influence of the companion on the wind particles, and the pressure gradients causing accelerations towards less dense regions are ignored. However, we show that this simplified approach suffices as a first-order estimator.

We measure the ratio $R_{\rm f}$ of our models by identifying the travelled distance in and perpendicular to the orbital plane of one spiral structure. 
These measured $R_{\rm f}$-values are given in Table~\ref{FlatRatioTable}, together with the analytically predicted ratios (Eq.~(\ref{R_f})).
To measure $R_{\rm f}$, we use the orbital-plane width and meridional plane height of a single arc, which we obtain from Fig.~\ref{1DFlratioPlots}.
The plots in this figure represent the residual density peaks after subtraction of the $1/r^2$ density profile on the positive and negative part of the $x$-axis in the orbital plane through the CoM (blue lines) and on axes perpendicular to the orbital plane through $x=50 \,\rm{au}$ and $x=-50 \,\rm{au}$ (orange lines). 
In the eccentric systems, the flattening ratio varies between apastron and periastron and is therefore calculated separately at both sides.

The strongest flattening, thus lowest $R_{\rm f}$, is measured in model v10e00, in which the spiral structures only reach about half the maximal travelled distance in the polar direction. 
Model v20e00 is less flattened, with $R_{\rm f} \approx 0.6$, due to the higher wind velocity. The eccentric models v20e25 and v20e50 are also flattened with measured ratios of about $R_{\rm f} \approx 0.55$ at apastron and respectively $0.6$ and $0.7$ at periastron. These measurements agree well with the analytically predicted ratios, although the measured flattening in these models is slightly stronger, which could be due to the gravitational impact of the companion which is ignored in the analytical prediction.
However, in model v05e00 no flattening is measured (measured $R_{\rm f} \approx 1$), whereas this model is predicted to have the strongest flattening (lowest analytical $R_{\rm f}$). 
Similarly, the analytical ratio predicts a strong flattening for models v05e25, v05e50, v10e25 and v10e50, which do not appear flattened in the meridional plane plots (in Figs.~\ref{MorphMedEcc} and \ref{MorphSlowEcc}), and for which no flattening ratio is measured due to a lack of recurrent arc structures. 
This discrepancy is caused by the irregular structures and energetic outflows that counteract the flattening and that arise due to the (in most cases periodic) launch of hot material in various directions as is discussed in Sect.~\ref{ch:results}. 

\begin{table}
    \caption{Flattening ratio.}
    \begin{center}
    \begin{tabular}{lcc}
    \hline
    \hline
     Model &Analytical $R_{\rm f}$ & Measured $R_{\rm f}$ \\
     &apa $|$ per & apa $|$ per \\
    \hline
    v05e00          & $0.39$ & $1$ \\
    v05e25          & $0.36$ $|$ $0.42$ & Irr \\
    v05e50          & $0.31$ $|$ $0.44$ & Irr \\
    v10e00          & $0.57$ & $0.5$ \\
    v10e25          & $0.53$ $|$ $0.59$ & Irr \\
    v10e50          & $0.48$ $|$ $0.61$ & Irr \\
    v20e00          & $0.72$ & $0.6$ \\
    v20e25          & $0.69$ $|$ $0.74$ &  $0.55$ $|$ $0.6$ \\
    v20e50          & $0.65$ $|$ $0.76$ &  $0.55$ $|$ $0.7$  \\
    \hline
    \end{tabular}
    \end{center}
    {\footnotesize{\textbf{Notes.} Analytically predicted and measured flattening ratio $R_{\rm f}$. The lower $R_{\rm f}$, the stronger the flattening. For the eccentric models the values are given both at the companion apastron and companion periastron side as $R_{\rm f}$(apa) $|$ $R_{\rm f}$(per). `Irr' indicates models for which the flattening ratio could not be measured due to irregular structures. For interpretation of this ratio see Fig.~\ref{theorFlatteningFig}.}}
    \label{FlatRatioTable}
\end{table}

\subsection{Equatorial Density Enhancement (EDE) }
\label{EDESection}

\begin{figure}
    \centering
    \includegraphics[width = 0.4\textwidth]{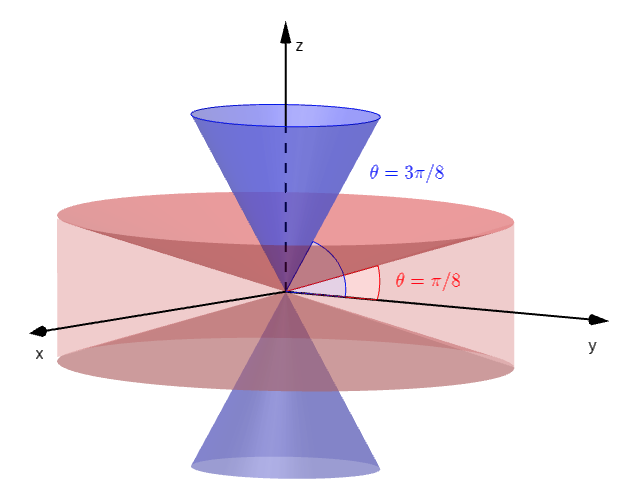}
    \caption{Conical torus around the orbital plane (red) and the cone around the polar axis (blue)  applied in the EDE ratio calculations in Table \ref{EDEtable}.}
    \label{EDEregions}
\end{figure}

A second, other possible origin for post-AGB circumbinary disks in the AGB phase is a density enhancement around the orbital plane. This accumulation of matter around the orbital plane is different from the flattening, since it does not originate from the orbital motion of the AGB star, but from the gravitational pull of the companion. This density enhancement is referred to as an equatorial density enhancement or EDE in short.
In our simulations, the equatorial plane coincides with the orbital plane. 
We assume the presence of an EDE when the density around the orbital plane is significantly larger than the density at large heights from the orbital plane and when this effect is not caused solely by the previously described global flattening.

To quantify the strength of the EDE, we subdivide the entire wind hemisphere into four regions, which are defined by equal ranges of the angle $\theta$ w.r.t. the orbital plane. This angle is defined as $\theta = 0$ for the orbital plane and $|\theta| = \pi/2$ for the polar axis. We construct a conical torus surrounding the orbital plane as the region defined by $|\theta| \leq \pi/8$, and a cone around the polar axis, that contains the gas particles with $3\pi/8 \leq |\theta| \leq \pi/2$, as illustrated in Fig.~\ref{EDEregions}. 
We define an EDE to be present when the mass in the conical torus, $M_{{\rm bin}}[{\rm orb}]$, is higher than in case of an isotropic density distribution, and the mass in the polar cone, $M_{{\rm bin}}[{\rm pol}]$, is lower: 
\begin{equation}
    M_{{\rm bin}}[{\rm orb}] \gg M_{{\rm iso}}[{\rm orb}] \\
    {\rm  and } \\
    M_{{\rm bin}}[{\rm pol}] \ll M_{{\rm iso}}[{\rm pol}].
\end{equation}
This is the case when
\begin{equation}
    \label{mu}
    \mu_{\rm orb} = \frac{M_{{\rm bin}}}{M_{{\rm iso}}}[{\rm orb}] \gg 1 \\
    {\rm  and } \\
    \mu_{\rm pol} = \frac{M_{{\rm bin}}}{M_{{\rm iso}}}[{\rm pol}] \ll 1, 
\end{equation}
where we define the EDE ratios $\mu_{\rm orb}$ and $\mu_{\rm pol}$ as the mass that is accumulated in the conical torus and polar cone in the binary models ($M_{\rm bin}$), respectively, scaled to the mass in the same region in a corresponding single-star model with an isotropic density distribution ($M_{\rm iso}$).
Table~\ref{EDEtable} provides the measured $\mu_{\rm orb}$ and $\mu_{\rm pol}$ ratios. 

The EDE ratios of the fast-wind models approach $1$, confirming that no EDE is formed and that the mass is distributed quasi-isotropically, as expected from the similar density profile on the polar axis and orbital plane axes presented in Fig.~\ref{1Dplots}.
The strongest EDE is identified in models v10e00 and v05e00. In model v10e00 the EDE is strongest, since about $1.4$ times more mass is accumulated in the conical torus around the orbital plane compared to the isotropic model and since the amount of mass within the polar cone is only about $40\%$ of that in the isotropic model. Note that we should take into account that the strong flattening observed in this model could slightly contribute to this EDE.
Increasing the eccentricity results in a weaker EDE in case of the $v_{\rm ini} = 10\,{\rm{km\,s^{-1}}}$ models, as the EDE ratios in Table~\ref{EDEtable} get closer to $1$. Contrary, a higher eccentricity results in a slightly stronger EDE for the models with $v_{\rm ini} = 5\,{\rm{km\,s^{-1}}}$.

We can conclude that an EDE is more likely to form in case of a lower wind velocity due to the stronger impact of the gravitational attraction of the companion on the wind material, and that no consistent relation is found between eccentricity and the strength of this EDE. 

\begin{table}
    \caption{EDE mass ratio.}
    \begin{center}
    \begin{tabular}{lcc}
        \hline
        \hline
         Model & $\mu_{\rm orb}$ & $\mu_{{\rm pol}}$ \\
        \hline
        v05e00          & $1.35$ & $0.49$ \\
        v05e25          & $1.35$ & $0.43$ \\
        v05e50          & $1.37$ & $0.40$ \\
        v10e00          & $1.42$ & $0.5$  \\
        v10e25          & $1.34$ & $0.37$ \\
        v10e50          & $1.25$ & $0.46$ \\
        v20e00          & $0.93$ & $1.05$ \\
        v20e25          & $0.95$ & $0.98$ \\
        v20e50          & $0.97$ & $0.86$ \\
        \hline
    \end{tabular}
    \end{center}
    {\footnotesize{\textbf{Notes.} Mass accumulated in a conical torus around the orbital plane and in a cone around the polar axis of the binary model $M_{{\rm bin}}$, scaled to the accumulated mass in the same region in a corresponding isotropic model $M_{{\rm iso}}$, as defined in Eq.~(\ref{mu}). The conical torus and polar cone are illustrated in Fig.~\ref{EDEregions}.}}
    \label{EDEtable}
\end{table}

\subsection{Energetic outflows caused by inefficient cooling}

The low wind-velocity models contain a high-pressure region around the companion where the wind particles are compressed, generating energetic outbursts, as described in detail in Sect.~\ref{ch:results}. 
We associate the formation of these high-energy outflows with inefficient cooling around the companion. Since cooling in our models is governed by the EOS of a polytropic gas (Eq.~(\ref{IdealGas})) with $\gamma = 1.2$, without any additional cooling factors, a more realistic cooling implementation may result in other morphologies than obtained in this work. In case of strong cooling, accretion disks may form instead of energetic outflows, as cooling limits the increase in pressure, allowing wind material to remain closer to the companion \citep{TheunsI}.

In two of our models, the inefficient cooling gives rise to bipolar outflows (see Sect.~\ref{sect:v05ecc}).
These bipolar outflows resemble the widely observed bipolar jets and outflows of binary post-AGB systems \citep{Bollen2017}. These outbursts are generally believed to be generated from a jet, that originates from an accretion disk around the companion, although the exact launching mechanism is still poorly understood \citep{Bollen2021}. 
Our simulations show that bipolar outflows could be present in the wind environment of AGB stars with a companion, when there is inefficient cooling near the companion, in combination with a compression of wind material. Moreover, the proposed physical conjecture of the possible origin of high-energy outflows due to inefficient cooling, could shed new light on the general understanding of the launching mechanisms of observed bipolar outflows in other type of systems.

\subsection{Morphology classification}
A classification of the global distribution of the gas according to the flattening and the presence of an EDE of each model can be found in Table~\ref{classificationTable}. The flattening is classified according to the measured $R_{\rm f}$ ratios (see Table~\ref{FlatRatioTable}) as follows: `Roughly spherical' in case $R_{\rm f} \approx 1$, `Flattened' in case $R_{\rm f} \leq 0.7$, `Irregular' if $R_{\rm f} =$ `Irr' and `(asymmetric)' if $R_{\rm f}$(apa)$ \neq R_{\rm f}$(per). 
In the literature no clear distinction has yet been made between an EDE and a flattened morphology. However, these features arise independently in different systems and are caused by different mechanisms. An EDE here arises for the models with input wind velocity $v_{\rm ini} \leq 10\,{\rm{km\,s^{-1}}}$, while the models with higher input wind velocities are flattened (limited to the restricted parameter selection of our simulations).
In order to present a complete classification, Table~\ref{classificationTable} also gives the characteristic meridional and orbital plane structures as described in Sect.~\ref{ch:results}.

\cite{Maes2021} investigated several classification parameters that attempt to predict the type of morphology of the AGB outflow given the input parameters of the simulation. They defined a parameter $\varepsilon$ as the ratio of the gravitational energy density of the companion to the kinetic energy of the wind:
\begin{equation}
    \label{vareps}
    \varepsilon = \frac{e_{{\rm grav}}}{e_{{\rm kin}}} = \frac{ \frac{G M_s \rho} {R_{\rm Hill}}}{\frac{1}{2} \rho v_{\rm w}^2},
\end{equation}
with $R_{\rm Hill} = a(1-e) \sqrt[3]{M_{\rm s} / (3 M_{\rm p})}$ \citep{HillSphere},
and wind velocity  
\begin{equation}
    v_{\rm w} = \sqrt{v_{\rm single}^2(r = a) + v_{\rm p}^2},
\end{equation}
with $v_{\rm single}(r)$ the wind velocity at radius $r$ in an isotropic, single-star model with the same input wind velocity and AGB star as the binary model.

They found that $\varepsilon \gg 1$ corresponds to wind morphologies with a bow shock spiral or high degree of complexity and irregularity, whereas for $\varepsilon \approx 1$, a regular Archimedes spiral results in the outflow.
Since in their work the set of simulations is limited to circular models, we here test if this classification parameter also holds for eccentric systems.
The complexity of our models increases for higher eccentricity and lower wind velocity. We calculated $\varepsilon$, given in Table \ref{classificationTable}, and find that (i) the highest $\varepsilon$ values correspond to models that reveal an unstable bow shock and strong irregularity, such as model v05e50, and that (ii) values of $\varepsilon \approx 1$ correspond to the most regular models v20e00, v20e25 and v20e50 with an approximate or perturbed Archimedes spiral in the orbital plane. This corresponds to the classification scheme found by \cite{Maes2021}. Hence, $\varepsilon$ also provides a good global estimate of the complexity of the morphology in eccentric systems.

\begin{table*}
    \caption{Morphology classification.}
    \begin{center}
    \begin{tabular}{clrllc}
    \hline
    \hline
    &&&&\\[-2ex]
    Model&   Global density distribution & & Meridional plane structure & Orbital Plane structure & $\varepsilon$\\
    &&&&\\[-2ex]
    \hline
    v05e00  & No flattening, Irregular & with EDE & Rose                                          & Spiral - Squared & $4.1$ \\
    v05e25  & No flattening, Irregular         & with EDE & Bipolar outflow                       & Irregular & $5.5$\\
    v05e50  & No flattening, Irregular          & with EDE & Bipolar outflow                       & Irregular & $8.3$\\
    
    v10e00  & Flattened           & with EDE & Bicentric rings - Peanut-shape             & Spiral - Archimedes & $2.6$ \\
    v10e25  & No flattening, Irregular           & with EDE & Irregular                             & Irregular & $3.4$\\
    v10e50  & No flattening, Irregular           & with EDE & Irregular                             & Irregular & $5.1$\\
 
    v20e00  & Flattened              & no EDE  & Concentric arcs                   & Spiral - Archimedes & $1.0$ \\
    v20e25  & Flattened, asymmetric & no EDE  & Arcs                              & Spiral - Perturbed & $1.3$\\
    v20e50  & Flattened, asymmetric & no EDE  & Ring-arcs                         & Spiral - Perturbed & $2.0$\\
    \hline
    \end{tabular}
    \end{center}
    {\footnotesize{\textbf{Notes.} 
    Morphology classification of the models, subdivided in: (i) global density distribution indicating if there is a flattening and/or EDE, (ii) structure in the meridional plane slice through the sink particles, (iii) wind structures in the orbital plane and (iv) classification parameter $\varepsilon$ defined as in Eq.~(\ref{vareps}).}}
    \label{classificationTable}
\end{table*}

%% file: conclusion.tex
\section{Summary}
\label{ch:conclusion}
 In this work, we performed high-resolution 3D hydrodynamic simulations with the SPH code \textsc{Phantom} of the outflow of a mass-loosing AGB star including a solar-mass companion. We constructed nine different models with a polytropic gas without additional cooling, and varied the input wind velocity and orbital eccentricity in order to unravel the impact of these parameters on the wind structure formation and on the resulting global morphology. We do this via the density, temperature, velocity and pressure distribution in 2D slices at different snapshots through the simulations. 
 We also present an analysis of the global morphology studying the degree of flattening and identifying which models have an equatorial density enhancement (EDE). 
 
 The wind structures of the circular models show increasing complexity with decreasing wind velocity. The global morphology of the highest wind velocity model harbours a spiral structure in the orbital plane, which we modelled as an Archimedes spiral, and concentric arcs in the meridional plane. This morphology results from a two-edged, broadening spiral structure attached to the companion, that originates from its (limited) gravitational attraction on the wind particles.
 Decreasing the wind velocity, the companion compresses more wind material around it, resulting in the formation of a high-pressure bubble surrounding the companion, delimited by a bow shock in the direction of orbital motion. This results in a regular spiral morphology with bicentric rings. Decreasing the wind velocity further, this high-pressure region becomes periodically unstable, with energetic outflows shaping the wind into irregular morphologies in three dimensions. We associate the formation of these energetic outflows to inefficient cooling around the companion.
 
 Introducing eccentricity increases the strength of the wind-companion interaction and makes it phase dependent, and thereby generates additional degrees of complexity. In the eccentric models with highest wind velocity, this leads to slightly perturbed spiral structures in the orbital plane and asymmetric arcs in the meridional plane. For lower wind velocity models, the increased gas compression around the companion generates an unstable, complex-structured bow shock spiral. Similar to the circular, low velocity model, this instability leads to escaping high-energy material. The three-dimensional directions at which high-energy matter escapes, determine the global morphology of these models. In the eccentric low-velocity models, the outburst are most energetic and are primarily directed towards low-density regions away from the orbital plane, in the form of high-speed bipolar outflows.
 
 The global distribution of gas can both be flattened by the orbital motion of the stars, and can harbour an EDE, resulting from the gravitational redistribution of wind particles into the orbital plane by the companion. For the models with regular morphologies, a stronger flattening arises for larger orbital velocities. 
 The theoretical prediction of the flattening, in which the gravitational impact of the companion is neglected, is found to underestimate the measured flattening. 
 When instabilities arise in the inner wind structures, the measurements strongly deviate from the theoretical predictions. This is caused by the energetic outflows that clear away the flattening. On the contrary, the formation of an EDE appears to be independent of this effect, as an EDE is detected in all lower-velocity models.
 
 At last, we provided an overview of the global density distribution and wind structures in the orbital plane and meridional planes of our models in a morphology classification.

%% file: appendix.tex
\appendix

\section{Radial density and velocity distribution in three directions}
\label{radial1Dplots}
\begin{figure*}
    \centering
    \includegraphics[width = 0.2875\textwidth]{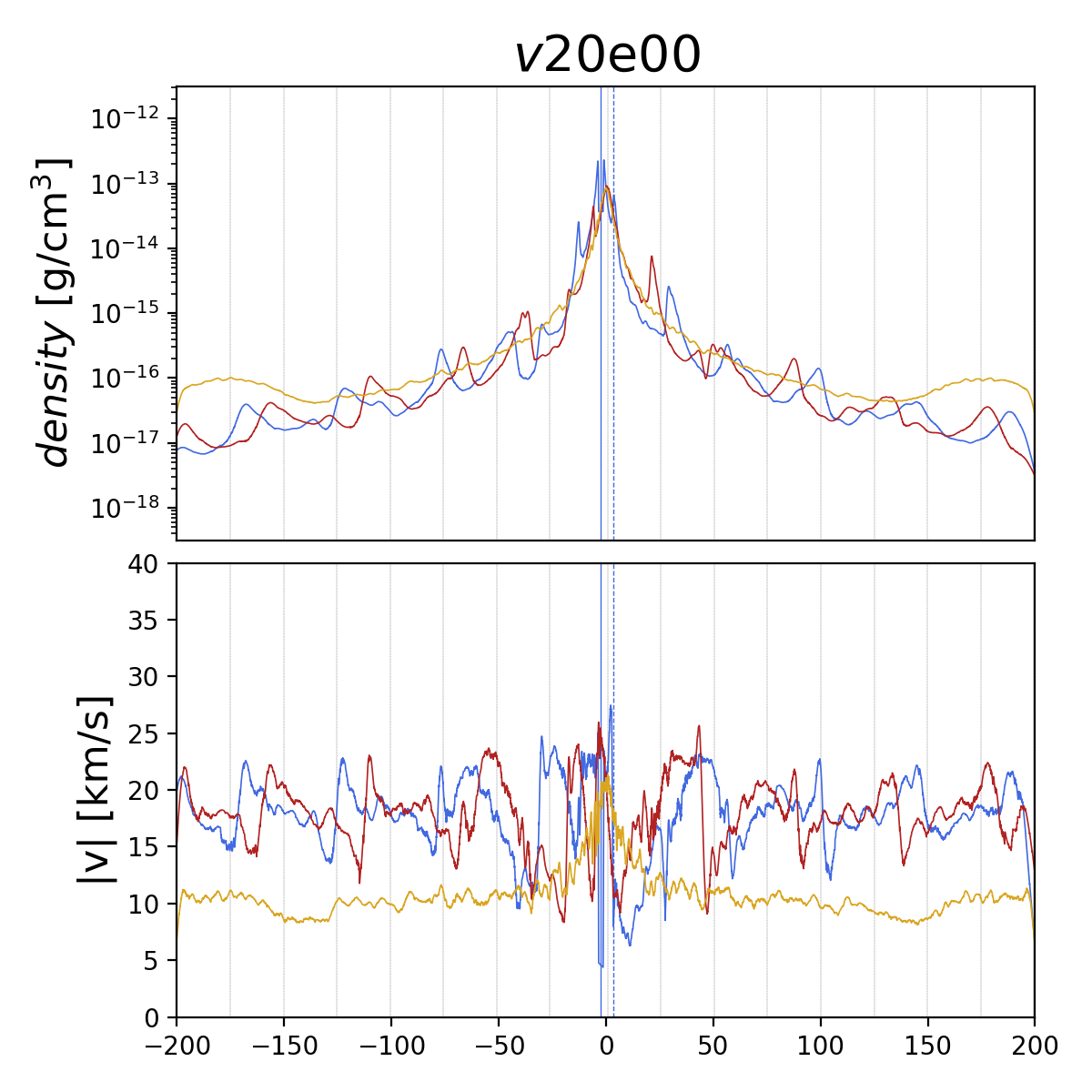}
    \includegraphics[width = 0.2875\textwidth]{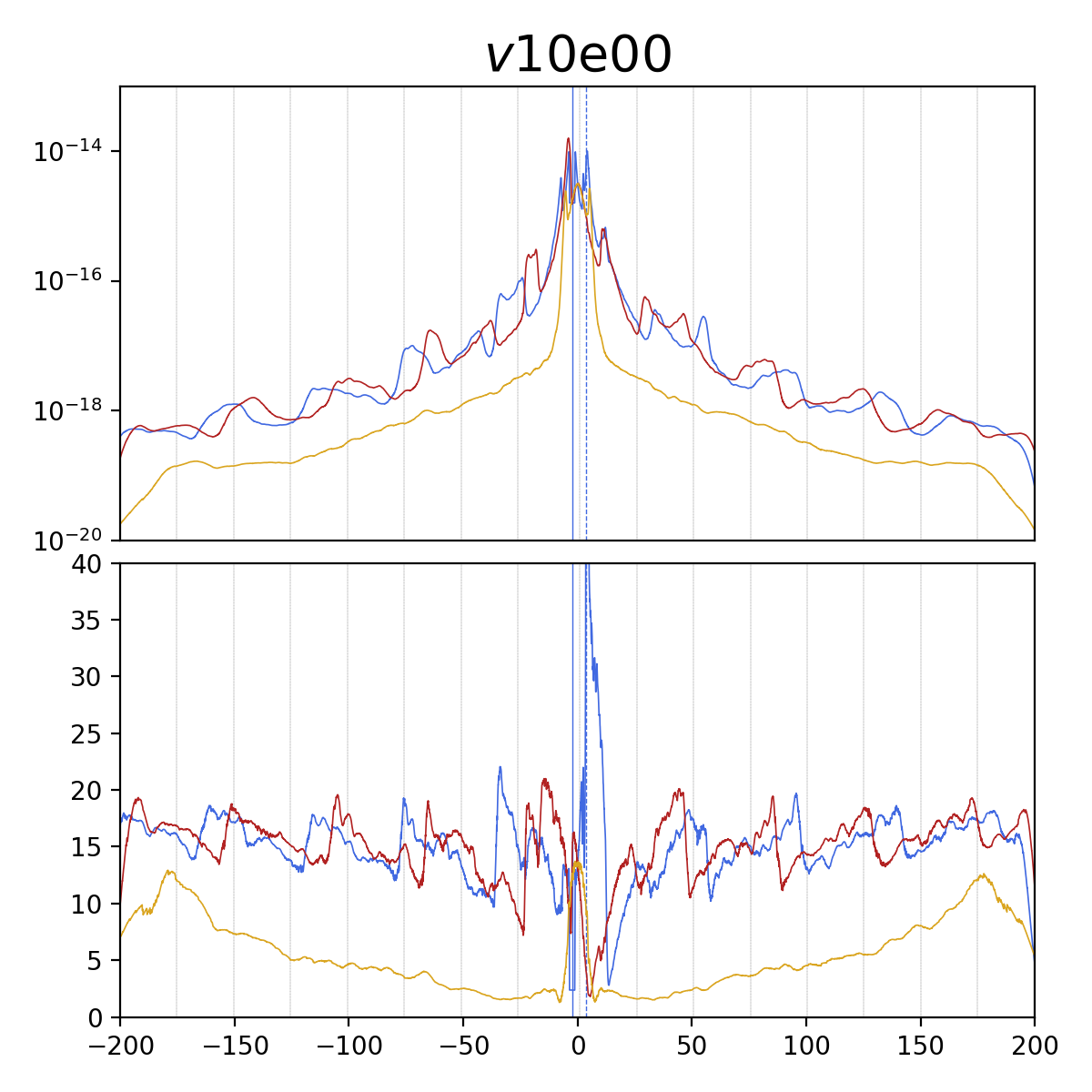}
    \includegraphics[width = 0.2875\textwidth]{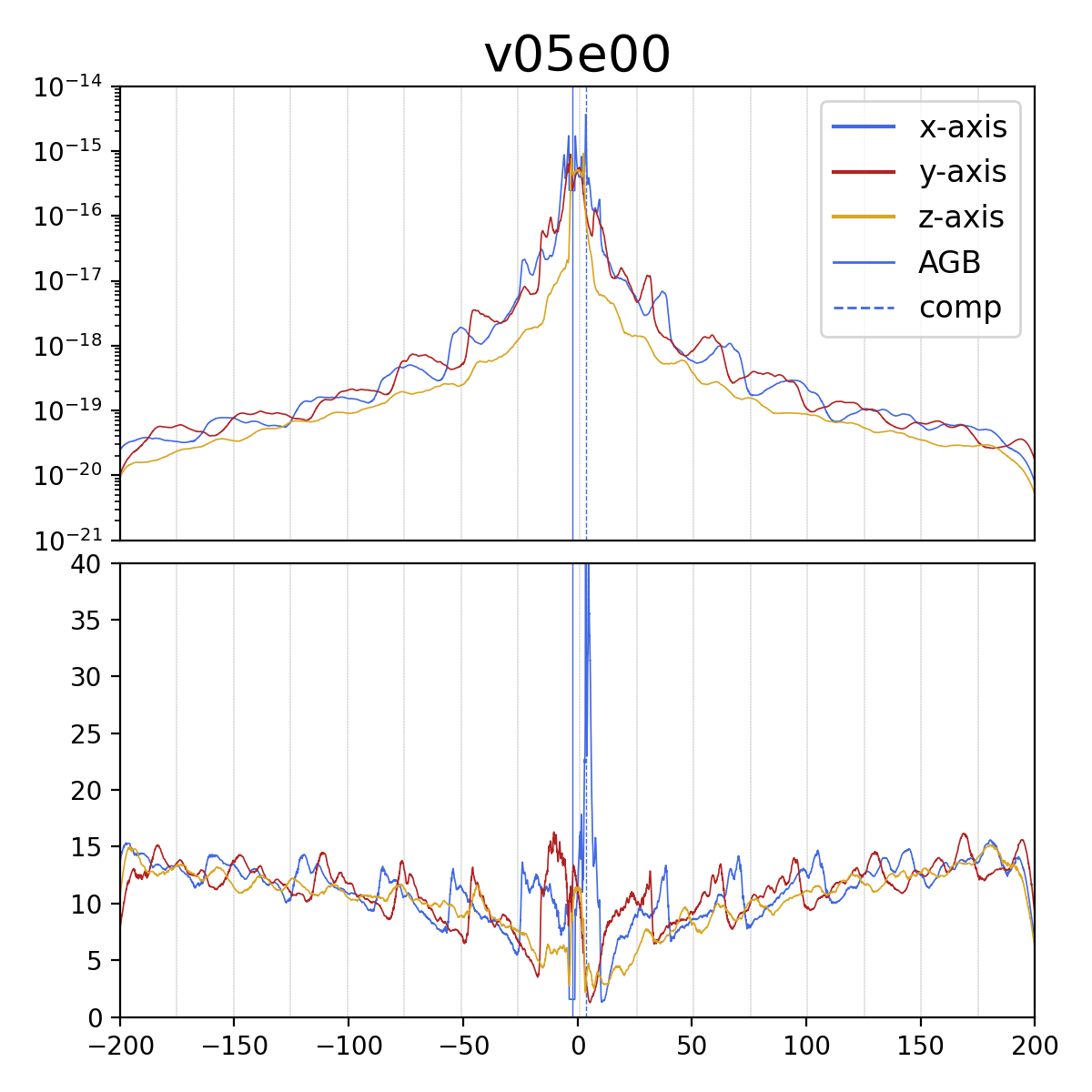}
    \includegraphics[width = 0.2875\textwidth]{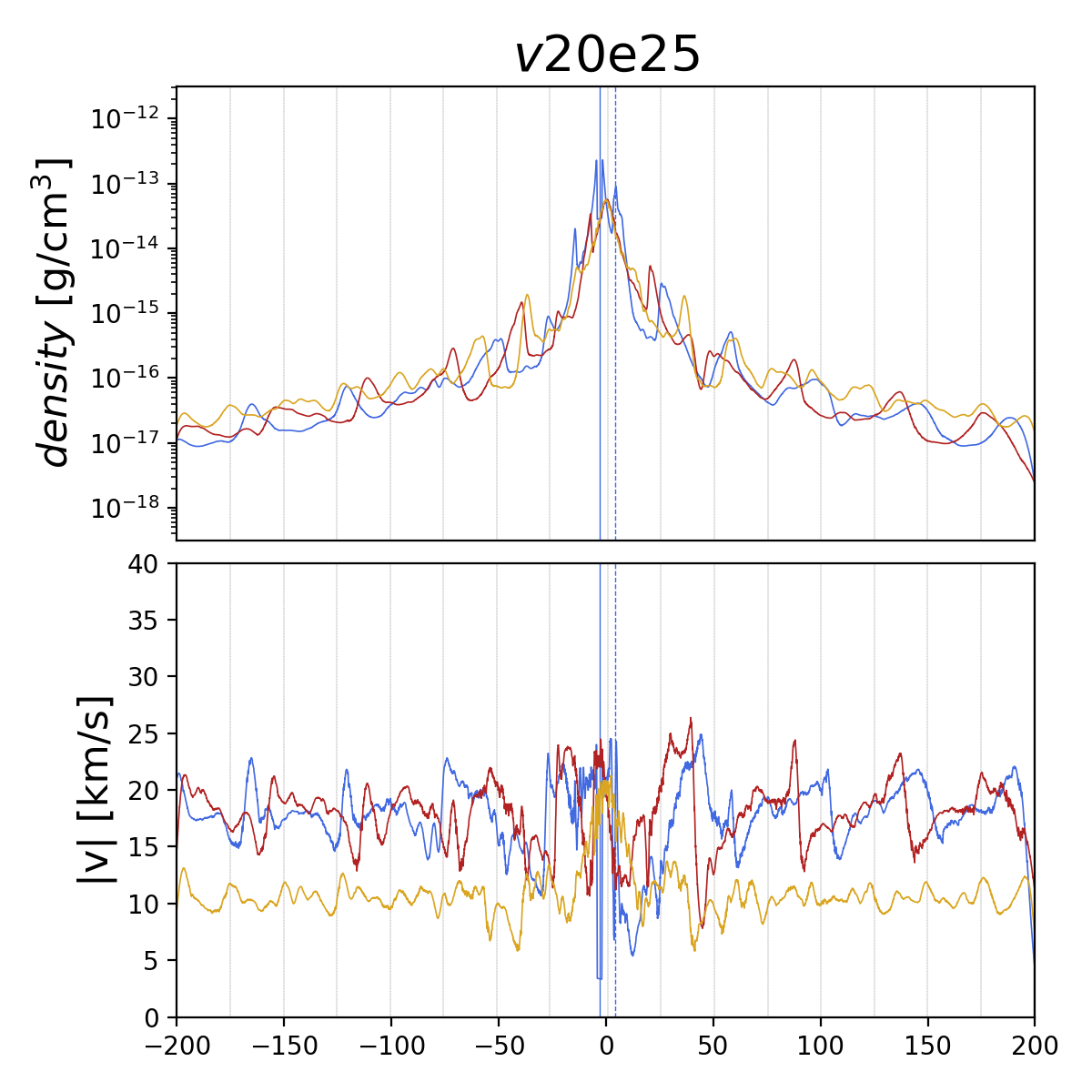}
    \includegraphics[width = 0.2875\textwidth]{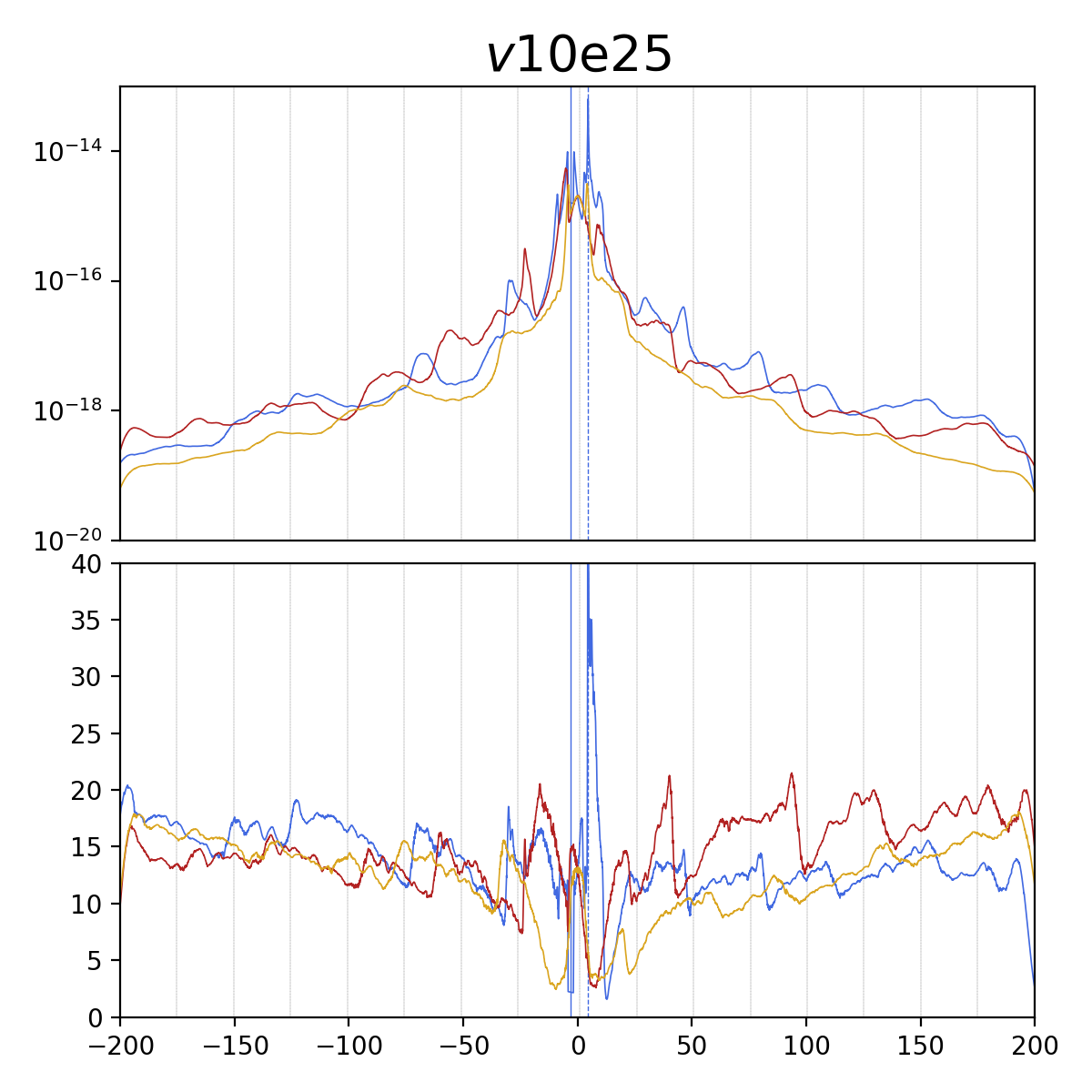}
    \includegraphics[width = 0.2875\textwidth]{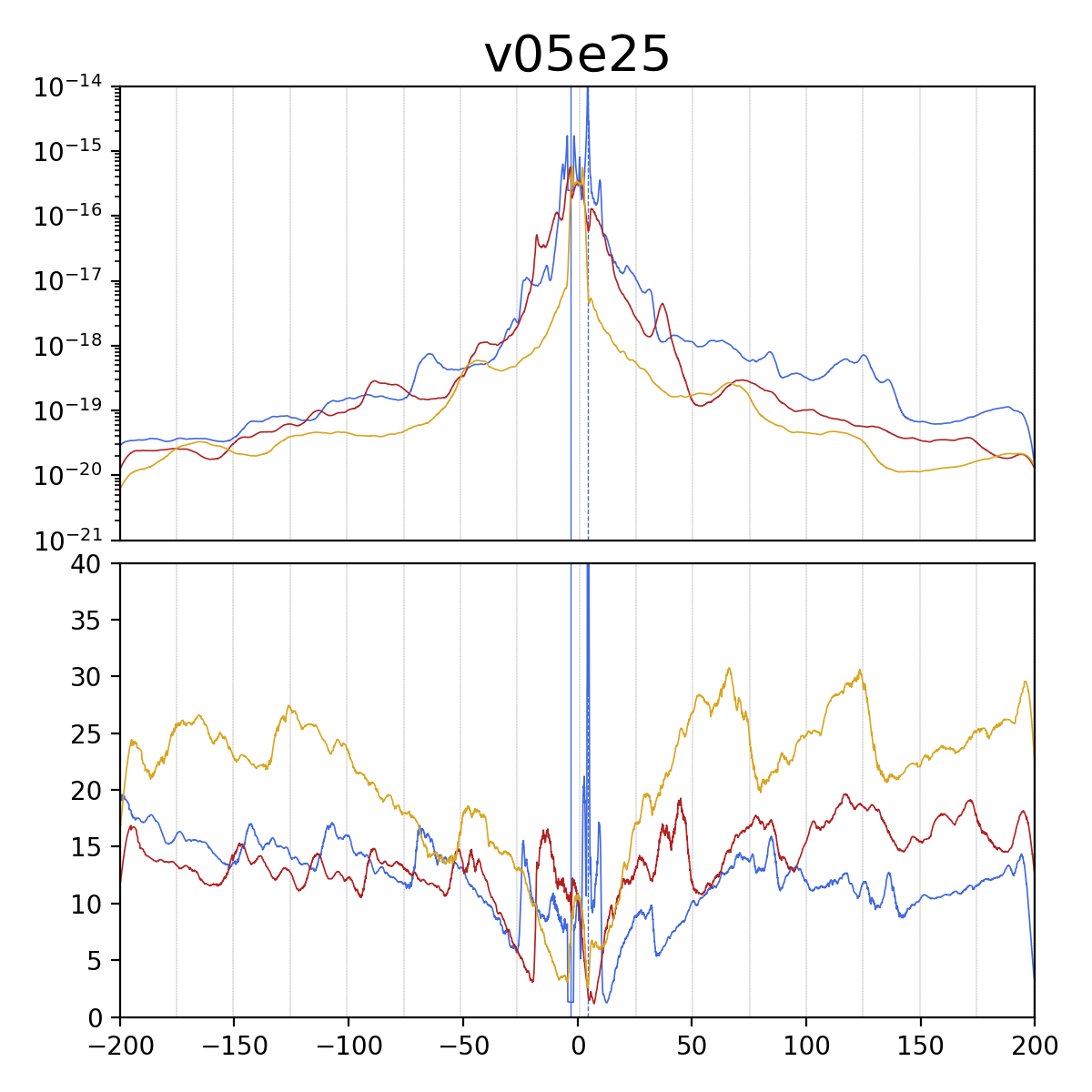}
    \includegraphics[width = 0.2875\textwidth]{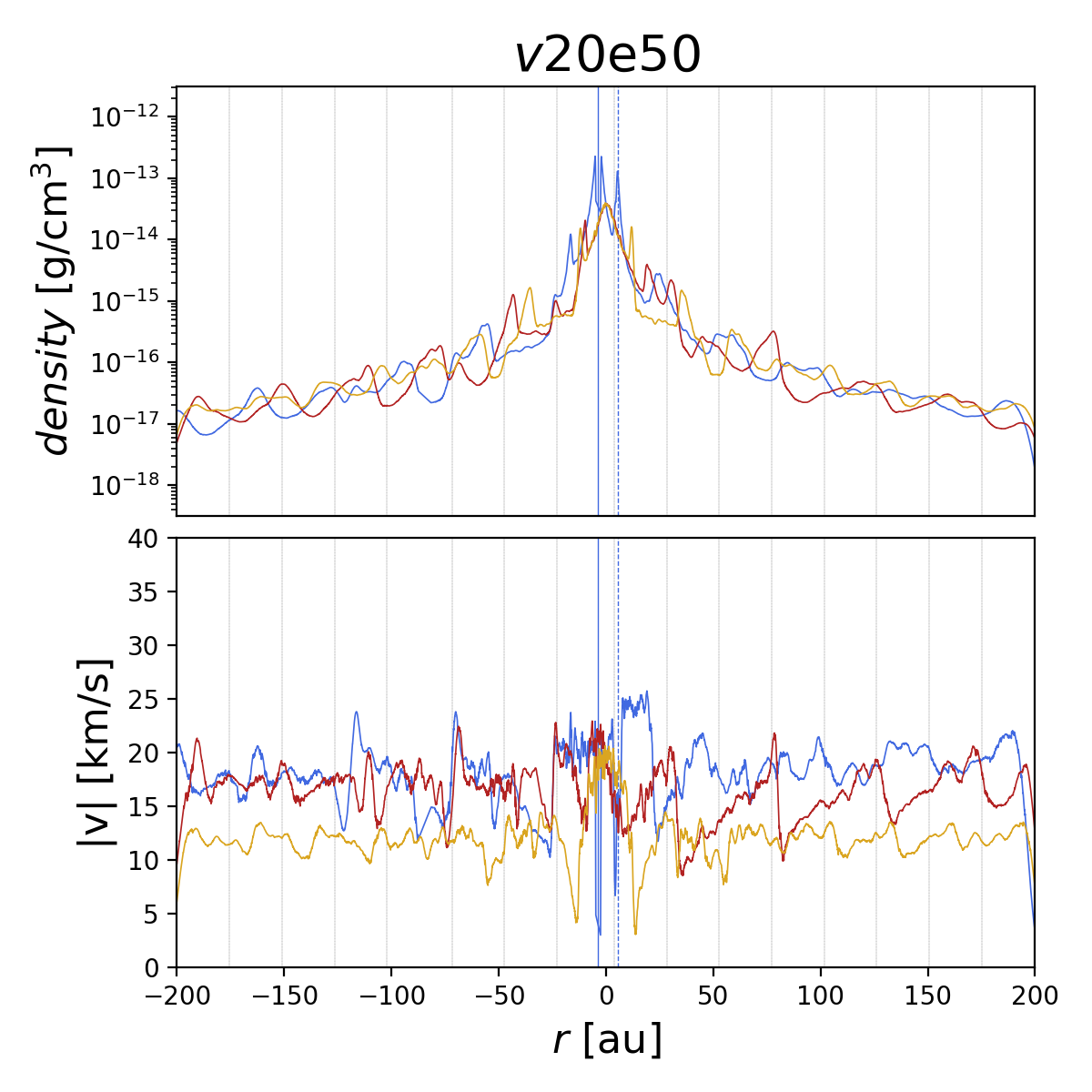}
    \includegraphics[width = 0.2875\textwidth]{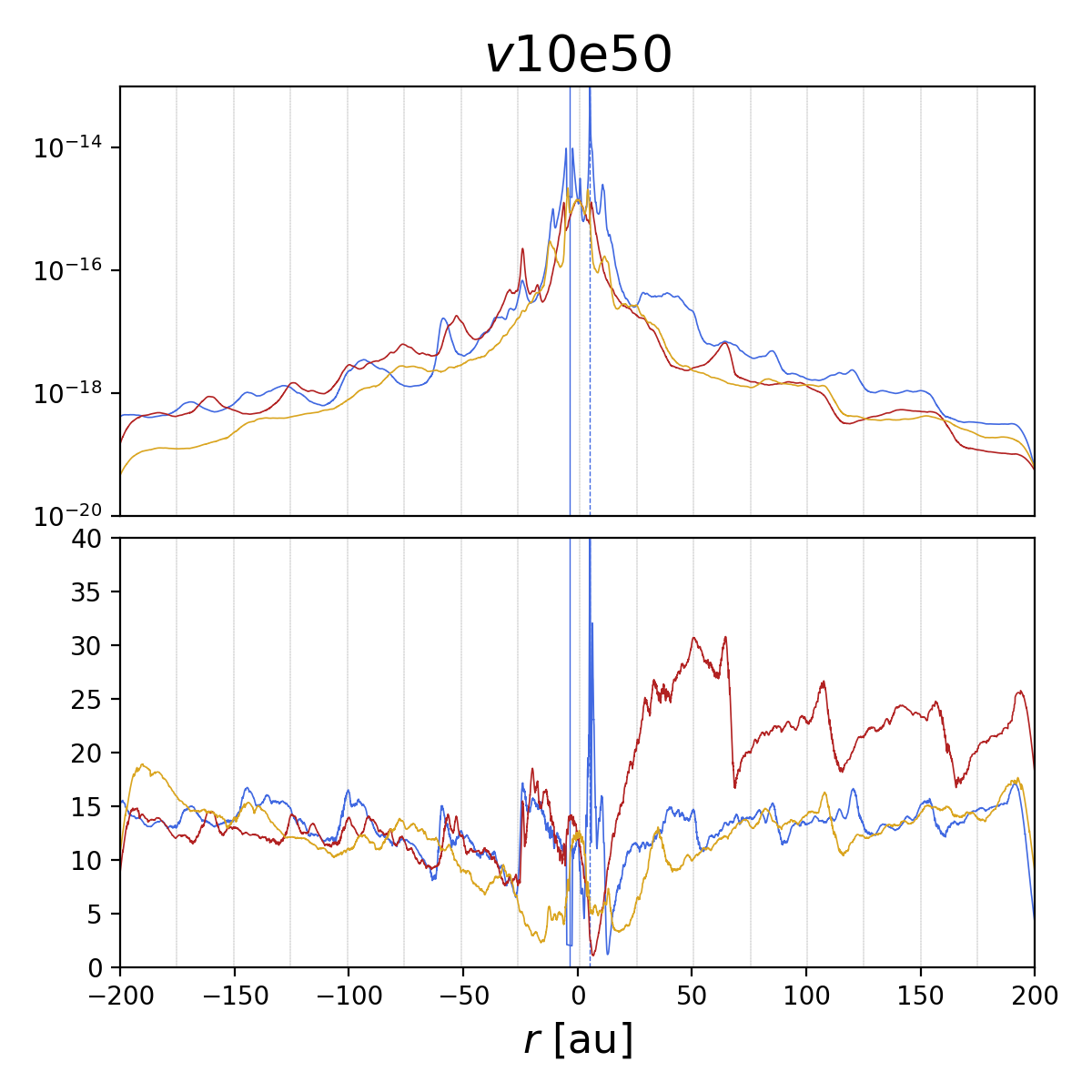}
    \includegraphics[width = 0.2875\textwidth]{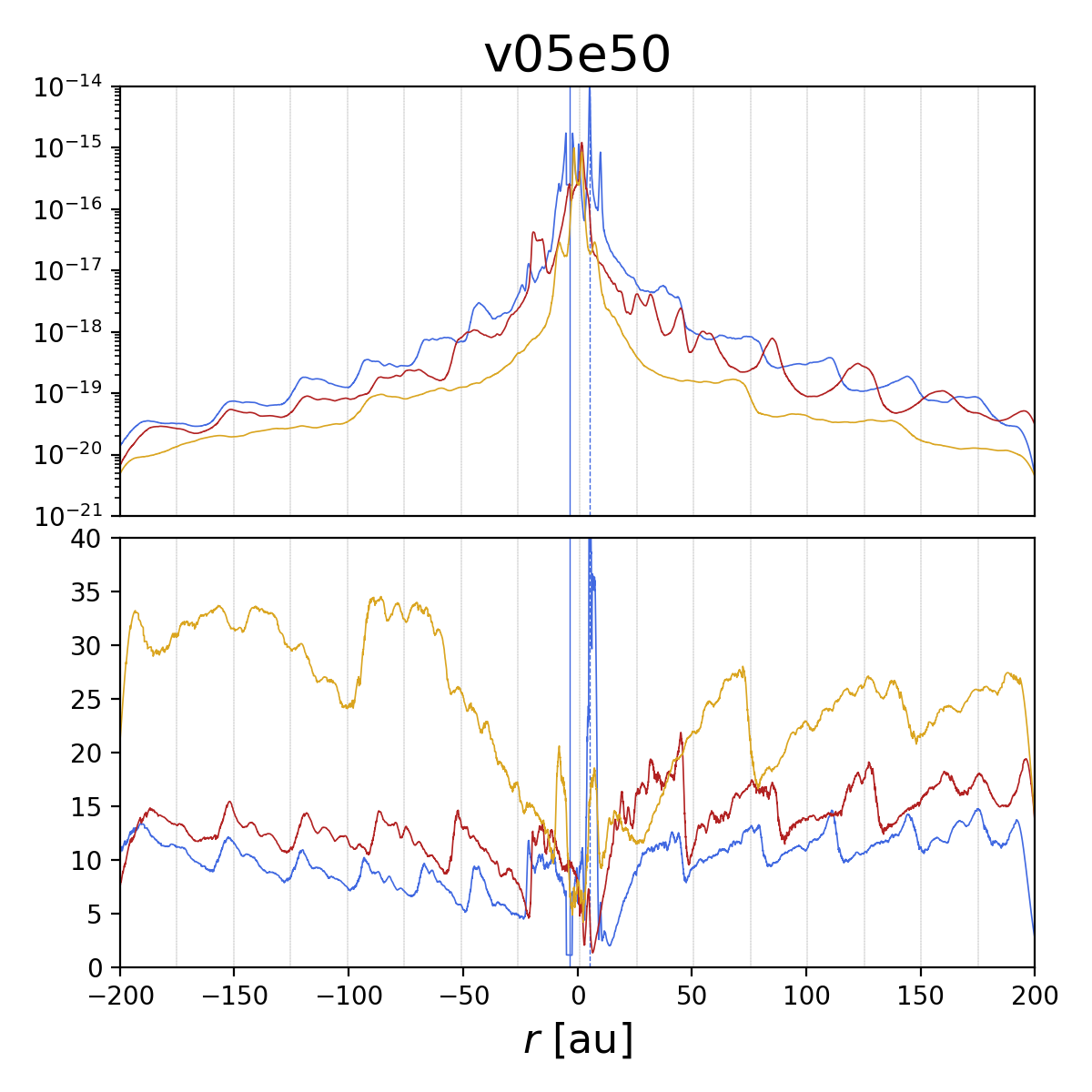}
    \caption{Density and velocity profiles on three perpendicular axes ($x$, $y$ and $z$) through the CoM for models with decreasing wind speed from left to right ($v_{\rm ini}=20\,{\rm{km\,s^{-1}}}$, $v_{\rm ini}=10\,{\rm{km\,s^{-1}}}$, $v_{\rm ini}=5\,{\rm{km\,s^{-1}}}$) and with increasing eccentricity from top to bottom ($e=0.00$, $e=0.25$, $e=0.50$). The solid and dotted vertical lines indicate the position of the AGB stars and of its companion, respectively. The density axes are scaled w.r.t. the input mass-loss rate for an optimal comparison.}
    \label{1Dplots}
\end{figure*}

Fig.~\ref{1Dplots} shows the density and velocity distribution of the models on three perpendicular axis through the CoM: the orbital plane $x$-axis through the two sink particles (with $y=z=0$), the orbital plane $y$-axis ($x=z=0$), and the polar $z$-axis ($x=y=0$). These plots provide a more detailed, quantitative view on the density and velocity distribution in different regions of the 3D models, and thereby provide more precise information on their morphological structures.

The first row of Fig.~\ref{1Dplots} displays the profiles for the circular models v20e00, v10e00 and v05e00. 
For these three systems, the density profiles are approximately symmetric w.r.t. $r=0$ on all axes. 
We first analyse the orbital plane structures, presented by the blue and red lines. 
The spiral structure in the orbital plane manifests itself as peaks on top of the density and velocity profiles that appear first on the $x$-axis and at slightly higher $r$-values reappear on the $y$-axis. After each peak, the density and velocity drop to a relative minimum. In models v20e00 and v10e00 the velocity in the orbital plane peaks and dips appear on top of a constant velocity profile with $v_{\rm w} \approx v_\infty \approx 18\,{\rm{km\,s^{-1}}}$ and $v_{\rm w} \approx v_\infty \approx 16\,{\rm{km\,s^{-1}}}$ respectively, with $v_\infty$ the terminal velocity. In model v05e00, the velocity increases for increasing radius up to $v_\infty \approx 12.5\,{\rm{km\,s^{-1}}}$. 
The inter-arm separation of a spiral is manifested as the distance between two velocity or density peaks on one axis. This separation decreases for lower wind speeds, since in that case the spiral structures travel more slowly in a radially outward direction, covering a smaller distance in one orbital period. 
The density contrast between the spiral structure and the inter-arm region is provided by the difference between the relative maxima and minima of the density profile. These density contrasts are smaller in models with lower wind velocity and gradually fade away at larger radii. 

The yellow lines in Fig.~\ref{1Dplots} represent the density and velocity along the polar axis of the system. In models v20e00 and v10e00, no significant structures appear on top of the unperturbed density and velocity profiles. This indicates that the spiral structure does not reach the polar axis. More structures appear on the polar axis in model v05e00 due to the rose-like shape in the meridional plane.
In model v20e00, the velocity on the polar axis drops to a roughly constant velocity of about $10\,{\rm{km\,s^{-1}}}$, which is about half the velocity in the orbital plane. This reveals that the wind material travels faster in the orbital plane than in the edge-on directions, which could imply that the morphology is flattened (see Sect.~\ref{flatteningSection}).
The wind particles on the polar axis of model v10e00 have extremely low velocities of about $5\,{\rm{km\,s^{-1}}}$ and may therefore lack kinetic energy to escape the system. This model has a remarkably low density on the polar axis compared to the density in the orbital plane. The density enhancement around the orbital plane can refer to the presence of an EDE, on which we elaborate in Sect.~\ref{EDESection}.
The two velocity peaks and density dips that arise at radii close to the outer boundary in model v10e00 are believed to be an artificial impact of the used boundary setup. 
At last, the velocity and density profile on the $z$-axis of model v05e00 are similar to its profile in the orbital plane. This indicates that of the three considered models so far, this model probably has the most spherical, i.e. least flattened, global gas distribution (see Sect.~\ref{flatteningSection}). 

The profiles of the eccentric models reveal more complexities than the circular models.
The density and velocity distribution of models v20e25 and v20e50 are given in the second and third row of the first column of Fig.~\ref{1Dplots}. These plots reveal asymmetric and more irregular structures on all axes compared to the profiles of the circular models.
Moreover, the asymmetry in the meridional plane makes that the spiral structure arises on the polar axis (yellow line), which was not the case for the symmetric circular system v20e00. Similar to model v20e00, the velocity on the polar axis is low compared to the velocity in the orbital plane, indicating that the wind morphology could appear elongated along the orbital plane (Sect.~\ref{flatteningSection}). 

\begin{figure*}[htp!]
    \centering
    \includegraphics[width = \textwidth]{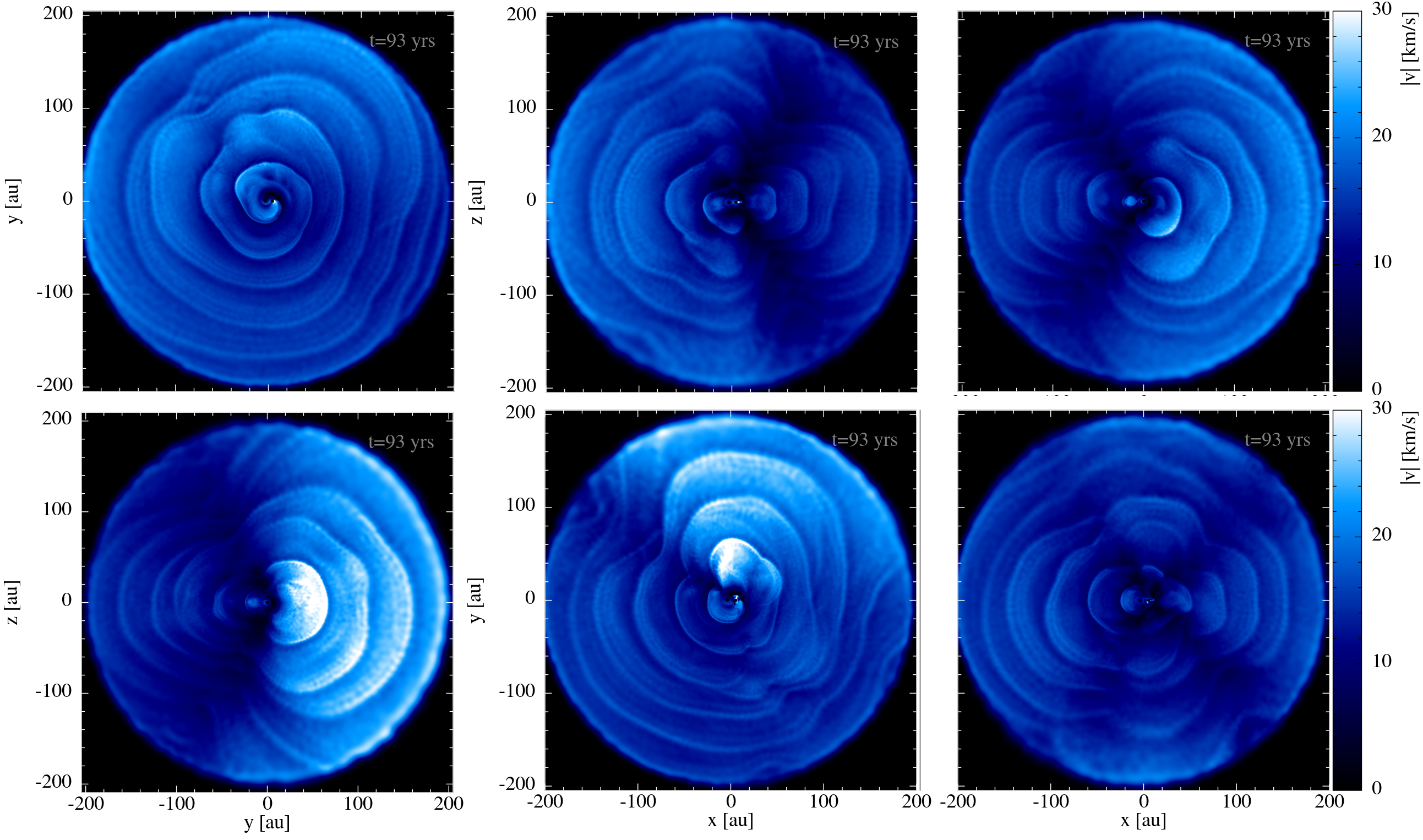}
    \caption{Velocity distribution of models v10e25 (top) and v10e50 (bottom), corresponding to the density distribution plots in Fig. \ref{MorphMedEcc}.}
    \label{VelMedEcc}
\end{figure*}

Models v10e25 and v10e50 are presented in the second and third row of the second column of Fig.~\ref{1Dplots}. These models reveal a stronger asymmetry and irregularity compared to the high wind speed and circular models. The peaks on top of the profiles do not reappear with a similar shape on the $x$- and $y$- axis. The most remarkable feature of these models is the increased velocity on the positive $y$-axis. This high velocity, that reaches up to $30\,{\rm{km\,s^{-1}}}$ in model v10e50, corresponds to energetic outflow in the upper region of the orbital plane (Figs.~\ref{MorphMedEcc} and \ref{VelMedEcc}) of which the origin is described in Sect.~\ref{Sectv10ecc}. The difference in density between the orbital plane and polar regions noticed for model v10e00 is less pronounced in models v10e25 and v10e50, indicating that they may have a less strong EDE (Sect.~\ref{EDESection}).

In the lower velocity models v05e25 and v05e50, which are presented in the second and third row of the last column, high-velocity polar streams are identified. An explanation for the appearance of these fast outflows is provided in Sect.~\ref{sect:v05ecc}. The velocity profiles reveal that the wind velocity of the roughly symmetric polar outflows reaches values up to $30-35\,{\rm{km\,s^{-1}}}$, which is twice the average wind velocity of particles located in the orbital plane. Similar to models v10e25 and v10e50, energetic outflows also appear on the $y>0$ axis. 

\section{Additional morphology slice plots}

\begin{figure*}[htp!]
    \centering
    \includegraphics[width = \textwidth]{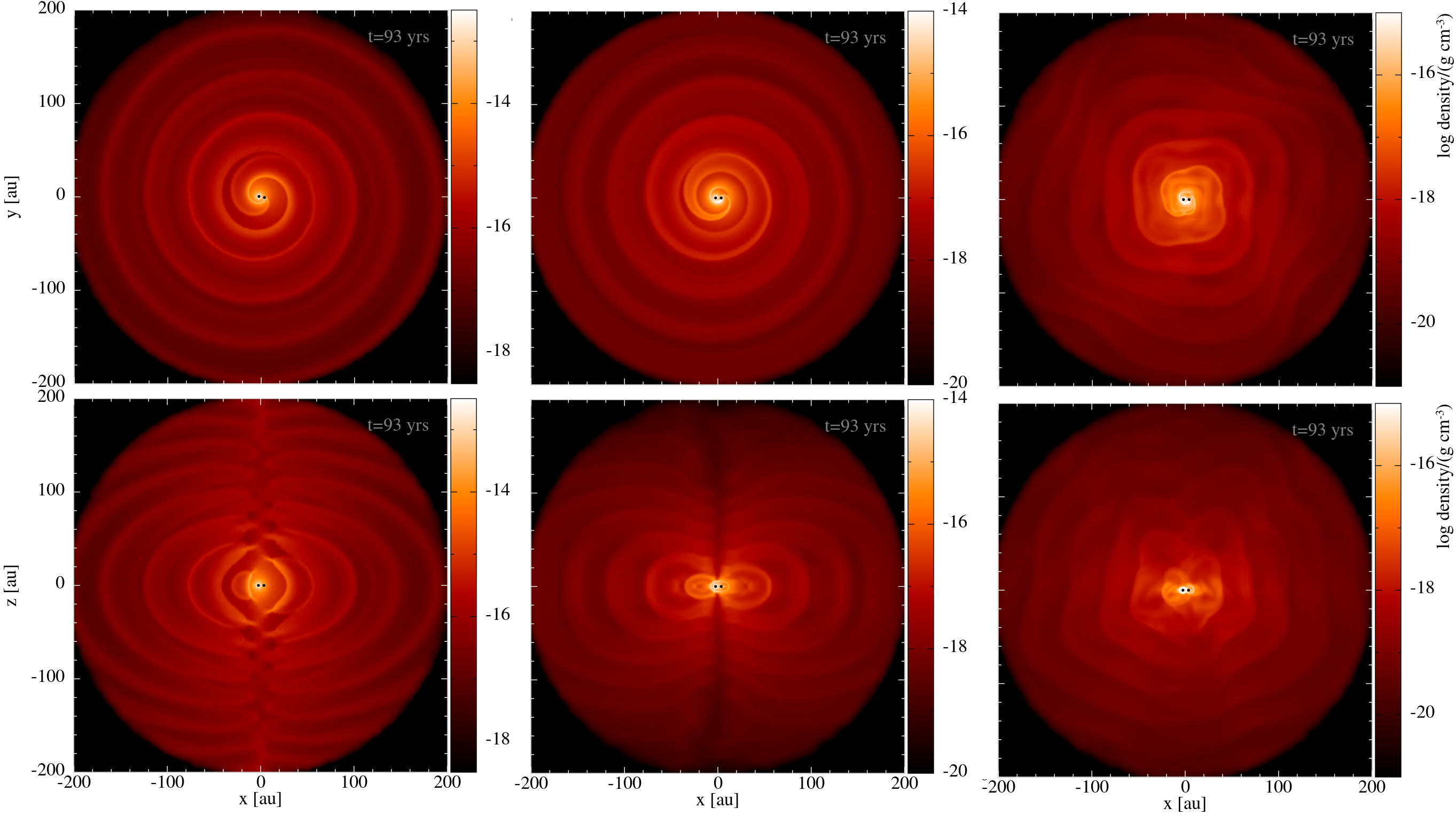}
    \caption{Similar plots as in Fig. \ref{velMorph} of models v20e00 (left), v10e00 (middle) and v05e00 (right) up to the outer boundary.}
    \label{fullMorphCirc}
\end{figure*}

\section{Additional analysis plots}

\begin{figure*}[htp!]
    \centering
    \includegraphics[width = \textwidth]{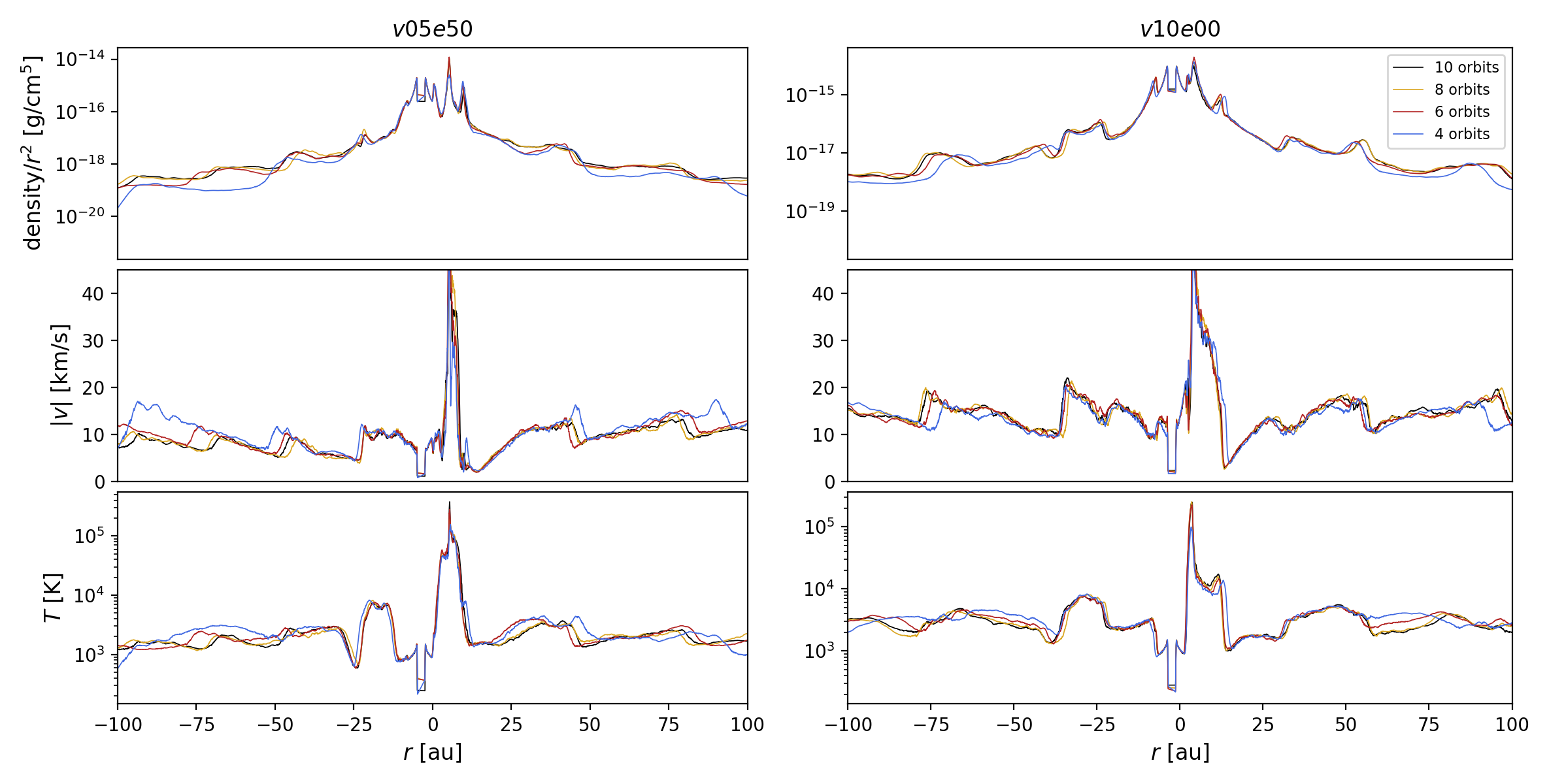}
    \caption{Density, velocity and temperature profile on $x$-axis through the CoM after $4$, $6$, $8$ and $10$ orbital periods for models v05e50 and v10e00, illustrating that self-similarity is reached.}
    \label{selfSimPlot}
\end{figure*}

\begin{figure}[htp!]
    \centering
    \includegraphics[width = 0.4 \textwidth]{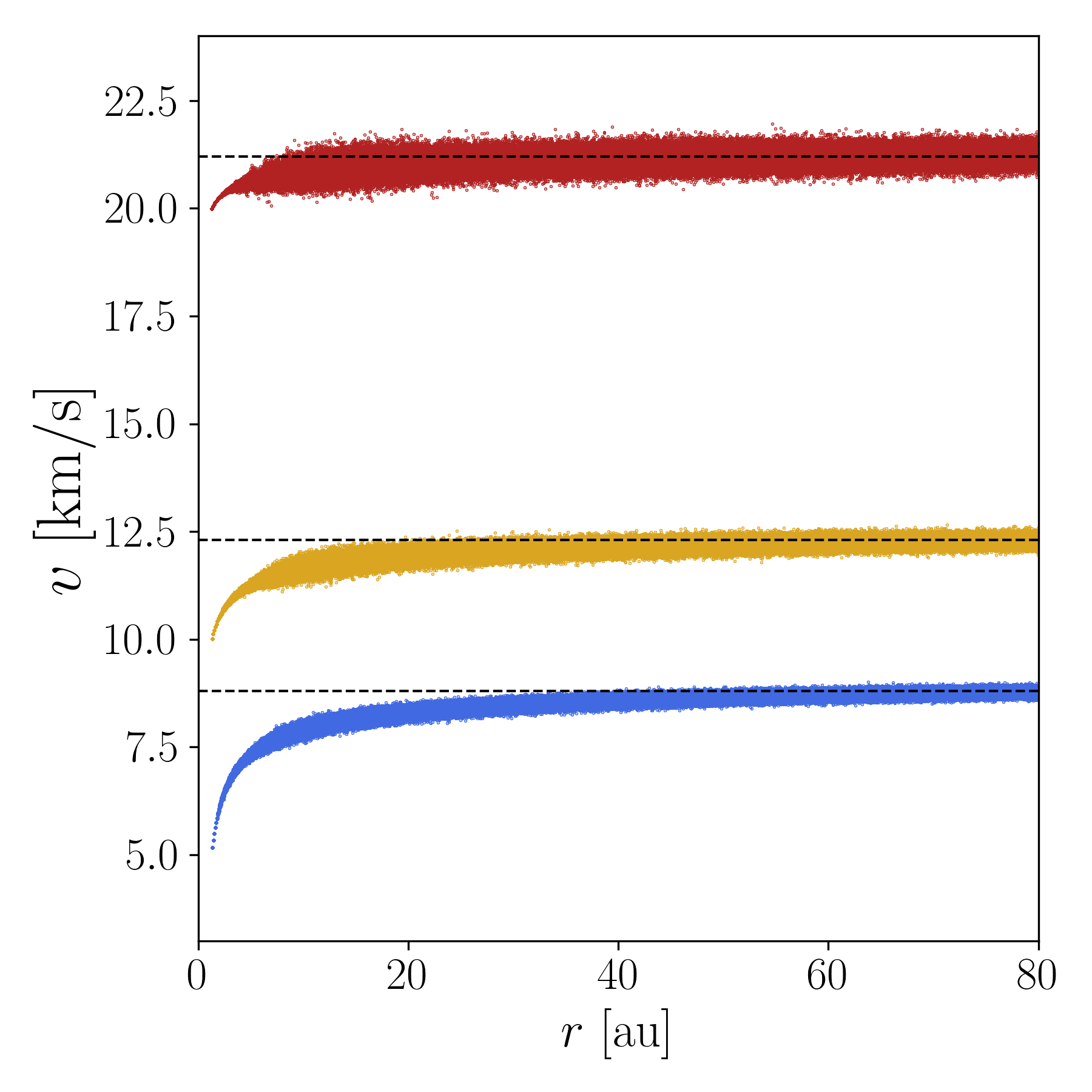}
    \caption{Wind velocity distribution profile in simulations without a companion and input wind velocity of $v_{\rm ini}$ = $5$ km/s (blue), $10$ km/s (yellow) and $20$ km/s (red). The estimated terminal velocities $v_\infty$ are indicated by horizontal dashed black lines. }
    \label{velProfileSS}
\end{figure}

\begin{figure*}[htp!]
    \centering
    \includegraphics[width = 0.2875\textwidth]{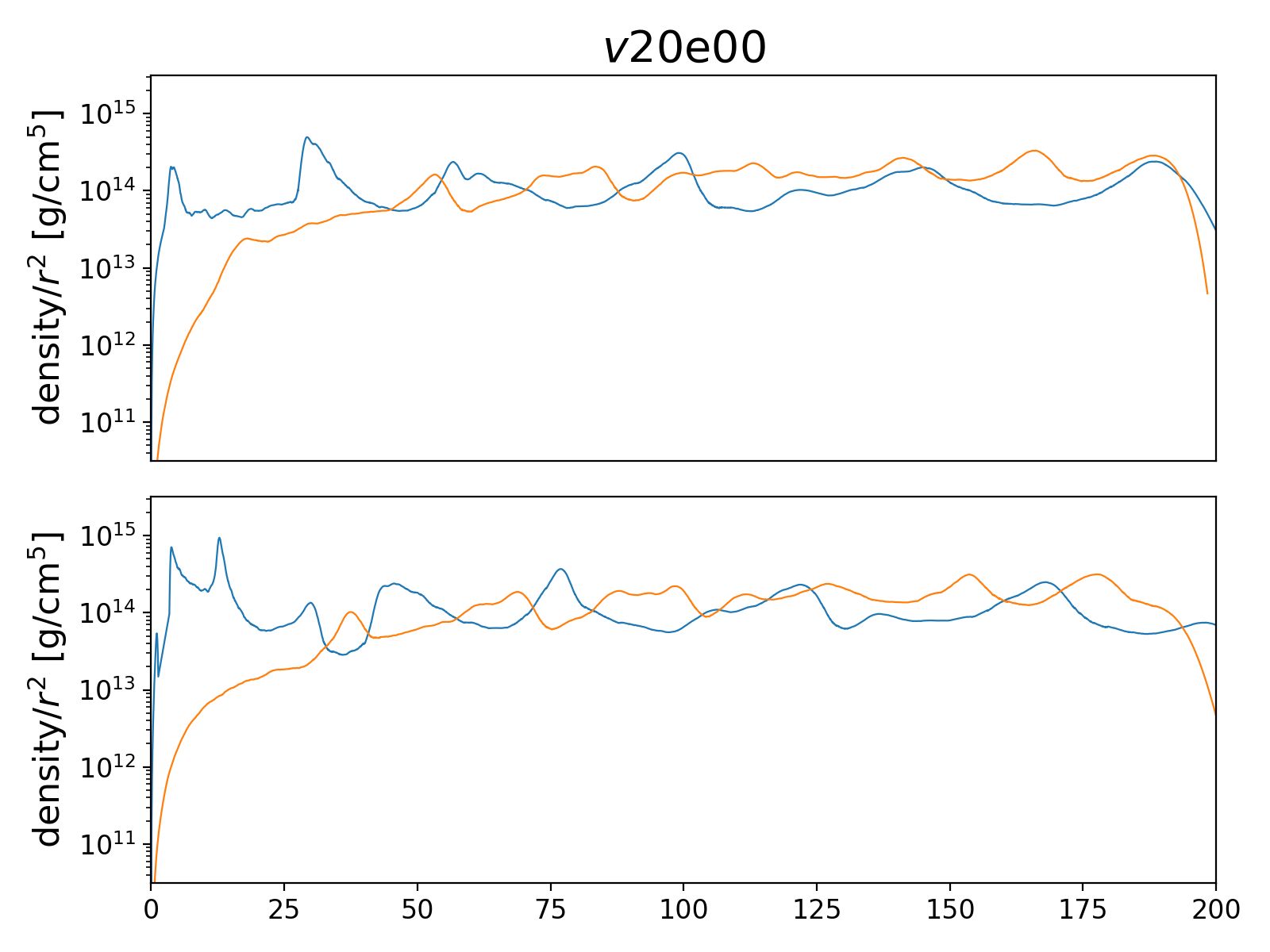}
    \includegraphics[width = 0.2875\textwidth]{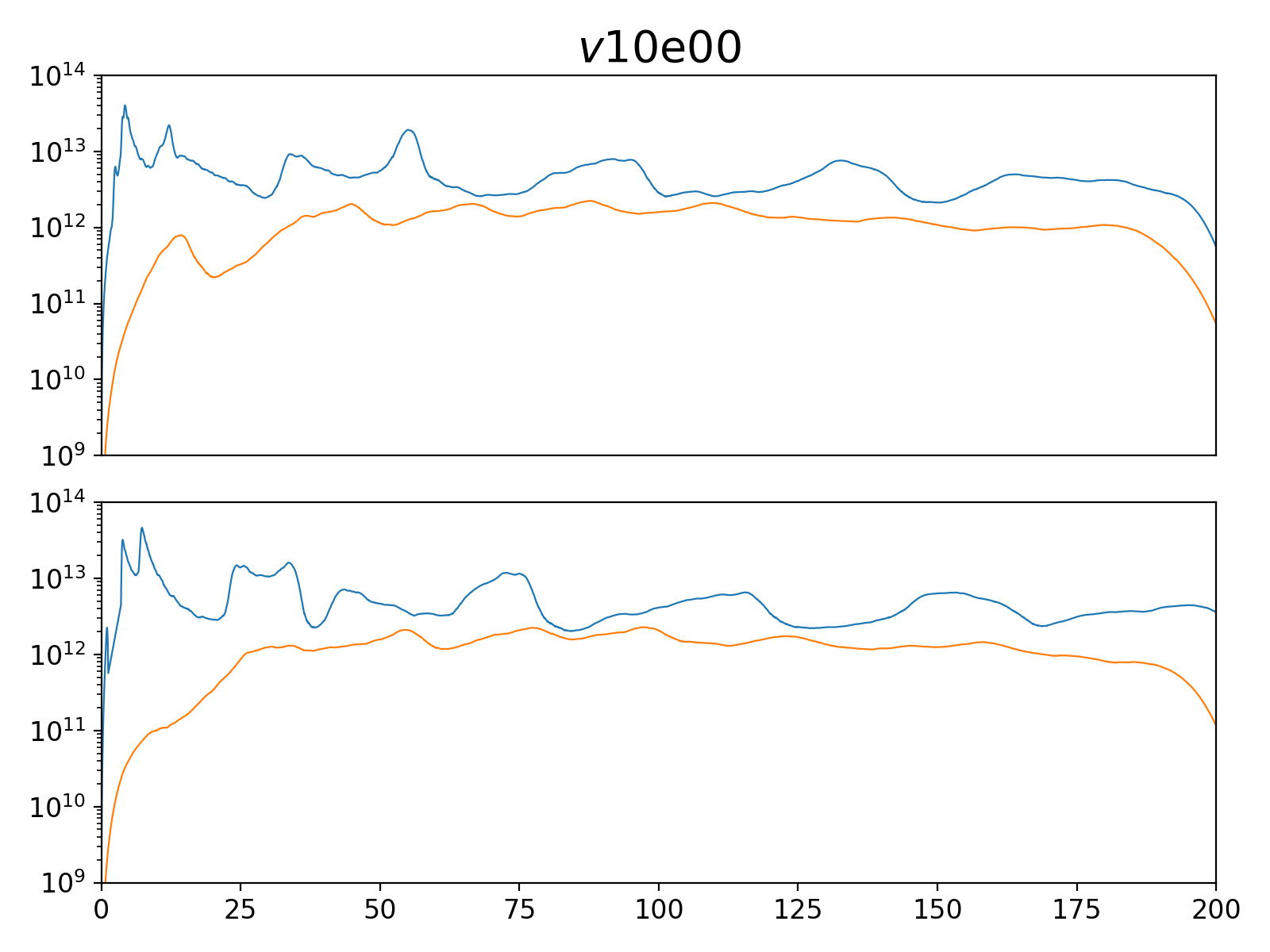}
    \includegraphics[width = 0.2875\textwidth]{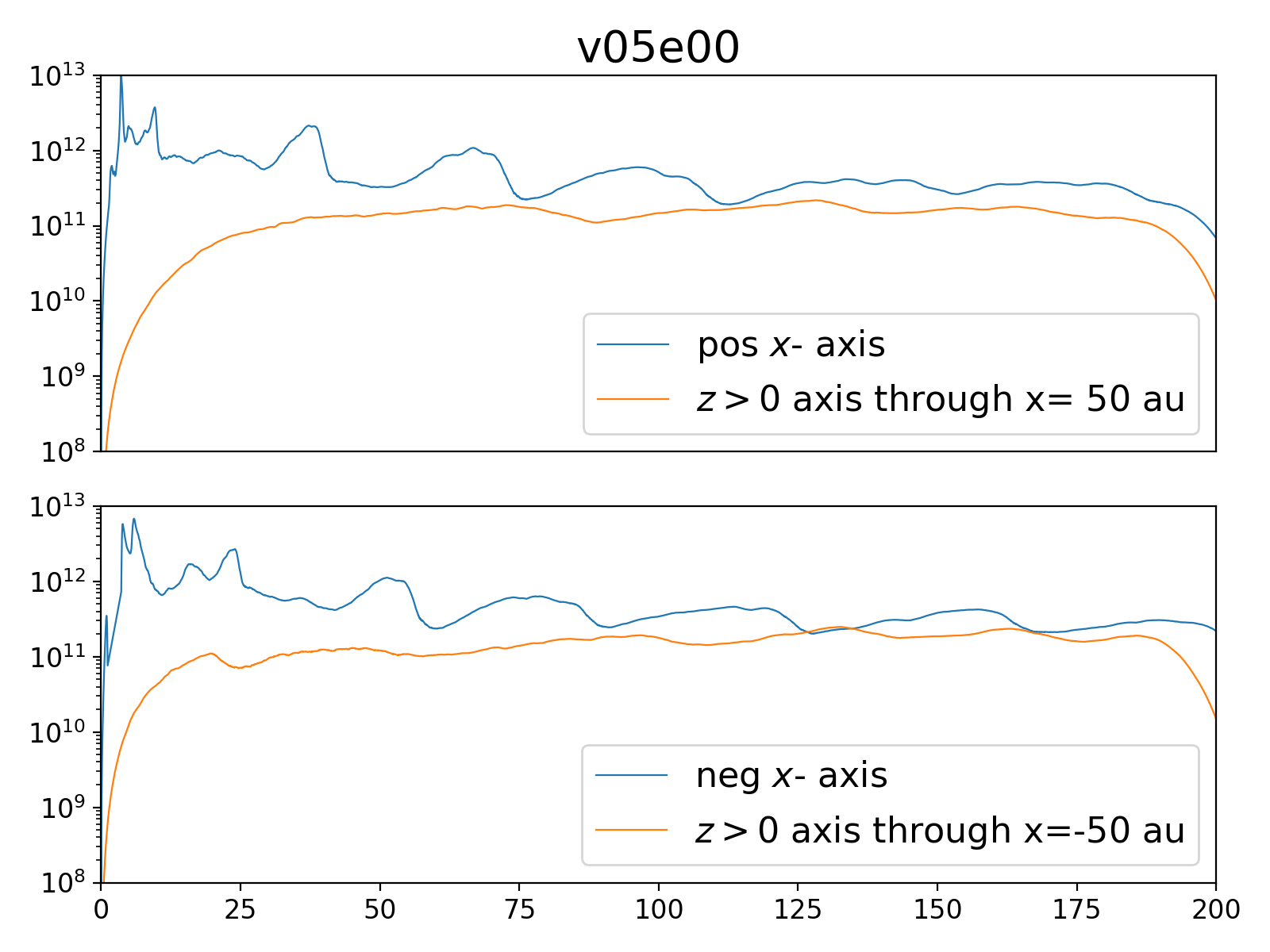}
    \includegraphics[width = 0.2875\textwidth]{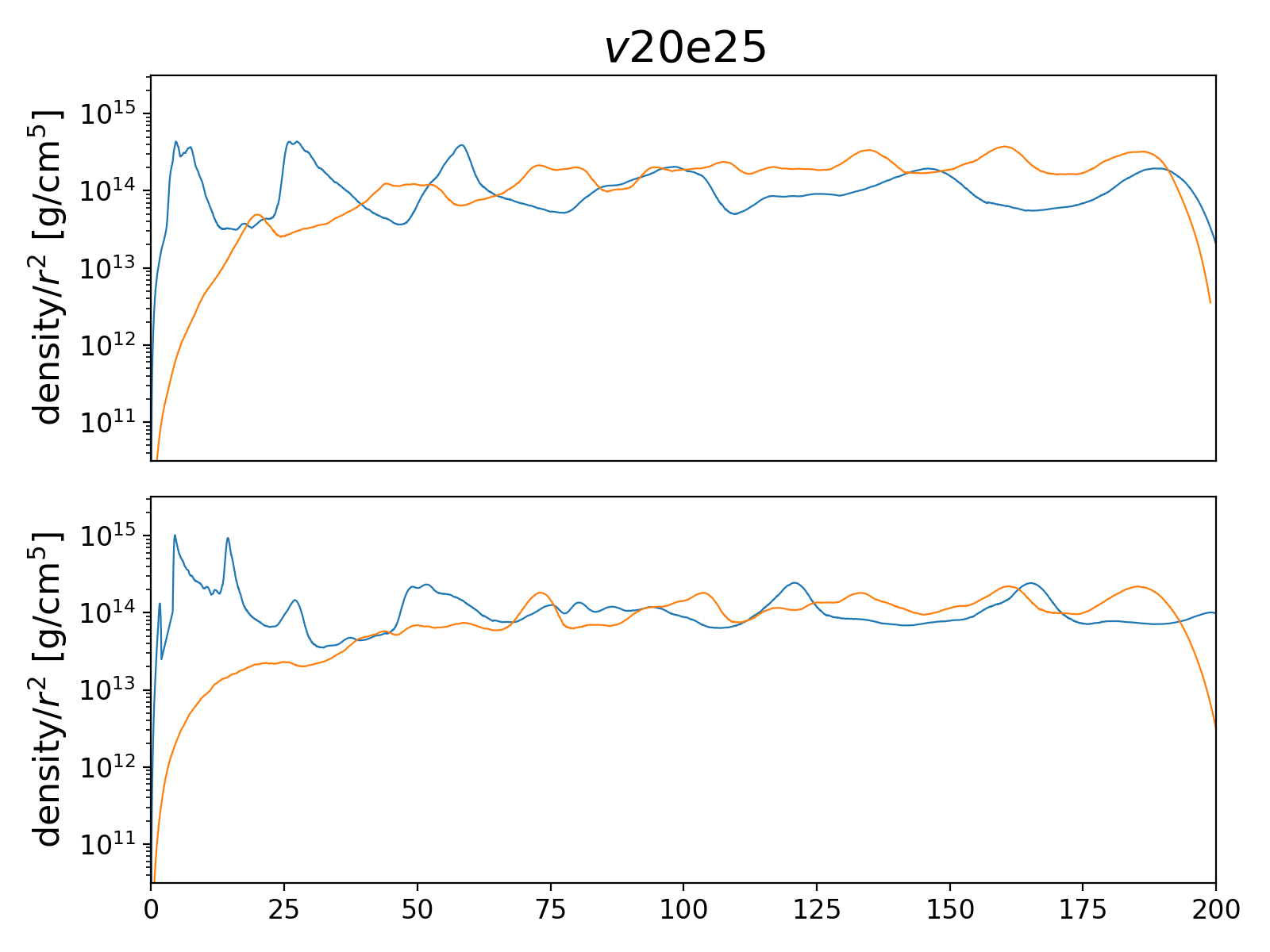}
    \includegraphics[width = 0.2875\textwidth]{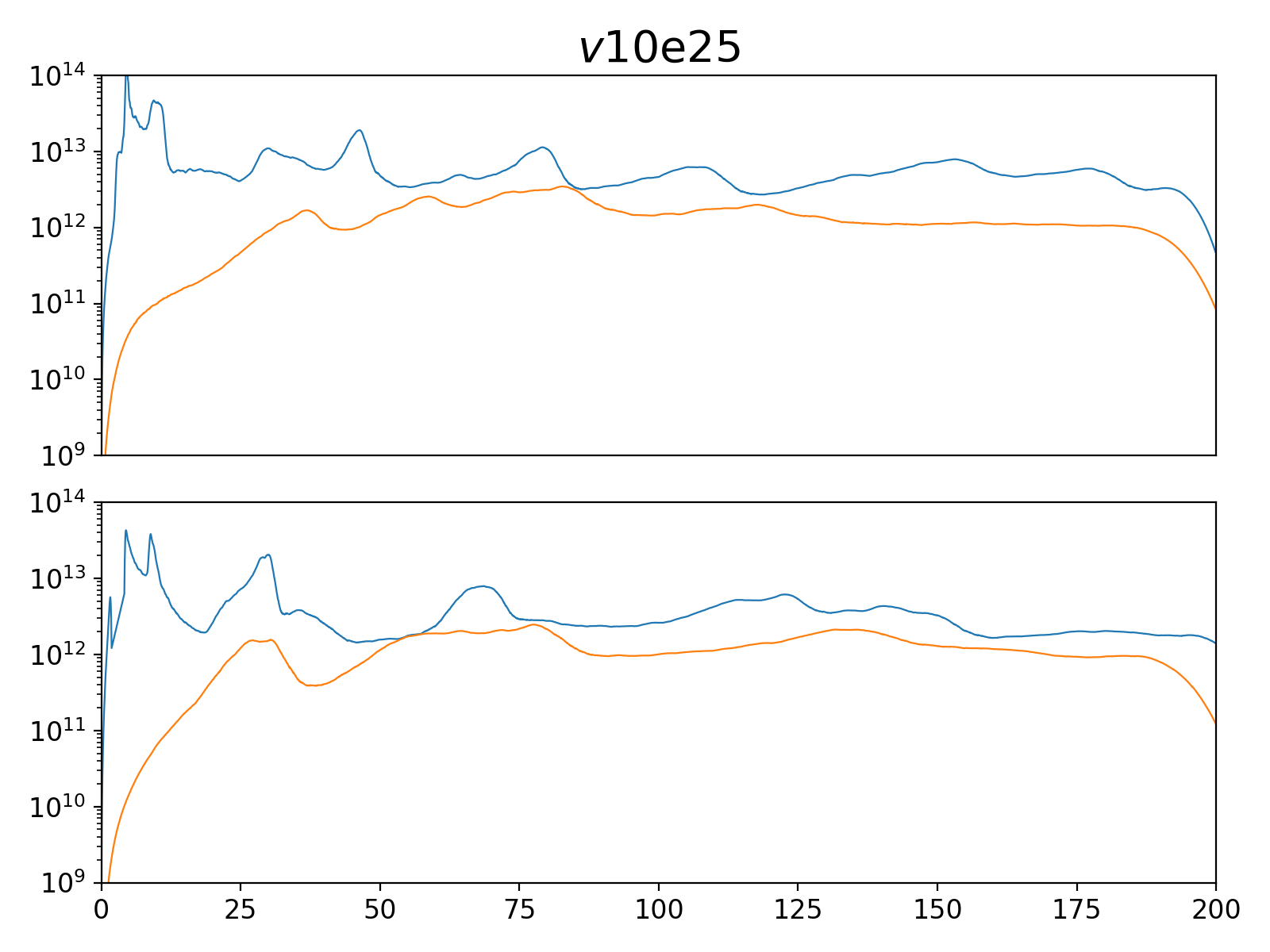}
    \includegraphics[width = 0.2875\textwidth]{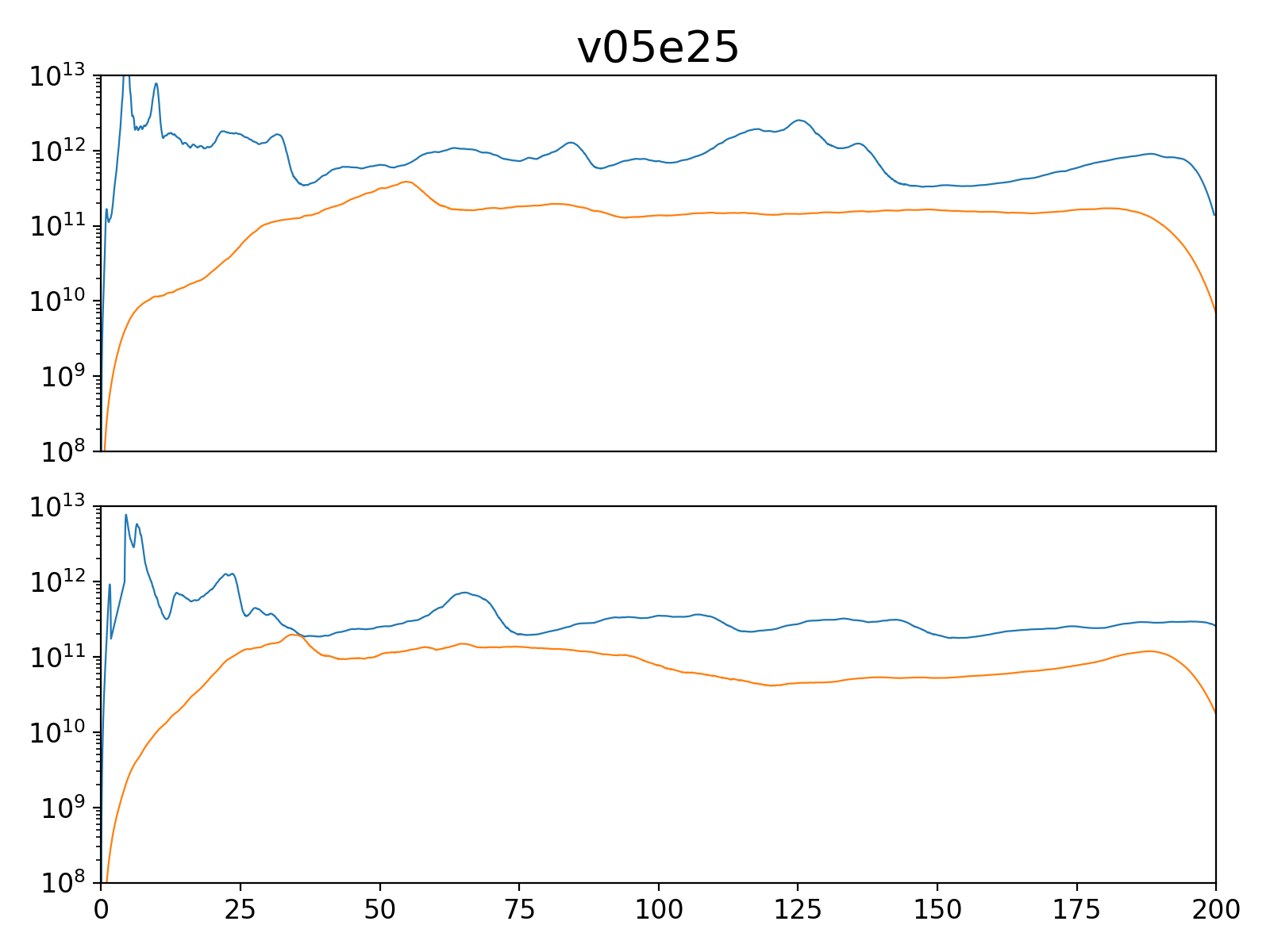}
    \includegraphics[width = 0.2875\textwidth]{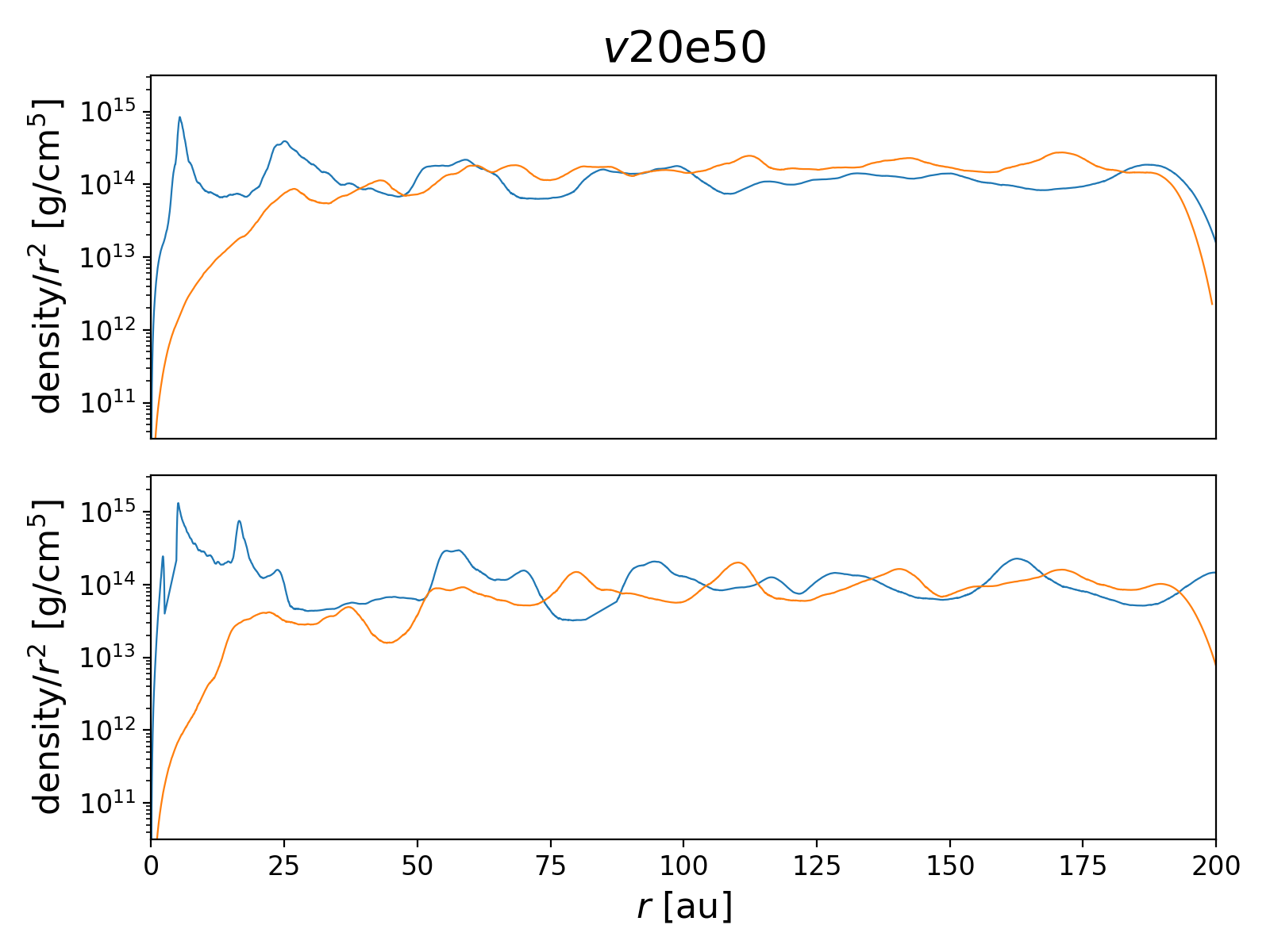}
    \includegraphics[width = 0.2875\textwidth]{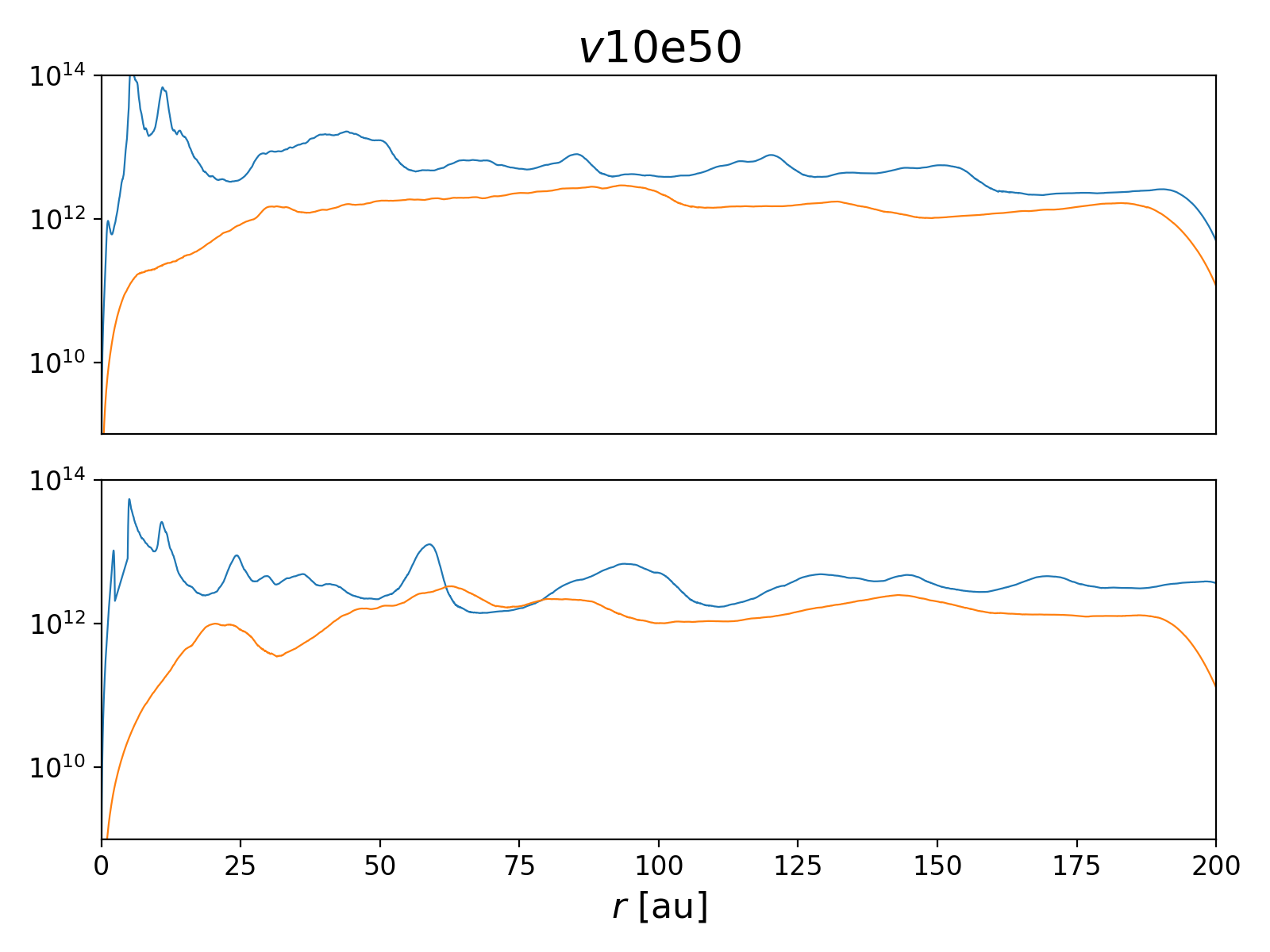}
    \includegraphics[width = 0.2875\textwidth]{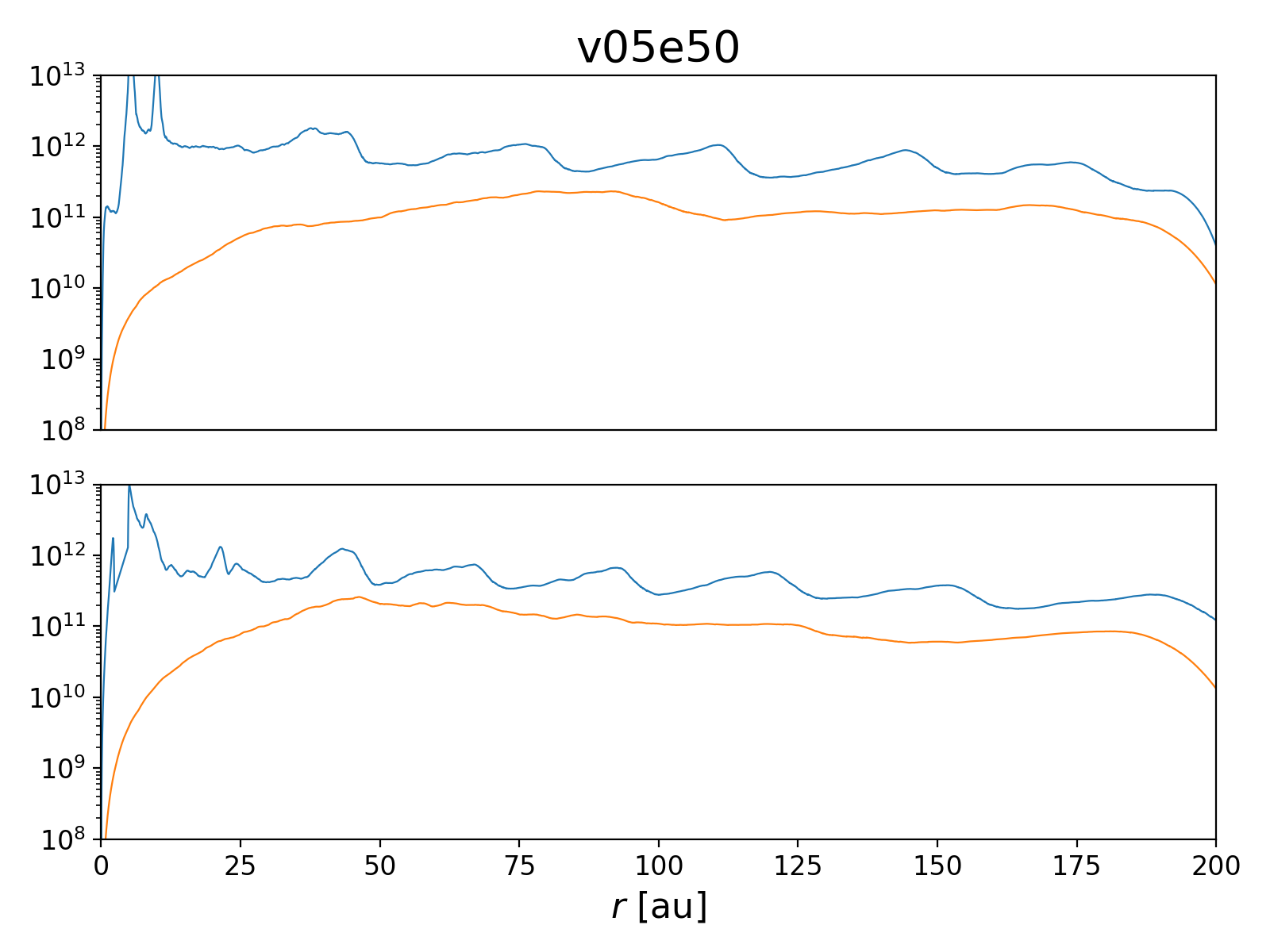}
    \caption{Residual density profiles on the $x$- axis through the CoM and on an axis parallel to the polar axis through $x=\pm 50 \, {\rm au}$ for models with decreasing wind velocity from left to right and with increasing eccentricity from top to bottom. Only the $z>0$ side of the axis through $x=\pm 50 \, {\rm au}$ is selected, because of the symmetry of the morphology with respect to the orbital plane. }
    \label{1DFlratioPlots}
\end{figure*}